\documentclass[aps,onecolumn,amsmath,showpacs,superscriptaddress,floatfix,nofootinbib,nopreprintnumbers]{revtex4-2}

\usepackage{verbatim}
\usepackage[T1]{fontenc}
\usepackage[utf8]{inputenc}
\usepackage[american]{babel}
\usepackage{epsfig}
\usepackage{graphicx}
\usepackage{booktabs}
\usepackage{multirow}
\usepackage{dcolumn}
\usepackage{amsmath}
\usepackage{amssymb}
\usepackage{mathtools}
\usepackage{amsfonts}
\usepackage{amssymb}
\usepackage{epstopdf}
\usepackage{bm}
\usepackage{siunitx}
\usepackage{braket}
\usepackage{enumitem}
\usepackage{soul}
\usepackage[table]{xcolor}
\usepackage{xcolor}
\usepackage{color}
\usepackage{transparent}
\usepackage{pifont}
\usepackage[normalem]{ulem}

\definecolor{navy}{HTML}{000080}
\definecolor{dodgerblue}{HTML}{1E90FF}

\usepackage[bookmarks=true,bookmarksnumbered=true,colorlinks=true,urlcolor=black, citecolor=black,linkcolor=black,breaklinks=true]{hyperref}

\hypersetup{
  colorlinks  = true,
  urlcolor    = navy,
  linkcolor   = navy,
  citecolor   = dodgerblue
}
\usepackage[capitalize]{cleveref}

\usepackage{booktabs}
\usepackage{multirow}
\usepackage{dcolumn}
\usepackage{colortbl}

\newcommand{\orcid}[1]{\begingroup
  \hypersetup{hidelinks}\href{https://orcid.org/#1}{\includegraphics[width=10pt]{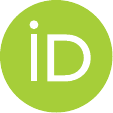}} \endgroup}

\begin{document}

\title{Dark energy and neutrinos along the cosmic expansion history}

\author{Pietro Ghedini\orcid{0009-0001-4861-4867}}
\email{pietro.ghedini@ific.uv.es}

\affiliation{Departament de F\'{i}sica Te\`{o}rica, Universitat de València, 46100 Burjassot, Spain}
\affiliation{Instituto de F\'{i}sica Corpuscular (IFIC), CSIC‐Universitat de València, Spain}

\author{Rasmi Hajjar\orcid{0000-0002-9227-5364}}
\email{hajjar.44@osu.edu}

\affiliation{Center for Cosmology and AstroParticle Physics (CCAPP), \href{https://ror.org/00rs6vg23}{\color{black}Ohio State University}, Columbus, OH 43210}
\affiliation{Department of Physics, \href{https://ror.org/00rs6vg23}{\color{black}Ohio State University}, Columbus, OH 43210}
\affiliation{Department of Astronomy, \href{https://ror.org/00rs6vg23}{\color{black}Ohio State University}, Columbus, OH 43210}

\author{Olga Mena\orcid{0000-0001-5225-975X}}
\email{omena@ific.uv.es}

\affiliation{Instituto de F\'{i}sica Corpuscular (IFIC), CSIC‐Universitat de València, Spain}

\date{\today}
\preprint{}

\begin{abstract}
Recent cosmological measurements are hinting that dark energy may evolve, with its equation of state, $w_\mathrm{DE}$, even showing oscillatory patterns. In this work, we employ a model-independent approach to jointly reconstruct $w_\mathrm{DE}$ and the sum of neutrino masses, $\sum m_\nu$, adopting the PCHIP method with seven fixed nodes in which we allow the two parameters to vary.
We employ CMB, Baryon Acoustic Oscillations and Supernovae Ia data to constrain the values of $w_\mathrm{DE}$ and $\sum m_\nu$ at each node. We conduct three different analyses in which we reconstruct $w_\mathrm{DE}$: one with fixed $\sum  m_\nu=0.06~\mathrm{eV}$; one in which we allow $\sum m_\nu$ to vary, and one in which we also reconstruct $\sum m_\nu$ using the PCHIP method. We find the dark energy equation of state to be consistent with the cosmological constant scenario, except when including DESI data and allowing for phantom crossing, where we find a $95\%$ CL deviation from $w_\mathrm{DE}=-1$ around $z\sim1.2$. For neutrino masses, we obtain looser constraints when focusing on phantom dark energy, that show further early and late relaxation when reconstructing the mass via the PCHIP method.
\end{abstract}

\maketitle

\section{Introduction}
\label{sec:intro}

The main current contribution to the energy density budget of the Universe is given by dark energy, parametrized by a cosmological constant, $\Lambda$, in the concordance $\Lambda$CDM cosmological model. Despite numerous observational evidences for the accelerated expansion of the Universe~\cite{SupernovaSearchTeam:1998fmf,SupernovaCosmologyProject:1998vns}, the nature of dark energy remains one of the main open questions in modern cosmology. From a quantum field theory perspective, it is difficult to understand the value of $\Lambda$, finding a discrepancy of $\sim$120 orders of magnitude between the predicted and the observed value (this is commonly known as the cosmological constant problem~\cite{Weinberg:1988cp,Martin:2012bt}). Additional theoretical issues challenge this particular dark energy interpretation (see, e.g.,~\cite{Silvestri:2009hh} and references therein), motivating the exploration of alternative models, ranging from modifications of gravity~\cite{Carroll:2003wy,Deffayet:2001pu,Dvali:2000hr,Freese:2002sq,Carroll:2006jn,Capozziello:2003tk,Deffayet:2000uy,Dvali:2003rk,Vollick:2003aw,Flanagan:2003rb,Flanagan:2003iw,Vollick:2003ic,Soussa:2003re,Nojiri:2003ni,Arkani-Hamed:2003pdi,Gabadadze:2003ck,Moffat:2004nw,Carroll:2004de,Clifton:2004st,Easson:2005ax,delaCruz-Dombriz:2006kob,Artola:2025fup, Capozziello:2024qol,Linder:2025pqt,Capozziello:2010uv,Capozziello:2005ku,Capozziello:2011et,Joyce:2016vqv,Bouhmadi-Lopez:2025lzm} to the addition of new interactions within the dark sectors~\cite{Bolotin_2015, Wang_2016,Ghedini:2024mdu,Amendola:1999er,Amendola:2003eq,Cai:2004dk,Huey:2004qv,Gumjudpai:2005ry,Brookfield:2005td,Berger:2006db,delCampo:2006vv,Barrow:2006hia,delCampo:2008sr,Valiviita:2008iv,Chongchitnan:2008ry,Gavela:2009cy,Xia:2009zzb,Gavela:2010tm,Mena:2010zz,Baldi:2010td,LopezHonorez:2010esq,Lee:2011tq,Beynon:2011hw,Amendola:2011ie,Farajollahi:2012zz,Pettorino:2012ts,Chimento:2012aea,Salvatelli:2013wra,Pettorino:2013oxa,Xia:2013nua,Li:2013bya,Yang:2014okp,Yang:2014gza,Salvatelli:2014zta,Wang:2014xca,Li:2014eba,Pan:2012ki,Li:2015vla,Cui:2015ueu,Landim:2015uda,Yang:2016evp,Wang:2016lxa,Nunes:2016dlj,vandeBruck:2016hpz,Xia:2016vnp,Pan:2016ngu,Fay:2016yow,Kumar:2017dnp,DiValentino:2017iww,Sharov:2017iue,Yang:2017yme,Yang:2017ccc,Pan:2017ent,Mifsud:2017fsy,VanDeBruck:2017mua,Yang:2017zjs,Yang:2018euj,Yang:2018xlt,Li:2018jiu,Barros:2019rdv,Teixeira:2019tfi,DiValentino:2019jae,Paliathanasis:2019hbi,Yang:2019uog,Pan:2019gop,Nakamura:2019phn,DiValentino:2019ffd,Yang:2019bpr,Kumar:2019wfs,Martinelli:2019dau,Costa:2019uvk,Arevalo:2019axj,Pan:2019jqh,Cheng:2019bkh,Oikonomou:2019nmm,Kase:2019veo,Pan:2020zza,Gomez-Valent:2020mqn,DiValentino:2020leo,Yang:2020uga,Lucca:2020zjb,DiValentino:2020kpf,Hogg:2020rdp,Yao:2020hkw,BeltranJimenez:2020qdu,Yao:2020pji,Lucca:2021dxo,Lucca:2021eqy,Nunes:2021zzi,Gariazzo:2021qtg,Potting:2021bje,Sa:2021eft,daFonseca:2021imp,Gao:2021xnk,Thipaksorn:2022yul,Gomez-Valent:2022bku,Harko:2022unn,Yengejeh:2022tpa,Nunes:2022bhn,Landim:2022jgr,Yao:2022kub,Li:2023gtu,Giare:2024ytc,Kritpetch:2024rgi,Giare:2024smz,Li:2024qso,Halder:2024gag,Carrion:2024jur,Giani:2024nnv,Tsedrik:2025cwc,Liu:2025pxy,Zhai:2025hfi,Li:2025owk,Silva:2025hxw,Chakraborty:2025syu,You:2025uon,Aoki:2025bmj,Yang:2025ume,Abedin:2025yru,Pan:2025qwy,vanderWesthuizen:2025iam,Liu:2025vda,Yang:2025boq,Paliathanasis:2025xxm,Li:2025ula,Yan:2025iga,Wang:2025znm} (for a more general review of dark energy models, see~\cite{Ghosh:2024ojq}).

Recent data releases from the Dark Energy Spectroscopic Instrument (DESI) collaboration~\cite{DESI:2024mwx, DESI:2025zgx} suggest that dark energy may be dynamical rather than a simple cosmological constant, $\Lambda$~\cite{DESI:2025wyn,DESI:2025fii}, with the second data release moving further away from $\Lambda$. In such scenarios, the dark energy equation of state parameter, $w_\mathrm{DE}=p_\mathrm{DE}/\rho_\mathrm{DE}$, is a function of the scale factor, $a$, in contrast to the $\Lambda$CDM cosmological model, where the equation of state is constant in time $w_\mathrm{\Lambda}=-1$.

The absence of a fundamental theoretical principle determining the functional form of $w_\mathrm{DE}$ in dynamical dark energy models has led to the development of several parameterizations in the literature~\cite{Efstathiou:1999tm, Barboza:2008rh, Dimakis:2016mip, Pan:2019brc, Jassal:2005qc, Cheng:2025yue, Alam:2025epg,Lee:2025pzo}, with the so-called CPL (Chevallier-Polarski-Linde)~\cite{Chevallier:2000qy, Linder:2002et} form, i.e. $w_\mathrm{DE}(a) = w_0+w_a(1-a)$, being the standard functional choice adopted to study dynamical dark energy.

Alternatively, one could study the nature of dark energy via a non-parametric approach. In this category are included all methods that aim to perform a data-driven reconstruction of $w_\mathrm{DE}$, without any specific assumption on the parameter itself or its functional form\footnote{Strictly speaking, this is not possible since each reconstruction method relies on some assumptions, but our main goal is to be as agnostic as possible about the underlying functional form.}. While parametric approaches are less computationally expensive, they can lead to biased results. Non-parametric reconstructions solve this issue but, on the other hand, require careful validation with simulations. After the DESI data have been released, these methods have regained attention, with several recent works applying various reconstruction methods~\cite{Gonzalez-Fuentes:2025lei,Wang:2025vfb,DESI:2024aqx,Lee:2025pzo,Ormondroyd:2025exu, Dinda:2024ktd,DESI:2025fii,deSouza:2025vdv,Berti:2025phi} in order to scrutinize the dynamical dark energy hypothesis. Works including these parametric approaches have recently focused on the late-time behavior of dark energy, with a surprising preference for an oscillating equation of state around $w_\mathrm{DE}=-1$ over the cosmological constant case~\cite{Wang:2025vfb, Calderon_2024, DESI:2025wyn,Nojiri:2025low}.

In the following, we shall consider a non-parametric approach to reconstruct the equation of state parameter of dark energy, $w_\mathrm{DE}$. Among all the different possible methods~\cite{Liu:2015mkm,Rasmussen2004,Gomez-Valent:2018hwc,Gomez-Valent:2018gvm}, we consider the so-called PCHIP one~\cite{doi:10.1137/0905021, doi:10.1137/0717021}. This method consists of a third degree polynomial that will transition smoothly between some function values at some nodes, avoiding spurious oscillatory behavior due to the interpolation method. This results in an improved data-driven approach, where the result is mainly determined by the values of the function under study at the chosen nodes.
It is important to note that the results within this approach depend on the number of nodes considered and we will justify our choices by the cosmological datasets employed in our analyses. 

Additionally, a better understanding of the nature of dark energy would help in constraining neutrino properties from cosmology, which is important given the emerging neutrino tension between cosmological and laboratory bounds~\cite{Gariazzo:2023joe,Bertolez-Martinez:2024wez,Naredo-Tuero:2024sgf,Craig:2024tky,Jiang:2024viw}. In fact, it is well known that there exist degeneracies between the dark energy and neutrino sectors~\cite{Hannestad:2005gj, Lorenz:2017fgo, Vagnozzi:2017ovm, Sutherland:2018ghu, Sahlen:2018cku, Yang:2020ope, Zhang:2015uhk, Khalifeh:2021ree, Nair:2025uyn}. For this reason, we focus on different neutrino scenarios when extracting  $w_\mathrm{DE}$. In particular, we aim to reconstruct at the same time the sum of neutrino masses $\sum m_\nu$ and $w_\mathrm{DE}$ using the PCHIP formalism, following a phenomenological approach~\cite{Zhao:2006zf, Lorenz:2021alz}.

The paper is structured as follows. \Cref{sec:formalism} describes the formalism adopted in the work when describing the dark energy and neutrino sectors. The cosmological datasets and inference methodology are presented in \Cref{sec:method}.
In \Cref{sec:results} we show and discuss our results and reconstructed relevant quantities. Finally, we present our conclusions and final remarks in Sec.~\ref{sec:summary_conclusions}. 

\section{Formalism}
\label{sec:formalism}

In the following, we shall introduce the formalism adopted to describe the dark energy and the neutrino masses. Our aim is to perform a joint phenomenological extraction of both the equation of state of dark energy and the total neutrino mass. Furthermore, we want to avoid any bias from fixed functional forms that may result in stronger deviations from the simplistic standard scenario and possible overfitting issues that may end up mimicking an oscillatory behavior.

\subsection{PCHIP reconstruction}
\label{sec:PCHIP}

One of the best methods to perform an agnostic data-driven determination of a particular function is via the Piecewise Cubic Hermite Interpolating Polynomial (PCHIP)~\cite{doi:10.1137/0905021, doi:10.1137/0717021}. Overall, this method consists of a combination of third order polynomials that ensure smoothness between a set of input nodes while avoiding potential spurious oscillating behaviors~\cite{Gariazzo:2014dla}. The PCHIP interpolation guarantees that the final function maintains local monotonicity and the shape of the input points. For more details on the PCHIP interpolation method, see~\Cref{sec:AppA_pchipformalism}.

To reconstruct the equation of state of dark energy, $w_\mathrm{DE}$, and the sum of neutrino masses, $\sum m_\nu$, in a model-independent way, we rely on the PCHIP interpolator across seven temporal nodes. In this way, the functional form is completely determined once the value of these parameters is introduced at the seven redshift values of our choice.

\begin{figure}[b]
\centering
\includegraphics[width=0.7\linewidth]{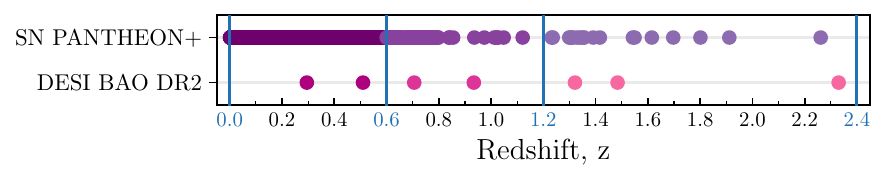}
\caption{Redshift points for each set of measurements considered in the analyses. We consider the PantheonPlus collection for Supernovae measurements and the second data release from the DESI collaboration as BAO measurements. The vertical lines depict the redshift points (nodes) we shall use for the PCHIP reconstruction.}
\label{fig:nodes}
\end{figure}

\Cref{fig:nodes} shows our late-time choice of temporal nodes. These particular values include a reasonable number of data points considered inside each interval for Baryon Acoustic Oscillations (BAO) and Supernovae Ia (SN Ia) measurements.
Additionally, to study the impact of the Cosmic Microwave Background (CMB) observations, we have included a node at recombination ($z\simeq 1100$). In order to avoid an extremely early dark energy component, we introduced an additional node at matter radiation equality ($z\simeq3400$), where we have imposed that $w_\mathrm{DE}=-1$. Finally, we also include the initial temporal point considered in \texttt{CLASS}, which corresponds to $a=10^{-14}$. For the same reasoning as the previous node, we fix $w_\mathrm{DE}=-1$ also in this node. The full set of nodes considered is shown in \Cref{tab:nodes}.

\begin{table}[h!]
\setlength{\tabcolsep}{5pt}
\renewcommand{\arraystretch}{1.4}
\centering
\begin{tabular} {c c c}
\textbf{Redshift} \boldmath{$z$} & \textbf{Scale factor} \boldmath{$a$} & \textbf{Parameter} \\
\hline
0.0 & 1.0 & $w_\mathrm{0}$ / $\sum m_{\nu,0}$\\
0.6 & 0.625 & $w_\mathrm{0.6}$ / $\sum m_{\nu,0.6}$\\
1.2 & $\approx$ 0.45 & $w_\mathrm{1.2}$ / $\sum m_{\nu,1.2}$\\
2.4 & $\approx$ 0.29 & $w_\mathrm{2.4}$ / $\sum m_{\nu,2.4}$\\
1100 & $\approx$ $9\times10^{-4}$ & $w_\mathrm{rec}$ / $\sum m_{\nu,\mathrm{rec}}$\\
3400 & $\approx$ $3\times10^{-4}$ & $w_\mathrm{mre}$ / $\sum m_{\nu,\mathrm{mre}}$\\
$10^{14}$ & $\approx$ $10^{-14}$ & $w_\mathrm{ini}$ / $\sum m_{\nu,\mathrm{ini}}$\\
\hline
\end{tabular}
\caption{Table summarizing the nodes chosen for the reconstruction. We show both the redshift and the equivalent scale factor, setting the nomenclature of the parameters that we sample.} 
\label{tab:nodes}
\end{table}

\subsection{Cosmological scenarios}
\label{sec:DE_and_neutrinos}

Focusing on dark energy, we define the equation of state of dark energy, $w_\mathrm{DE}$, as a PCHIP function with seven redshift nodes, as defined in \Cref{tab:nodes}. We fix the two initial nodes to $w_\mathrm{ini} = w_\mathrm{mre} = -1$ in order to avoid an early dark energy domination epoch. The rest of the nodes are allowed to freely vary, but we study two different scenarios depending on the range in which $w_\mathrm{DE}$ is allowed to vary. Specifically, we allow $w_{\mathrm{DE, }\,i}\in \mathcal{U}[-1.0,0.0]$ in the so-called \emph{non-phantom} scenario (\textbf{NP}) and $w_{\mathrm{DE, }\,i}\in \mathcal{U}[-2.0,0.0]$ in the \emph{phantom} case (\textbf{P}).

The results of this work have been obtained focusing in three different neutrino mass treatments, studied within both the phantom and non-phantom dark energy scenarios:

\textbf{\emph{Case $\boldsymbol{w_\mathrm{DE} + \sum m_{\nu,\,\mathrm{NO}}}$ }$\,-\,$} In this case we do not vary the neutrino sector, focusing only on the equation of state of dark energy. The sum of neutrino masses is fixed to the lowest allowed value by oscillation data~\cite{Esteban:2024eli,deSalas:2020pgw,Capozzi:2021fjo}, $\sum m_{\nu,\,\mathrm{NO}} = 0.06~\mathrm{eV}$, and we reconstruct the dark energy equation of state parameter, $w_\mathrm{DE}$, using the PCHIP prescription with five free nodes.

\textbf{\emph{Case $\boldsymbol{w_\mathrm{DE}+\sum m_{\nu,\,\mathrm{free}}}$}$\,-\,$} We reconstruct $w_\mathrm{DE}$ following the PCHIP method with five free nodes, but allowing $\sum m_\nu$ to vary in the MCMC analysis. We impose a uniform prior, $\sum m_\nu\in \mathcal{U}[0.0,3.0]~\mathrm{eV}$, allowing neutrinos to have masses higher than the current upper bound set by KATRIN~\cite{KATRIN:2024cdt}. 

\textbf{\emph{Case $\boldsymbol{w_\mathrm{DE}+\sum m_{\nu,\,\mathrm{PCHIP}}}$}$\,-\,$} Finally, in the last analysis, we reconstruct both parameters using the PCHIP formalism with seven redshift nodes, imposing only $\sum m_{\nu,\,\mathrm{ini}} = \sum m_{\nu,\,\mathrm{mre}}$, leaving the function with only six free parameters. Here, we impose a flat prior over each neutrino mass at the node to $\sum m_{\nu,\,i}\in \mathcal{U}[0.0,3.0]~\mathrm{eV}$.
Introducing time-varying neutrino masses typically involves convoluted modeling~\cite{Lorenz:2021alz}. Models involving phase transitions~\cite{Dvali:2016uhn, Lorenz:2018fzb}, topological defects~\cite{Dvali:2021uvk} or coupling to a scalar field that could represent dark energy~\cite{Franca:2009xp} or ultra-light dark matter~\cite{Huang:2022wmz,Bertolez-Martinez:2025pgm} have been proposed in order to have a neutrino mass which effectively changes in time. Instead, our aim is to be as model independent as possible. Thus, we follow a phenomenological approach similar to the one used in Refs.~\cite{Zhao:2006zf, Lorenz:2021alz}. In this case, we vary the total neutrino mass assuming that the additional dark sector coupled to neutrinos has no other observational consequences, or that those changes are somehow included in our free variation of the dark energy species. We emphasize that this is not a conservative approach, but a model-independent one. We will study the viability of this general scenario for realistic models, such as MaVaNs~\cite{Hung:2000yg, Gu:2003er, Fardon:2003eh, Franca:2009xp,Brookfield:2005bz}, in future work~\cite{Ghedini2026}.

\section{Methodology and data sets}
\label{sec:method}

The statistical methodology adopted in this work makes use of a modified version of the Cosmic Linear Anisotropy Solving System code (\texttt{CLASS})~\cite{Lesgourgues:2011re, Blas:2011rf} to include the PCHIP reconstruction. The cosmological parameter inference is performed using \texttt{Cobaya}~\cite{Torrado:2020dgo, 2019ascl.soft10019T} with convergence set as a Gelman-Rubin test~\cite{Gelman:1992zz} of $R-1 \leq 0.01$.
In addition to the base $\Lambda$CDM parameters, i.e. $\{\log(10^{10}A_\mathrm{s}), n_\mathrm{s}, 100\theta_\mathrm{s}, w_\mathrm{b}, w_\mathrm{dm},\tau\}$, we sample the nodes for the $w_\mathrm{DE}$ reconstruction, $w_{\mathrm{DE},\,i}$, plus $\sum m_\nu$ or $\sum m_{\nu,\,i}$ for their respective scenarios. We obtain statistics for the chains and plots with \texttt{Getdist}~\cite{Lewis:2019xzd}.
The priors imposed on the whole set of parameters for the different analyses are shown in \Cref{tab:priors}. 

\begin{table}[h!]
\centering
\small
\setlength{\tabcolsep}{6pt}
\renewcommand{\arraystretch}{1.35}
\begin{minipage}[t]{0.3\textwidth}
\centering
\begin{tabular}{cc}
\textbf{Parameters} & \textbf{Priors} \\
\hline
{\boldmath$\log(10^{10} A_\mathrm{s})$} & $\mathcal{U}[1.61, 3.91]$ \\
{\boldmath$n_\mathrm{s}$}              & $\mathcal{U}[0.8, 1.2]$ \\
{\boldmath$100\theta_\mathrm{s}$}      & $\mathcal{U}[0.5, 10]$ \\
{\boldmath$\Omega_\mathrm{b} h^2$}     & $\mathcal{U}[0.005, 0.1]$ \\
{\boldmath$\Omega_\mathrm{dm} h^2$}    & $\mathcal{U}[0.001, 0.99]$ \\
{\boldmath$\tau_\mathrm{reio}$}        & $\mathcal{U}[0.01, 0.8]$ \\
\hline
\end{tabular}
\end{minipage}
\hspace{0.02\textwidth}
\begin{minipage}[t]{0.6\textwidth}
\centering
\begin{tabular}{lcc}
\hspace{10mm}\textbf{Case} & $\boldsymbol{\sum m_\nu}$ \textbf{prior} & $\boldsymbol{w_{\mathrm{DE},i}}$ \textbf{prior} \\
\hline
$\boldsymbol{w_\mathrm{DE,\,NP} + \sum m_{\nu,\,\mathrm{NO}}}$ & Fixed 0.06 eV & \multirow{3}{*}{$\mathcal{U}[-1.0, 0.0]$} \\
$\boldsymbol{w_\mathrm{DE,\,NP} + \sum m_{\nu,\,\mathrm{free}}}$ & $\mathcal{U}[0.0, 3.0]$ eV &  \\
$\boldsymbol{w_\mathrm{DE,\,NP} + \sum m_{\nu,\,\mathrm{PCHIP}}}$ & $\sum m_{\nu,i} \in \mathcal{U}[0.0, 3.0]$ eV & \\
\hline
$\boldsymbol{w_\mathrm{DE,\,P} + \sum m_{\nu,\,\mathrm{NO}}}$  & Fixed 0.06 eV & \multirow{3}{*}{$\mathcal{U}[-2.0, 0.0]$} \\
$\boldsymbol{w_\mathrm{DE,\,P} + \sum m_{\nu,\,\mathrm{free}}}$  & $\mathcal{U}[0.0, 3.0]$ eV & \\
$\boldsymbol{w_\mathrm{DE,\,P} + \sum m_{\nu,\,\mathrm{PCHIP}}}$  & $\sum m_{\nu,i} \in \mathcal{U}[0.0, 3.0]$ eV & \\
\hline
\end{tabular}
\end{minipage}
\vspace{0.5em}
\caption{\textbf{Left}: Uniform priors on the standard base cosmological parameters varied in these analyses. \textbf{Right}: Priors on the dark energy and neutrino parameters for each of the six analyses performed. The \textbf{NP} subindex refers to the non-phantom dark energy case, while \textbf{P} refers to the phantom one. For neutrino masses, the \textbf{NO} subindex indicates the scenario in which the sum of neutrino masses is fixed to the lowest value allowed in the case of normal ordering by oscillation data, \textbf{free} refers to the case where we leave it as a free parameter in the cosmological inference and, finally, \textbf{PCHIP} refers to the particular method in which the sum of neutrino masses is defined through a set of nodes using the PCHIP interpolator.}
\label{tab:priors}
\end{table}

The datasets and likelihoods used to derive the constraints on our model are:

\textbf{\emph{CMB measurements}$\,-\,$} The Planck mission~\cite{Planck:2018nkj, Planck:2018vyg, Planck:2019nip} has achieved exceptionally precise measurements of the power spectra of CMB anisotropies. We use as our baseline dataset the temperature (TT) and polarization (EE) auto-spectra, plus their cross-spectra (TE), as incorporated the native \texttt{Cobaya} implementation of \texttt{Commander} (for multipoles $\ell<30$) and the original \texttt{plik} (for multipoles $\ell>30$) likelihoods from the PR3 release~\cite{Planck:2019nip}. In addition to the primary temperature and polarization anisotropy power spectra, we combine the information on the power spectrum of the gravitational lensing potential from Planck18~\cite{Planck:2018lbu} and from the Atacama Cosmology Telescope (ACT) DR6~\cite{ACT:2025fju,ACT:2023dou}, using in particular the \texttt{ACTDR6LensLike} likelihood.
The results obtained using Planck temperature, polarization and lensing measurements, combined with lensing measurements from ACT will be referred as \textbf{\textit{CMB}}.

\textbf{\emph{Supernovae measurements}$\,-\,$}
SN Ia serve as standard candles that can be used to measure the expansion of the Universe. We consider the PantheonPlus supernova type Ia sample~\cite{Brout:2022vxf}, which consists of 1701 lightcurves of 1550 unique SNeIa usable for cosmological constraints, covering a range of 0.001 $<$ z $<$ 2.26 (see \Cref{fig:nodes}). In particular, we use the public likelihood of Ref.~\cite{Brout:2022vxf}, included in \texttt{Cobaya}, which does not include SH0ES. We choose specifically this SN collection because is the one reaching higher redshifts. In the following, we shall denote this dataset as \textbf{\textit{SN}}.

\textbf{\emph{BAO measurements}$\,-\,$} As Baryon Acoustic Oscillations (BAO) measurements we consider the three-year data collection of DESI, presented in the second data release~\cite{DESI:2025zgx}. In DESI DR2, the accuracy of the measurements has been improved thanks to a larger dataset compared to the first version. Furthermore, separate measurements of the transverse comoving distance $D_\mathrm{M}(z)$ and the Hubble distance $D_\mathrm{H}(z)$ are now provided. These measurements are summarized in
Table IV of Ref.~\cite{DESI:2025zgx}. In the following, we shall denote this dataset as \textbf{\textit{BAO}}.

We conduct, for each of the aforementioned cases, analyses with the following combination of measurements: \textbf{CMB}, \textbf{CMB + BAO}, \textbf{CMB + SN} and \textbf{CMB + BAO + SN}. 

\section{Results}
\label{sec:results}

In this section we present the results of our analyses. We show the reconstructed quantities and the impact of our model on the parameters that currently exhibit tensions, namely $H_0$, $\sigma_8$ and $\Omega_\mathrm{m}$.

In Appendix \ref{sec:nutensions} we provide additional details on the $H_0$-$\sigma_8$ plots, highlighting the correlation of the value of $\sum m_\nu$ or $\sum m_{\nu,\,i}$ with these quantities. 
Finally, in Appendix \ref{sec:FullResult} we show the triangular plots showing the constraints on the values of $w_\mathrm{DE}$ and on $\sum m_\nu$ and their 2D posteriors, as well as complete tables showing the corresponding 95\% C.L. bounds. We also show in the tables the constraints on $H_0$, $\sigma_8$, and $\Omega_\mathrm{m}$ for the different analyses performed.

\subsection{Fixed {\boldmath$\sum m_\nu$} and PCHIP {\boldmath$w_\mathrm{DE}$}}
\label{sec:nu_fixed}

\begin{figure}[t]
\centering
\includegraphics[width=\linewidth]{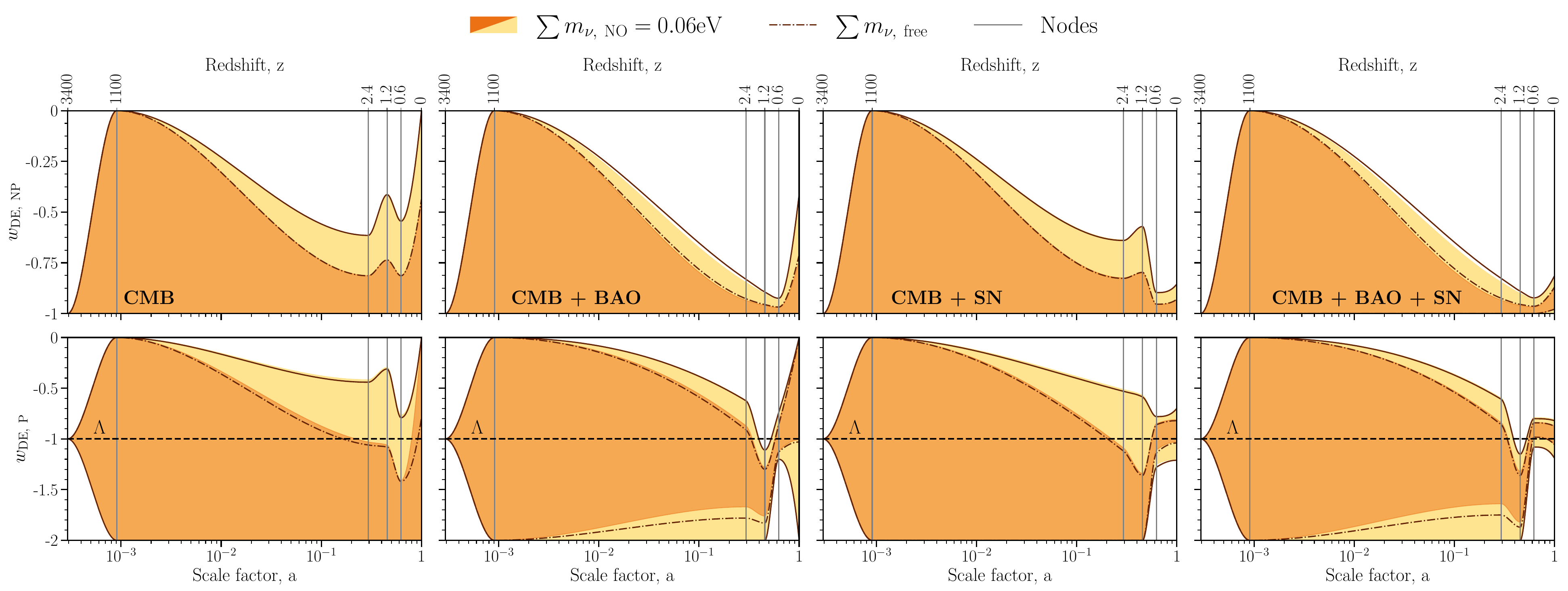}
\caption{Reconstruction of $w_\mathrm{DE}$ obtained using the PCHIP method. We show the 68 and 95 \% C.L. bounds for the case in which we fix $\sum m_\nu$ (shaded regions) and for the case in which we leave it free in the MCMC (lines: dot-dashed is 68 \% C.L.; solid is 95 \% C.L.).}
\label{fig:rec_case1_and_case2}
\end{figure}

\Cref{fig:rec_case1_and_case2} colored regions show the reconstruction of $w_\mathrm{DE}$ with $\sum m_\nu$ fixed to the normal ordering value ($\sum m_\nu = 0.06~\mathrm{eV}$). Orange contours correspond to the $1\sigma$ allowed region, while the yellow band shows the $2\sigma$ one. On the top row $w_\mathrm{DE}$ is not allowed to cross the phantom divide ($w_\mathrm{DE}=-1$), while on the bottom row it can take values up to $w_\mathrm{DE}\geq-2$ .
The nodes in which we vary the equation of state of dark energy are highlighted as vertical lines.

\textbf{\emph{Non-phantom DE}$\,-\,$} CMB data alone has enough constraining power such that $w_\mathrm{DE}<-1/3$ values are preferred at late times. The oscillatory pattern of the upper bound of the equation of state depicts the late-time constraining power of CMB data alone. Once one includes into account BAO measurements, the constraints on $w_\mathrm{DE}$ between $z=0.6$ and $z=2.4$ get more stringent, pointing towards the cosmological constant scenario. In turn, when including SN data, we obtain tighter limits for today's value at $z=0$. Thus, when combining all three datasets we mainly recover the CMB+BAO result with a tighter limit on $w_\mathrm{DE}(z=0)$.
In all cases the node at recombination ($z=1100$) remains unconstrained, as expected due to the subdominant contribution of the dark energy to the energy density budget at that epoch.

\textbf{\emph{Phantom DE}$\,-\,$}
When allowing for phantom crossing, CMB data alone shows a small ($\sim1\sigma$) preference for $w_\mathrm{DE}<-1$ near $z=0.6$. Adding DESI BAO data strengthens this preference to the $2\sigma$ level and shifts it to an earlier time, $z=1.2$. Within our choice of nodes, we do not find compelling evidence for an oscillatory evolution of $w_\mathrm{DE}$.
Previous studies~\cite{DESI:2025fii,Silva:2025twg,Li:2025vuh,Capozziello:2025qmh} reported a preferred phantom crossing around $z\simeq0.6$, whereas our results point to a crossing at $z=1.2$, with the adjacent $z=0.6$ and $z=2.4$ nodes remaining compatible with the cosmological constant scenario. As in the non-phantom case, SN measurements play the leading role in tightening the constraints on the present-day value of $w_\mathrm{DE}$.

\subsection{Free {\boldmath$\sum m_\nu$} and PCHIP {\boldmath$w_\mathrm{DE}$}}
\label{sec:nu_free}

The second analysis conducted performed a reconstruction of $w_\mathrm{DE}$ in the case in which the parameter $\sum m_\nu$ is left as a free parameter. \Cref{fig:rec_case1_and_case2} dot-dashed and solid lines show the $1\sigma$ and $2\sigma$ results when allowing the sum of neutrino masses to be a free parameter. \Cref{tab:summnufree} contains the 95 \% C.L. constraints on $\sum m_\nu$ associated to the different dark energy cases and cosmological datasets taken into account.

\begin{table}[b]
\centering
\small\addtolength{\tabcolsep}{+3pt}
\def\arraystretch{1.25}
\begin{tabular}{c c c c c c}
\textbf{Case} & \textbf{Parameter} & \textbf{CMB} & \textbf{CMB + BAO} & \textbf{CMB + SN} & \textbf{CMB + BAO + SN} \\ \hline
\textbf{Non-phantom DE} & \multirow{2}{*}{{\boldmath$\sum m_\nu$}} & $< 0.202$ eV & $< 0.0532$ eV & $< 0.138$ eV & $< 0.0545$ eV \\
\textbf{Phantom DE} & & $< 0.264$ eV & $< 0.163$ eV & $< 0.267$ eV & $< 0.163$ eV \\ \hline
\end{tabular}
\caption{95 \% C.L. constraints on $\sum m_\nu$ for the non-phantom and phantom dark energy species cases and for all the different combinations of cosmological datasets considered along this work.}
\label{tab:summnufree}
\end{table}

\textbf{\emph{Dark energy}$\,-\,$} There are no significant differences with respect to the previous case in which we fixed $\sum m_\nu = 0.06$ eV, both forbidding and allowing dark energy to cross the phantom divide. However, the overall trend has the opposite behavior in the two scenarios: while in the non-phantom case slightly less negative values of $w_\mathrm{DE}$ are preferred, when we allow for the phantom crossing, more negative values are allowed. The reason behind this behavior is the degeneracy between dark energy and neutrinos: indeed, increasing $\sum m_\nu$ suppresses the growth of structures, therefore modifying the expansion, reflected in a shift in the value of $w_\mathrm{DE}$. Moreover, allowing for $w_\mathrm{DE}<-1$ implies more freedom for dark energy to mimic the effects of massive neutrinos. All these effects are more evident when we introduce BAO data in the analysis: this is indeed a consequence of the preference of the DESI data for lower $\Omega_\mathrm{m}$ values.

\textbf{\emph{Neutrino mass constraints}$\,-\,$} It has been extensively studied in the literature that the constrains on $\sum m_\nu$ get relaxed when we allow for dark energy to cross the phantom regime~\cite{Vagnozzi:2018jhn, Nair:2025uyn,Sharma:2022ifr,Du:2024pai,RoyChoudhury:2018vnm,RoyChoudhury:2019hls}. Our results are consistent with this statement, as can be seen in \Cref{tab:summnufree}, where the phantom constraints are relaxed by a factor of $\sim2$ when considering SN data and $\sim3$ when including BAO data. As it has been shown lately, our findings are consistent with DESI BAO data driving the most stringent constraints on $\sum m_\nu$. 

\subsection{PCHIP {\boldmath$\sum m_\nu$} and PCHIP {\boldmath$w_\mathrm{DE}$}}
\label{sec:nu_pchip}

\begin{figure}[t]
\centering
\includegraphics[width=\linewidth]{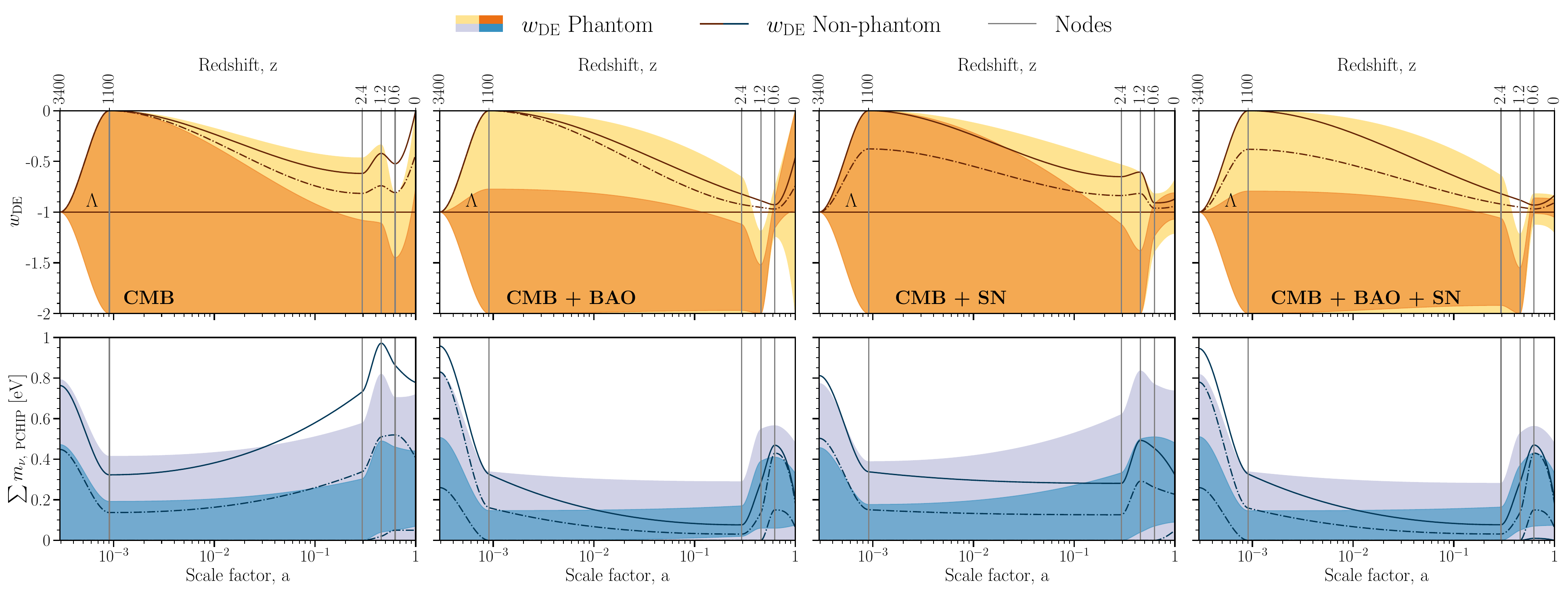}
\caption{Reconstruction of $w_\mathrm{DE}$ and $\sum m_\nu$ obtained using the PCHIP method. We show the 68 and 95 \% C.L. bounds, both for the case in which we allow for phantom dark energy (shaded regions) and in which we don't allow for it (lines: dot-dashed is 68 \% C.L.; solid is 95 \% C.L.).}
\label{fig:rec_case3}
\end{figure}

\Cref{fig:rec_case3} illustrates the outcome of the analysis in which we reconstruct both $w_\mathrm{DE}$ and $\sum m_\nu$ using the PCHIP formalism. In this case the top row depicts the DE equation of state for both the non-phantom (solid and dot-dashed lines) and phantom (shaded regions) scenarios. In turn, the bottom row now shows the reconstructed sum of the neutrino mass parameter via the PCHIP function with six free nodes. The value of $\sum m_\nu$ at $z=3400$ will remain constant up to the initial time considered by construction. This is not a conservative scenario since we are assuming that either the dark sector inducing changes on the value of the neutrino masses has no observational consequences for the cosmological observables that we use, or that those changes lie within our generic dark energy species. Nevertheless, once one focuses on a particular theory-motivated model, the dark energy-neutrino mass relation will be fixed, and constraints on that scenario may differ significantly from our generic case. As mentioned, we will devote a companion paper to study the viability of this model-independent scenario in the context of theory-specific models~\cite{Ghedini2026}.

\textbf{\emph{Dark energy}$\,-\,$} The results obtained in this analysis are in general agreement with the ones obtained in the previous cases, except when we include more data with respect to CMB. In that cases, we find that there is a 1$\sigma$ preference for for $w_\mathrm{DE}<0$ at the recombination node $z=1100$.
Also when phantom crossing is allowed, the results do not differ much from the other two cases considered in this work. It is interesting to note that now we find a non-zero value at 1$\sigma$ for $w_\mathrm{DE}$ at the recombination node $z=1100$ when more data are included in the analysis with respect to CMB, while it is only when considering also SN that we find this 1$\sigma$ preference when we force $w_\mathrm{DE}\geq-1$. 
Additionally, when we reconstruct also the $\sum m_\nu$ parameter, we obtain that more negative values for $w_\mathrm{DE}$ are preferred with respect to previous cases, meaning that in this case $w_\mathrm{DE}$ is preferred to lie in the phantom region.

\textbf{\emph{Neutrinos}$\,-\,$} It is still true that the constraints in the phantom scenario are wider than the ones obtained in the non-phantom scenario. When considering the combination of CMB + BAO + SN in the non-phantom dark energy scenario, we find a 2$\sigma$ deviation from $\sum m_\nu = 0$ at the $z=0.6$ node. This arises from the well-known degeneracy present between dark energy and neutrino parameters. Moreover, as expected, the bounds get relaxed in the phantom case, since phantom dark energy can accommodate higher values of neutrino masses. In both cases, we note that BAO data are the ones that lead to tighter constraints even at low redshift, highlighting a small preference for non-zero $\sum m_\nu$ at early times when we do not allow phantom crossing. For all data combinations and for both dark energy scenarios, we observe a tiny 1$\sigma$ preference for non-zero $\sum m_\nu$ between redshifts $z=0$ and $z\simeq2.4$. This is again a sign of the degeneracy between dark energy and neutrinos. In the phantom scenario this corresponds to the same redshift range in which we obtain deviations of $w_\mathrm{DE}$ from the cosmological constant scenario.

\subsection{Impact on cosmological tensions}
\label{sec:tensions}

\begin{figure}[t]
\centering
\includegraphics[width=\linewidth]{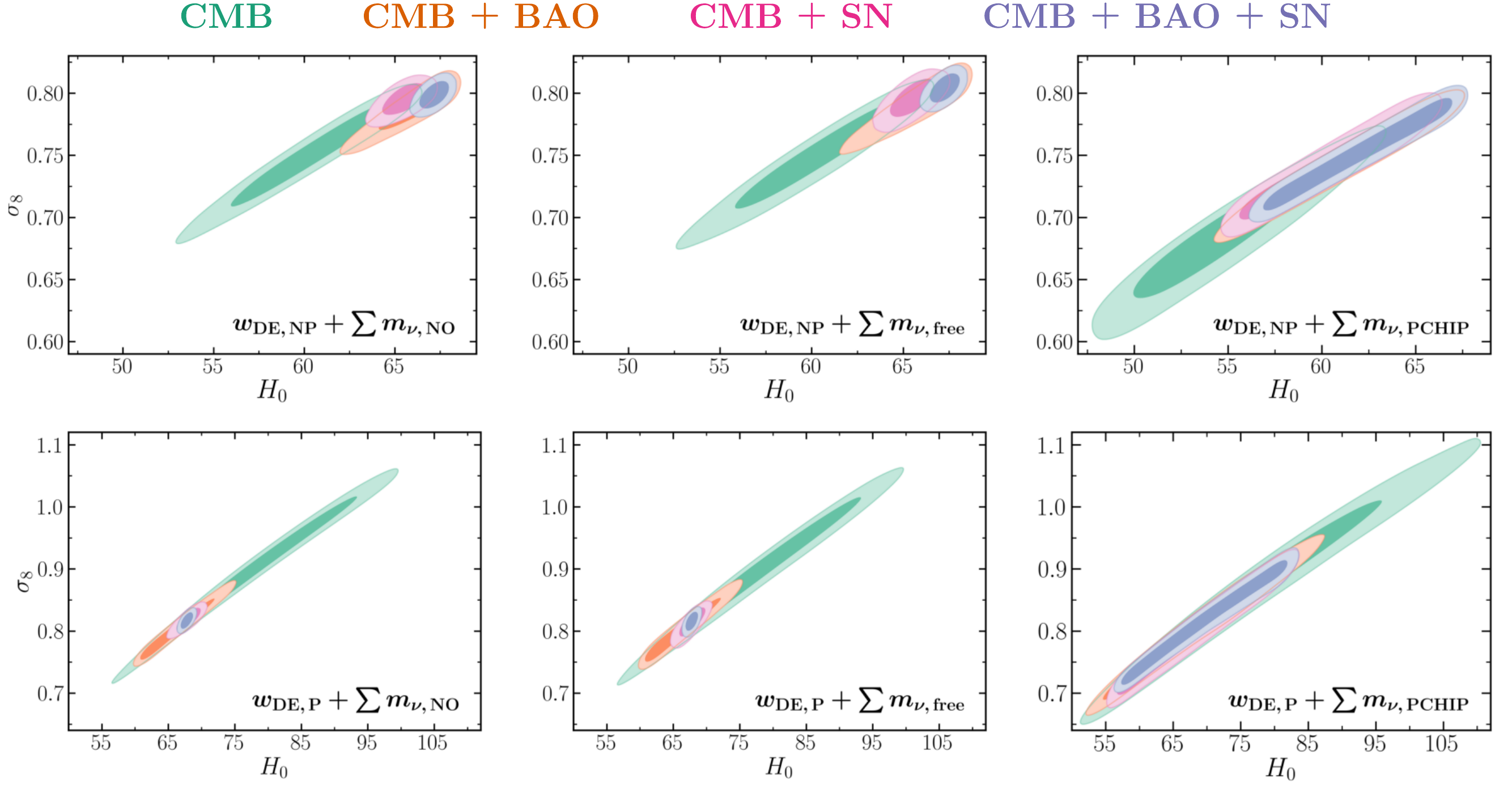}
\caption{Contour plot showing the 68 and 95 \% C.L. for the two parameters $H_0$ and $\sigma_8$, for all the analyses considered. Notice the different x axis values for the two rows.}
\label{fig:Hosigma8full}
\end{figure}

In this final subsection, we present the 2D contours in the $H_0-\sigma_8$ and $H_0-\Omega_m$ planes, for all the scenarios considered along this manuscript.  \Cref{fig:Hosigma8full} illustrates the degeneracy between $H_0$ and $\sigma_8$. Higher values of one parameter are allowed if the other one increases consequently. This behavior is well-understood in the  literature, where models that tend to solve the Hubble tension usually fail to alleviate the $\sigma_8$ one.
Minor changes are found when considering $\sum m_\nu$ free with respect to $\sum m_\nu = 0.06~\mathrm{eV}$. Instead, when we reconstruct $w_\mathrm{DE}$ and $\sum m_\nu$ simultaneously, the lower degeneracy direction for the non-phantom case is enlarged, while both directions in the phantom case are bigger. The combination of CMB+BAO+SN is the most constraining one in all cases, where SN data is the one driving the most stringent constraints for these parameters, while CMB data alone provides the looser constraints.

\begin{figure}[t]
\centering
\includegraphics[width=\linewidth]{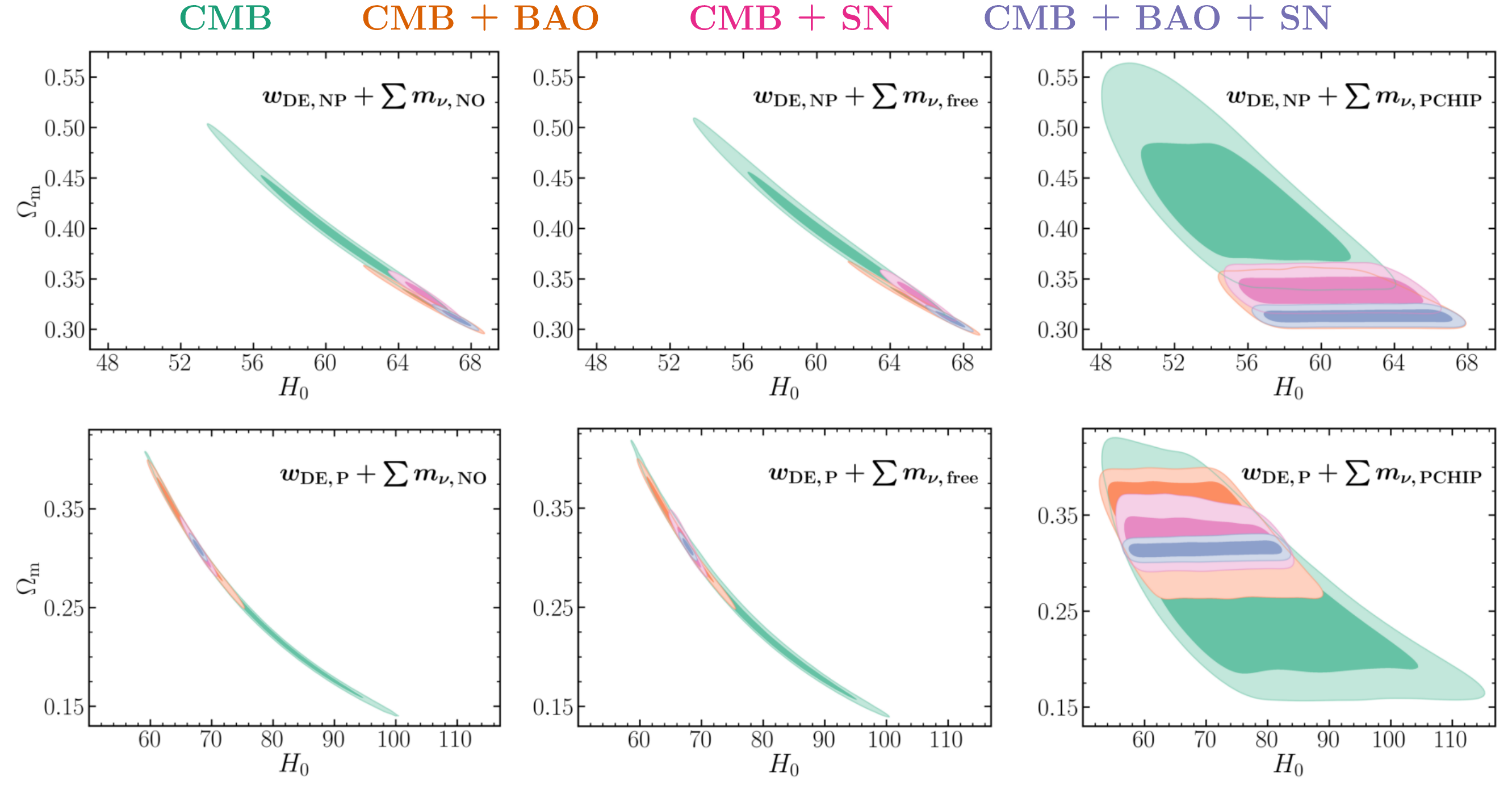}
\caption{Contour plot showing the 68 and 95 \% C.L. for the two parameters $H_0$ and $\Omega_\mathrm{m}$, for all the analyses considered. Notice the different x axis values for the two rows.}
\label{fig:HoOmegamfull}
\end{figure}

\Cref{fig:HoOmegamfull} shows instead the degeneracy between $H_0$ and $\Omega_m$. Here, the correlation between both parameters is inverse.
As before, the only difference between the first two columns is a small enlargement of the contours when $\sum m_\nu$ is allowed to vary.
However, when reconstructing $w_\mathrm{DE}$ and $\sum m_\nu$ simultaneously, we observe an almost perfect horizontal expansion of the contours, meaning that for a single value of $\Omega_m$, a larger range of values of $H_0$ is allowed. 
This is a consequence of not imposing a fixed relation between neutrinos and dark energy. However, if one considers that neutrino mass changes are induced by dark energy, this could potentially introduce significant changes on the behavior in this plane. The combination of CMB+BAO+SN is still the most constraining one in all cases, where SN data drives the strongest constraints. Finally, we recover the expected DESI preference for lower $\Omega_m$ values in the phantom regime.

\section{Summary and conclusions}
\label{sec:summary_conclusions}

Recent observations from the DESI collaboration prefer a dynamical nature of dark energy over a cosmological constant. In light of these new results, we followed a model independent approach to study the equation of state parameter of dark energy, $w_\mathrm{DE}$, at some relevant redshifts. We used the so-called PCHIP formalism to reconstruct the shape of $w_\mathrm{DE}$ along the cosmic expansion history, using a set of seven time nodes. In addition, we considered two different scenarios: one in which we force $w_\mathrm{DE}\geq-1$ during the whole evolution and one in which instead we allow $w_\mathrm{DE}$ to enter the phantom region.

Recent works focused on the possibility of an oscillatory behavior for $w_\mathrm{DE}$ at late times. However, due to the limited number of redshift measurements covered by DESI and SN observations, we wanted to test if this behavior could be related to the reconstruction method used and if it could be seen at larger timescales. For this reason, we chose the PCHIP formalism, which removes spurious oscillations. Indeed, in our reconstruction we do not find evidence of an oscillatory behavior on the late-time nodes, but indication for a $95\%$ C.L. deviation at $z=1.2$ once DESI BAO data are taken into account within the phantom dark energy scenario. If one does not include DESI BAO data, results are perfectly consistent with the cosmological constant hypothesis.

Given the well-known degeneracy between neutrino and dark energy parameters in cosmological analyses, we also explored three different neutrino scenarios: in the first one, we fix $\sum m_\nu = 0.06~\mathrm{eV}$, in the second we leave it as a free parameter in the MCMC analysis, and in the last one we reconstruct it in the same fashion than $w_\mathrm{DE}$. In this last case, we followed a phenomenological approach to perform the reconstruction of $\sum m_\nu$~\cite{Lorenz:2021alz}, leaving as future work a more realistic, theory-motivated treatment of mass-varying neutrinos~\cite{Ghedini2026}.

We obtain that, in both phantom and non-phantom dark energy and for all data combinations, the $95\%$ C.L. constrains on $w_\mathrm{DE}$ do not change significantly among the different neutrino mass scenarios. In all cases, the recombination node is unconstrained, validating that dark energy has only a significant contribution to energy density at late times.
The impact of varying $\sum m_\nu$ is to shift to more negative values in the phantom region the dark energy equation of state $w_\mathrm{DE}$. At the same time, by allowing dark energy to be phantom, the bounds on $\sum m_\nu$ become less tight.

When we also reconstruct the neutrino mass, we recover that $\sum m_\nu=0$ is consistent at 2$\sigma$ at all times, being the only exception the bound on the $z=0.6$ node in the CMB+BAO+SN non-phantom scenario. In this particular case, $\sum m_\nu (z=0.6) = 0.27^{+0.20}_{-0.26}$ at 95\% C.L., a deviation which is not statistically significant. Our findings are in agreement with previous findings concerning the DESI points around $z=0.6$.

Our work also proves how different measurements help in constraining different combinations of parameters, highlighting the non-trivial behavior of the $95\%$ C.L. bounds of $w_\mathrm{DE}$ and $\sum m_\nu$ obtained from the different combinations. We find that SN data impacts the $z=0$ node mainly (also $z=0.6$), while BAO data has more constraining power on the $z=0.24$, $z=0.12$, and $z=0.6$ nodes.

Overall, our analyses state that more data are needed to understand the nature of dark energy and that, using the PCHIP method, the cosmological constant prescription is not yet to be excluded, since we do not find robust evidence for an oscillatory or dynamical behavior of the dark energy component. Nevertheless, a $95\%$ CL deviation from the cosmological constant is found around $z=1.2$, hinting a consistent behavior with DESI dynamical dark energy studies.
Physically motivated analyses of time-dependent neutrino masses, incorporating general dark energy dynamics, will be required to test whether the DESI discrepancy with the cosmological-constant case could be linked to the neutrino sector. In addition, future late-time observations together with updated CMB measurements will further improve cosmological analyses of neutrinos and dark energy, where evolving dark energy and time-varying neutrino masses will be tested, even constraining potential interactions between the two sectors.

\begin{acknowledgments}

The authors thank Stefano Gariazzo for help with the PCHIP implementation. This article is based upon work from the COST Action CA21136 - “Addressing observational tensions in cosmology with systematics and fundamental physics (CosmoVerse)”, supported by COST - “European Cooperation in Science and Technology”. This work has received financial support from the Spanish MCIN/AEI/10.13039/501100011033 grants PID2020-113644GB-I00 (RH and OM), the SO project CEX2023-001292-S funded by MCIU/AEI/10.13039/501100011033 (PG), and by the European Union’s Horizon 2020 research and innovation programme under the Marie Skłodowska-Curie grants H2020-MSCA-ITN-2019/860881-HIDDeN and HORIZON-MSCA-2021-SE-01/101086085-ASYMMETRY. OM acknowledges the financial support from the MCIU with funding from the European Union NextGenerationEU (PRTR-C17.I01) and Generalitat Valenciana (ASFAE/2022/020). The authors also acknowledge support from the Generalitat Valenciana grants PROMETEO/2019/083 and CIPROM/2022/69 (RH and OM). PG and RH would like to thank the the Fermi National Accelerator Laboratory FERMILAB for its hospitality during the completion of this work.

\end{acknowledgments}

\bibliography{bib}

\begin{thebibliography}{244}%
\makeatletter
\providecommand \@ifxundefined [1]{%
 \@ifx{#1\undefined}
}%
\providecommand \@ifnum [1]{%
 \ifnum #1\expandafter \@firstoftwo
 \else \expandafter \@secondoftwo
 \fi
}%
\providecommand \@ifx [1]{%
 \ifx #1\expandafter \@firstoftwo
 \else \expandafter \@secondoftwo
 \fi
}%
\providecommand \natexlab [1]{#1}%
\providecommand \enquote  [1]{``#1''}%
\providecommand \bibnamefont  [1]{#1}%
\providecommand \bibfnamefont [1]{#1}%
\providecommand \citenamefont [1]{#1}%
\providecommand \href@noop [0]{\@secondoftwo}%
\providecommand \href [0]{\begingroup \@sanitize@url \@href}%
\providecommand \@href[1]{\@@startlink{#1}\@@href}%
\providecommand \@@href[1]{\endgroup#1\@@endlink}%
\providecommand \@sanitize@url [0]{\catcode `\\12\catcode `\$12\catcode
  `\&12\catcode `\#12\catcode `\^12\catcode `\_12\catcode `\%12\relax}%
\providecommand \@@startlink[1]{}%
\providecommand \@@endlink[0]{}%
\providecommand \url  [0]{\begingroup\@sanitize@url \@url }%
\providecommand \@url [1]{\endgroup\@href {#1}{\urlprefix }}%
\providecommand \urlprefix  [0]{URL }%
\providecommand \Eprint [0]{\href }%
\providecommand \doibase [0]{https://doi.org/}%
\providecommand \selectlanguage [0]{\@gobble}%
\providecommand \bibinfo  [0]{\@secondoftwo}%
\providecommand \bibfield  [0]{\@secondoftwo}%
\providecommand \translation [1]{[#1]}%
\providecommand \BibitemOpen [0]{}%
\providecommand \bibitemStop [0]{}%
\providecommand \bibitemNoStop [0]{.\EOS\space}%
\providecommand \EOS [0]{\spacefactor3000\relax}%
\providecommand \BibitemShut  [1]{\csname bibitem#1\endcsname}%
\let\auto@bib@innerbib\@empty
\bibitem [{\citenamefont {Riess}\ \emph {et~al.}(1998)\citenamefont {Riess}
  \emph {et~al.}}]{SupernovaSearchTeam:1998fmf}%
  \BibitemOpen
  \bibfield  {author} {\bibinfo {author} {\bibfnamefont {A.~G.}\ \bibnamefont
  {Riess}} \emph {et~al.} (\bibinfo {collaboration} {Supernova Search Team}),\
  }\href {https://doi.org/10.1086/300499} {\bibfield  {journal} {\bibinfo
  {journal} {Astron. J.}\ }\textbf {\bibinfo {volume} {116}},\ \bibinfo {pages}
  {1009} (\bibinfo {year} {1998})},\ \Eprint
  {https://arxiv.org/abs/astro-ph/9805201} {arXiv:astro-ph/9805201}
  \BibitemShut {NoStop}%
\bibitem [{\citenamefont {Perlmutter}\ \emph {et~al.}(1999)\citenamefont
  {Perlmutter} \emph {et~al.}}]{SupernovaCosmologyProject:1998vns}%
  \BibitemOpen
  \bibfield  {author} {\bibinfo {author} {\bibfnamefont {S.}~\bibnamefont
  {Perlmutter}} \emph {et~al.} (\bibinfo {collaboration} {Supernova Cosmology
  Project}),\ }\href {https://doi.org/10.1086/307221} {\bibfield  {journal}
  {\bibinfo  {journal} {Astrophys. J.}\ }\textbf {\bibinfo {volume} {517}},\
  \bibinfo {pages} {565} (\bibinfo {year} {1999})},\ \Eprint
  {https://arxiv.org/abs/astro-ph/9812133} {arXiv:astro-ph/9812133}
  \BibitemShut {NoStop}%
\bibitem [{\citenamefont {Weinberg}(1989)}]{Weinberg:1988cp}%
  \BibitemOpen
  \bibfield  {author} {\bibinfo {author} {\bibfnamefont {S.}~\bibnamefont
  {Weinberg}},\ }\href {https://doi.org/10.1103/RevModPhys.61.1} {\bibfield
  {journal} {\bibinfo  {journal} {Rev. Mod. Phys.}\ }\textbf {\bibinfo {volume}
  {61}},\ \bibinfo {pages} {1} (\bibinfo {year} {1989})}\BibitemShut {NoStop}%
\bibitem [{\citenamefont {Martin}(2012)}]{Martin:2012bt}%
  \BibitemOpen
  \bibfield  {author} {\bibinfo {author} {\bibfnamefont {J.}~\bibnamefont
  {Martin}},\ }\href {https://doi.org/10.1016/j.crhy.2012.04.008} {\bibfield
  {journal} {\bibinfo  {journal} {Comptes Rendus Physique}\ }\textbf {\bibinfo
  {volume} {13}},\ \bibinfo {pages} {566} (\bibinfo {year} {2012})},\ \Eprint
  {https://arxiv.org/abs/1205.3365} {arXiv:1205.3365 [astro-ph.CO]}
  \BibitemShut {NoStop}%
\bibitem [{\citenamefont {Silvestri}\ and\ \citenamefont
  {Trodden}(2009)}]{Silvestri:2009hh}%
  \BibitemOpen
  \bibfield  {author} {\bibinfo {author} {\bibfnamefont {A.}~\bibnamefont
  {Silvestri}}\ and\ \bibinfo {author} {\bibfnamefont {M.}~\bibnamefont
  {Trodden}},\ }\href {https://doi.org/10.1088/0034-4885/72/9/096901}
  {\bibfield  {journal} {\bibinfo  {journal} {Rept. Prog. Phys.}\ }\textbf
  {\bibinfo {volume} {72}},\ \bibinfo {pages} {096901} (\bibinfo {year}
  {2009})},\ \Eprint {https://arxiv.org/abs/0904.0024} {arXiv:0904.0024
  [astro-ph.CO]} \BibitemShut {NoStop}%
\bibitem [{\citenamefont {Carroll}\ \emph {et~al.}(2004)\citenamefont
  {Carroll}, \citenamefont {Duvvuri}, \citenamefont {Trodden},\ and\
  \citenamefont {Turner}}]{Carroll:2003wy}%
  \BibitemOpen
  \bibfield  {author} {\bibinfo {author} {\bibfnamefont {S.~M.}\ \bibnamefont
  {Carroll}}, \bibinfo {author} {\bibfnamefont {V.}~\bibnamefont {Duvvuri}},
  \bibinfo {author} {\bibfnamefont {M.}~\bibnamefont {Trodden}},\ and\ \bibinfo
  {author} {\bibfnamefont {M.~S.}\ \bibnamefont {Turner}},\ }\href
  {https://doi.org/10.1103/PhysRevD.70.043528} {\bibfield  {journal} {\bibinfo
  {journal} {Phys. Rev. D}\ }\textbf {\bibinfo {volume} {70}},\ \bibinfo
  {pages} {043528} (\bibinfo {year} {2004})},\ \Eprint
  {https://arxiv.org/abs/astro-ph/0306438} {arXiv:astro-ph/0306438}
  \BibitemShut {NoStop}%
\bibitem [{\citenamefont {Deffayet}\ \emph {et~al.}(2002)\citenamefont
  {Deffayet}, \citenamefont {Dvali},\ and\ \citenamefont
  {Gabadadze}}]{Deffayet:2001pu}%
  \BibitemOpen
  \bibfield  {author} {\bibinfo {author} {\bibfnamefont {C.}~\bibnamefont
  {Deffayet}}, \bibinfo {author} {\bibfnamefont {G.~R.}\ \bibnamefont
  {Dvali}},\ and\ \bibinfo {author} {\bibfnamefont {G.}~\bibnamefont
  {Gabadadze}},\ }\href {https://doi.org/10.1103/PhysRevD.65.044023} {\bibfield
   {journal} {\bibinfo  {journal} {Phys. Rev. D}\ }\textbf {\bibinfo {volume}
  {65}},\ \bibinfo {pages} {044023} (\bibinfo {year} {2002})},\ \Eprint
  {https://arxiv.org/abs/astro-ph/0105068} {arXiv:astro-ph/0105068}
  \BibitemShut {NoStop}%
\bibitem [{\citenamefont {Dvali}\ \emph {et~al.}(2000)\citenamefont {Dvali},
  \citenamefont {Gabadadze},\ and\ \citenamefont {Porrati}}]{Dvali:2000hr}%
  \BibitemOpen
  \bibfield  {author} {\bibinfo {author} {\bibfnamefont {G.~R.}\ \bibnamefont
  {Dvali}}, \bibinfo {author} {\bibfnamefont {G.}~\bibnamefont {Gabadadze}},\
  and\ \bibinfo {author} {\bibfnamefont {M.}~\bibnamefont {Porrati}},\ }\href
  {https://doi.org/10.1016/S0370-2693(00)00669-9} {\bibfield  {journal}
  {\bibinfo  {journal} {Phys. Lett. B}\ }\textbf {\bibinfo {volume} {485}},\
  \bibinfo {pages} {208} (\bibinfo {year} {2000})},\ \Eprint
  {https://arxiv.org/abs/hep-th/0005016} {arXiv:hep-th/0005016} \BibitemShut
  {NoStop}%
\bibitem [{\citenamefont {Freese}\ and\ \citenamefont
  {Lewis}(2002)}]{Freese:2002sq}%
  \BibitemOpen
  \bibfield  {author} {\bibinfo {author} {\bibfnamefont {K.}~\bibnamefont
  {Freese}}\ and\ \bibinfo {author} {\bibfnamefont {M.}~\bibnamefont {Lewis}},\
  }\href {https://doi.org/10.1016/S0370-2693(02)02122-6} {\bibfield  {journal}
  {\bibinfo  {journal} {Phys. Lett. B}\ }\textbf {\bibinfo {volume} {540}},\
  \bibinfo {pages} {1} (\bibinfo {year} {2002})},\ \Eprint
  {https://arxiv.org/abs/astro-ph/0201229} {arXiv:astro-ph/0201229}
  \BibitemShut {NoStop}%
\bibitem [{\citenamefont {Carroll}\ \emph {et~al.}(2006)\citenamefont
  {Carroll}, \citenamefont {Sawicki}, \citenamefont {Silvestri},\ and\
  \citenamefont {Trodden}}]{Carroll:2006jn}%
  \BibitemOpen
  \bibfield  {author} {\bibinfo {author} {\bibfnamefont {S.~M.}\ \bibnamefont
  {Carroll}}, \bibinfo {author} {\bibfnamefont {I.}~\bibnamefont {Sawicki}},
  \bibinfo {author} {\bibfnamefont {A.}~\bibnamefont {Silvestri}},\ and\
  \bibinfo {author} {\bibfnamefont {M.}~\bibnamefont {Trodden}},\ }\href
  {https://doi.org/10.1088/1367-2630/8/12/323} {\bibfield  {journal} {\bibinfo
  {journal} {New J. Phys.}\ }\textbf {\bibinfo {volume} {8}},\ \bibinfo {pages}
  {323} (\bibinfo {year} {2006})},\ \Eprint
  {https://arxiv.org/abs/astro-ph/0607458} {arXiv:astro-ph/0607458}
  \BibitemShut {NoStop}%
\bibitem [{\citenamefont {Capozziello}\ \emph {et~al.}(2003)\citenamefont
  {Capozziello}, \citenamefont {Carloni},\ and\ \citenamefont
  {Troisi}}]{Capozziello:2003tk}%
  \BibitemOpen
  \bibfield  {author} {\bibinfo {author} {\bibfnamefont {S.}~\bibnamefont
  {Capozziello}}, \bibinfo {author} {\bibfnamefont {S.}~\bibnamefont
  {Carloni}},\ and\ \bibinfo {author} {\bibfnamefont {A.}~\bibnamefont
  {Troisi}},\ }\href@noop {} {\bibfield  {journal} {\bibinfo  {journal} {Recent
  Res. Dev. Astron. Astrophys.}\ }\textbf {\bibinfo {volume} {1}},\ \bibinfo
  {pages} {625} (\bibinfo {year} {2003})},\ \Eprint
  {https://arxiv.org/abs/astro-ph/0303041} {arXiv:astro-ph/0303041}
  \BibitemShut {NoStop}%
\bibitem [{\citenamefont {Deffayet}(2001)}]{Deffayet:2000uy}%
  \BibitemOpen
  \bibfield  {author} {\bibinfo {author} {\bibfnamefont {C.}~\bibnamefont
  {Deffayet}},\ }\href {https://doi.org/10.1016/S0370-2693(01)00160-5}
  {\bibfield  {journal} {\bibinfo  {journal} {Phys. Lett. B}\ }\textbf
  {\bibinfo {volume} {502}},\ \bibinfo {pages} {199} (\bibinfo {year}
  {2001})},\ \Eprint {https://arxiv.org/abs/hep-th/0010186}
  {arXiv:hep-th/0010186} \BibitemShut {NoStop}%
\bibitem [{\citenamefont {Dvali}\ and\ \citenamefont
  {Turner}(2003)}]{Dvali:2003rk}%
  \BibitemOpen
  \bibfield  {author} {\bibinfo {author} {\bibfnamefont {G.}~\bibnamefont
  {Dvali}}\ and\ \bibinfo {author} {\bibfnamefont {M.~S.}\ \bibnamefont
  {Turner}},\ }\href@noop {} {\  (\bibinfo {year} {2003})},\ \Eprint
  {https://arxiv.org/abs/astro-ph/0301510} {arXiv:astro-ph/0301510}
  \BibitemShut {NoStop}%
\bibitem [{\citenamefont {Vollick}(2003)}]{Vollick:2003aw}%
  \BibitemOpen
  \bibfield  {author} {\bibinfo {author} {\bibfnamefont {D.~N.}\ \bibnamefont
  {Vollick}},\ }\href {https://doi.org/10.1103/PhysRevD.68.063510} {\bibfield
  {journal} {\bibinfo  {journal} {Phys. Rev. D}\ }\textbf {\bibinfo {volume}
  {68}},\ \bibinfo {pages} {063510} (\bibinfo {year} {2003})},\ \Eprint
  {https://arxiv.org/abs/astro-ph/0306630} {arXiv:astro-ph/0306630}
  \BibitemShut {NoStop}%
\bibitem [{\citenamefont {Flanagan}(2004)}]{Flanagan:2003rb}%
  \BibitemOpen
  \bibfield  {author} {\bibinfo {author} {\bibfnamefont {E.~E.}\ \bibnamefont
  {Flanagan}},\ }\href {https://doi.org/10.1103/PhysRevLett.92.071101}
  {\bibfield  {journal} {\bibinfo  {journal} {Phys. Rev. Lett.}\ }\textbf
  {\bibinfo {volume} {92}},\ \bibinfo {pages} {071101} (\bibinfo {year}
  {2004})},\ \Eprint {https://arxiv.org/abs/astro-ph/0308111}
  {arXiv:astro-ph/0308111} \BibitemShut {NoStop}%
\bibitem [{\citenamefont {Flanagan}(2003)}]{Flanagan:2003iw}%
  \BibitemOpen
  \bibfield  {author} {\bibinfo {author} {\bibfnamefont {E.~E.}\ \bibnamefont
  {Flanagan}},\ }\href {https://doi.org/10.1088/0264-9381/21/2/006} {\bibfield
  {journal} {\bibinfo  {journal} {Class. Quant. Grav.}\ }\textbf {\bibinfo
  {volume} {21}},\ \bibinfo {pages} {417} (\bibinfo {year} {2003})},\ \Eprint
  {https://arxiv.org/abs/gr-qc/0309015} {arXiv:gr-qc/0309015} \BibitemShut
  {NoStop}%
\bibitem [{\citenamefont {Vollick}(2004)}]{Vollick:2003ic}%
  \BibitemOpen
  \bibfield  {author} {\bibinfo {author} {\bibfnamefont {D.~N.}\ \bibnamefont
  {Vollick}},\ }\href {https://doi.org/10.1088/0264-9381/21/15/N01} {\bibfield
  {journal} {\bibinfo  {journal} {Class. Quant. Grav.}\ }\textbf {\bibinfo
  {volume} {21}},\ \bibinfo {pages} {3813} (\bibinfo {year} {2004})},\ \Eprint
  {https://arxiv.org/abs/gr-qc/0312041} {arXiv:gr-qc/0312041} \BibitemShut
  {NoStop}%
\bibitem [{\citenamefont {Soussa}\ and\ \citenamefont
  {Woodard}(2004)}]{Soussa:2003re}%
  \BibitemOpen
  \bibfield  {author} {\bibinfo {author} {\bibfnamefont {M.~E.}\ \bibnamefont
  {Soussa}}\ and\ \bibinfo {author} {\bibfnamefont {R.~P.}\ \bibnamefont
  {Woodard}},\ }\href {https://doi.org/10.1023/B:GERG.0000017037.92729.69}
  {\bibfield  {journal} {\bibinfo  {journal} {Gen. Rel. Grav.}\ }\textbf
  {\bibinfo {volume} {36}},\ \bibinfo {pages} {855} (\bibinfo {year} {2004})},\
  \Eprint {https://arxiv.org/abs/astro-ph/0308114} {arXiv:astro-ph/0308114}
  \BibitemShut {NoStop}%
\bibitem [{\citenamefont {Nojiri}\ and\ \citenamefont
  {Odintsov}(2004)}]{Nojiri:2003ni}%
  \BibitemOpen
  \bibfield  {author} {\bibinfo {author} {\bibfnamefont {S.}~\bibnamefont
  {Nojiri}}\ and\ \bibinfo {author} {\bibfnamefont {S.~D.}\ \bibnamefont
  {Odintsov}},\ }\href {https://doi.org/10.1023/B:GERG.0000035950.40718.48}
  {\bibfield  {journal} {\bibinfo  {journal} {Gen. Rel. Grav.}\ }\textbf
  {\bibinfo {volume} {36}},\ \bibinfo {pages} {1765} (\bibinfo {year}
  {2004})},\ \Eprint {https://arxiv.org/abs/hep-th/0308176}
  {arXiv:hep-th/0308176} \BibitemShut {NoStop}%
\bibitem [{\citenamefont {Arkani-Hamed}\ \emph {et~al.}(2004)\citenamefont
  {Arkani-Hamed}, \citenamefont {Cheng}, \citenamefont {Luty},\ and\
  \citenamefont {Mukohyama}}]{Arkani-Hamed:2003pdi}%
  \BibitemOpen
  \bibfield  {author} {\bibinfo {author} {\bibfnamefont {N.}~\bibnamefont
  {Arkani-Hamed}}, \bibinfo {author} {\bibfnamefont {H.-C.}\ \bibnamefont
  {Cheng}}, \bibinfo {author} {\bibfnamefont {M.~A.}\ \bibnamefont {Luty}},\
  and\ \bibinfo {author} {\bibfnamefont {S.}~\bibnamefont {Mukohyama}},\ }\href
  {https://doi.org/10.1088/1126-6708/2004/05/074} {\bibfield  {journal}
  {\bibinfo  {journal} {JHEP}\ }\textbf {\bibinfo {volume} {05}},\ \bibinfo
  {pages} {074}},\ \Eprint {https://arxiv.org/abs/hep-th/0312099}
  {arXiv:hep-th/0312099} \BibitemShut {NoStop}%
\bibitem [{\citenamefont {Gabadadze}\ and\ \citenamefont
  {Shifman}(2004)}]{Gabadadze:2003ck}%
  \BibitemOpen
  \bibfield  {author} {\bibinfo {author} {\bibfnamefont {G.}~\bibnamefont
  {Gabadadze}}\ and\ \bibinfo {author} {\bibfnamefont {M.}~\bibnamefont
  {Shifman}},\ }\href {https://doi.org/10.1103/PhysRevD.69.124032} {\bibfield
  {journal} {\bibinfo  {journal} {Phys. Rev. D}\ }\textbf {\bibinfo {volume}
  {69}},\ \bibinfo {pages} {124032} (\bibinfo {year} {2004})},\ \Eprint
  {https://arxiv.org/abs/hep-th/0312289} {arXiv:hep-th/0312289} \BibitemShut
  {NoStop}%
\bibitem [{\citenamefont {Moffat}(2004)}]{Moffat:2004nw}%
  \BibitemOpen
  \bibfield  {author} {\bibinfo {author} {\bibfnamefont {J.~W.}\ \bibnamefont
  {Moffat}},\ }\href@noop {} {\  (\bibinfo {year} {2004})},\ \Eprint
  {https://arxiv.org/abs/astro-ph/0403266} {arXiv:astro-ph/0403266}
  \BibitemShut {NoStop}%
\bibitem [{\citenamefont {Carroll}\ \emph {et~al.}(2005)\citenamefont
  {Carroll}, \citenamefont {De~Felice}, \citenamefont {Duvvuri}, \citenamefont
  {Easson}, \citenamefont {Trodden},\ and\ \citenamefont
  {Turner}}]{Carroll:2004de}%
  \BibitemOpen
  \bibfield  {author} {\bibinfo {author} {\bibfnamefont {S.~M.}\ \bibnamefont
  {Carroll}}, \bibinfo {author} {\bibfnamefont {A.}~\bibnamefont {De~Felice}},
  \bibinfo {author} {\bibfnamefont {V.}~\bibnamefont {Duvvuri}}, \bibinfo
  {author} {\bibfnamefont {D.~A.}\ \bibnamefont {Easson}}, \bibinfo {author}
  {\bibfnamefont {M.}~\bibnamefont {Trodden}},\ and\ \bibinfo {author}
  {\bibfnamefont {M.~S.}\ \bibnamefont {Turner}},\ }\href
  {https://doi.org/10.1103/PhysRevD.71.063513} {\bibfield  {journal} {\bibinfo
  {journal} {Phys. Rev. D}\ }\textbf {\bibinfo {volume} {71}},\ \bibinfo
  {pages} {063513} (\bibinfo {year} {2005})},\ \Eprint
  {https://arxiv.org/abs/astro-ph/0410031} {arXiv:astro-ph/0410031}
  \BibitemShut {NoStop}%
\bibitem [{\citenamefont {Clifton}\ \emph {et~al.}(2005)\citenamefont
  {Clifton}, \citenamefont {Mota},\ and\ \citenamefont
  {Barrow}}]{Clifton:2004st}%
  \BibitemOpen
  \bibfield  {author} {\bibinfo {author} {\bibfnamefont {T.}~\bibnamefont
  {Clifton}}, \bibinfo {author} {\bibfnamefont {D.~F.}\ \bibnamefont {Mota}},\
  and\ \bibinfo {author} {\bibfnamefont {J.~D.}\ \bibnamefont {Barrow}},\
  }\href {https://doi.org/10.1111/j.1365-2966.2005.08831.x} {\bibfield
  {journal} {\bibinfo  {journal} {Mon. Not. Roy. Astron. Soc.}\ }\textbf
  {\bibinfo {volume} {358}},\ \bibinfo {pages} {601} (\bibinfo {year}
  {2005})},\ \Eprint {https://arxiv.org/abs/gr-qc/0406001}
  {arXiv:gr-qc/0406001} \BibitemShut {NoStop}%
\bibitem [{\citenamefont {Easson}\ \emph {et~al.}(2005)\citenamefont {Easson},
  \citenamefont {Schuller}, \citenamefont {Trodden},\ and\ \citenamefont
  {Wohlfarth}}]{Easson:2005ax}%
  \BibitemOpen
  \bibfield  {author} {\bibinfo {author} {\bibfnamefont {D.~A.}\ \bibnamefont
  {Easson}}, \bibinfo {author} {\bibfnamefont {F.~P.}\ \bibnamefont
  {Schuller}}, \bibinfo {author} {\bibfnamefont {M.}~\bibnamefont {Trodden}},\
  and\ \bibinfo {author} {\bibfnamefont {M.~N.~R.}\ \bibnamefont {Wohlfarth}},\
  }\href {https://doi.org/10.1103/PhysRevD.72.043504} {\bibfield  {journal}
  {\bibinfo  {journal} {Phys. Rev. D}\ }\textbf {\bibinfo {volume} {72}},\
  \bibinfo {pages} {043504} (\bibinfo {year} {2005})},\ \Eprint
  {https://arxiv.org/abs/astro-ph/0506392} {arXiv:astro-ph/0506392}
  \BibitemShut {NoStop}%
\bibitem [{\citenamefont {de~la Cruz-Dombriz}\ and\ \citenamefont
  {Dobado}(2006)}]{delaCruz-Dombriz:2006kob}%
  \BibitemOpen
  \bibfield  {author} {\bibinfo {author} {\bibfnamefont {A.}~\bibnamefont
  {de~la Cruz-Dombriz}}\ and\ \bibinfo {author} {\bibfnamefont
  {A.}~\bibnamefont {Dobado}},\ }\href
  {https://doi.org/10.1103/PhysRevD.74.087501} {\bibfield  {journal} {\bibinfo
  {journal} {Phys. Rev. D}\ }\textbf {\bibinfo {volume} {74}},\ \bibinfo
  {pages} {087501} (\bibinfo {year} {2006})},\ \Eprint
  {https://arxiv.org/abs/gr-qc/0607118} {arXiv:gr-qc/0607118} \BibitemShut
  {NoStop}%
\bibitem [{\citenamefont {Artola}\ \emph {et~al.}(2025)\citenamefont {Artola},
  \citenamefont {Ayuso}, \citenamefont {Lazkoz}, \citenamefont {Olmo},\ and\
  \citenamefont {Salzano}}]{Artola:2025fup}%
  \BibitemOpen
  \bibfield  {author} {\bibinfo {author} {\bibfnamefont {M.}~\bibnamefont
  {Artola}}, \bibinfo {author} {\bibfnamefont {I.}~\bibnamefont {Ayuso}},
  \bibinfo {author} {\bibfnamefont {R.}~\bibnamefont {Lazkoz}}, \bibinfo
  {author} {\bibfnamefont {G.}~\bibnamefont {Olmo}},\ and\ \bibinfo {author}
  {\bibfnamefont {V.}~\bibnamefont {Salzano}},\ }\href@noop {} {\  (\bibinfo
  {year} {2025})},\ \Eprint {https://arxiv.org/abs/2510.27415}
  {arXiv:2510.27415 [gr-qc]} \BibitemShut {NoStop}%
\bibitem [{\citenamefont {Capozziello}\ \emph {et~al.}(2024)\citenamefont
  {Capozziello}, \citenamefont {Mazumdar},\ and\ \citenamefont
  {Meluccio}}]{Capozziello:2024qol}%
  \BibitemOpen
  \bibfield  {author} {\bibinfo {author} {\bibfnamefont {S.}~\bibnamefont
  {Capozziello}}, \bibinfo {author} {\bibfnamefont {A.}~\bibnamefont
  {Mazumdar}},\ and\ \bibinfo {author} {\bibfnamefont {G.}~\bibnamefont
  {Meluccio}},\ }\href {https://doi.org/10.1016/j.dark.2024.101517} {\bibfield
  {journal} {\bibinfo  {journal} {Phys. Dark Univ.}\ }\textbf {\bibinfo
  {volume} {45}},\ \bibinfo {pages} {101517} (\bibinfo {year} {2024})},\
  \Eprint {https://arxiv.org/abs/2403.11301} {arXiv:2403.11301 [gr-qc]}
  \BibitemShut {NoStop}%
\bibitem [{\citenamefont {Linder}(2025)}]{Linder:2025pqt}%
  \BibitemOpen
  \bibfield  {author} {\bibinfo {author} {\bibfnamefont {E.~V.}\ \bibnamefont
  {Linder}},\ }\href@noop {} {\  (\bibinfo {year} {2025})},\ \Eprint
  {https://arxiv.org/abs/2512.03139} {arXiv:2512.03139 [astro-ph.CO]}
  \BibitemShut {NoStop}%
\bibitem [{\citenamefont {Capozziello}\ \emph {et~al.}(2010)\citenamefont
  {Capozziello}, \citenamefont {Matsumoto}, \citenamefont {Nojiri},\ and\
  \citenamefont {Odintsov}}]{Capozziello:2010uv}%
  \BibitemOpen
  \bibfield  {author} {\bibinfo {author} {\bibfnamefont {S.}~\bibnamefont
  {Capozziello}}, \bibinfo {author} {\bibfnamefont {J.}~\bibnamefont
  {Matsumoto}}, \bibinfo {author} {\bibfnamefont {S.}~\bibnamefont {Nojiri}},\
  and\ \bibinfo {author} {\bibfnamefont {S.~D.}\ \bibnamefont {Odintsov}},\
  }\href {https://doi.org/10.1016/j.physletb.2010.08.030} {\bibfield  {journal}
  {\bibinfo  {journal} {Phys. Lett. B}\ }\textbf {\bibinfo {volume} {693}},\
  \bibinfo {pages} {198} (\bibinfo {year} {2010})},\ \Eprint
  {https://arxiv.org/abs/1004.3691} {arXiv:1004.3691 [hep-th]} \BibitemShut
  {NoStop}%
\bibitem [{\citenamefont {Capozziello}\ \emph {et~al.}(2005)\citenamefont
  {Capozziello}, \citenamefont {Cardone},\ and\ \citenamefont
  {Troisi}}]{Capozziello:2005ku}%
  \BibitemOpen
  \bibfield  {author} {\bibinfo {author} {\bibfnamefont {S.}~\bibnamefont
  {Capozziello}}, \bibinfo {author} {\bibfnamefont {V.~F.}\ \bibnamefont
  {Cardone}},\ and\ \bibinfo {author} {\bibfnamefont {A.}~\bibnamefont
  {Troisi}},\ }\href {https://doi.org/10.1103/PhysRevD.71.043503} {\bibfield
  {journal} {\bibinfo  {journal} {Phys. Rev. D}\ }\textbf {\bibinfo {volume}
  {71}},\ \bibinfo {pages} {043503} (\bibinfo {year} {2005})},\ \Eprint
  {https://arxiv.org/abs/astro-ph/0501426} {arXiv:astro-ph/0501426}
  \BibitemShut {NoStop}%
\bibitem [{\citenamefont {Capozziello}\ and\ \citenamefont
  {De~Laurentis}(2011)}]{Capozziello:2011et}%
  \BibitemOpen
  \bibfield  {author} {\bibinfo {author} {\bibfnamefont {S.}~\bibnamefont
  {Capozziello}}\ and\ \bibinfo {author} {\bibfnamefont {M.}~\bibnamefont
  {De~Laurentis}},\ }\href {https://doi.org/10.1016/j.physrep.2011.09.003}
  {\bibfield  {journal} {\bibinfo  {journal} {Phys. Rept.}\ }\textbf {\bibinfo
  {volume} {509}},\ \bibinfo {pages} {167} (\bibinfo {year} {2011})},\ \Eprint
  {https://arxiv.org/abs/1108.6266} {arXiv:1108.6266 [gr-qc]} \BibitemShut
  {NoStop}%
\bibitem [{\citenamefont {Joyce}\ \emph {et~al.}(2016)\citenamefont {Joyce},
  \citenamefont {Lombriser},\ and\ \citenamefont {Schmidt}}]{Joyce:2016vqv}%
  \BibitemOpen
  \bibfield  {author} {\bibinfo {author} {\bibfnamefont {A.}~\bibnamefont
  {Joyce}}, \bibinfo {author} {\bibfnamefont {L.}~\bibnamefont {Lombriser}},\
  and\ \bibinfo {author} {\bibfnamefont {F.}~\bibnamefont {Schmidt}},\ }\href
  {https://doi.org/10.1146/annurev-nucl-102115-044553} {\bibfield  {journal}
  {\bibinfo  {journal} {Ann. Rev. Nucl. Part. Sci.}\ }\textbf {\bibinfo
  {volume} {66}},\ \bibinfo {pages} {95} (\bibinfo {year} {2016})},\ \Eprint
  {https://arxiv.org/abs/1601.06133} {arXiv:1601.06133 [astro-ph.CO]}
  \BibitemShut {NoStop}%
\bibitem [{\citenamefont {Bouhmadi-L{\'o}pez}\ \emph
  {et~al.}(2025)\citenamefont {Bouhmadi-L{\'o}pez}, \citenamefont {Chiang},
  \citenamefont {Boiza},\ and\ \citenamefont {Chen}}]{Bouhmadi-Lopez:2025lzm}%
  \BibitemOpen
  \bibfield  {author} {\bibinfo {author} {\bibfnamefont {M.}~\bibnamefont
  {Bouhmadi-L{\'o}pez}}, \bibinfo {author} {\bibfnamefont {H.-W.}\ \bibnamefont
  {Chiang}}, \bibinfo {author} {\bibfnamefont {C.~G.}\ \bibnamefont {Boiza}},\
  and\ \bibinfo {author} {\bibfnamefont {P.}~\bibnamefont {Chen}},\ }\href@noop
  {} {\  (\bibinfo {year} {2025})},\ \Eprint {https://arxiv.org/abs/2512.09991}
  {arXiv:2512.09991 [astro-ph.CO]} \BibitemShut {NoStop}%
\bibitem [{\citenamefont {Bolotin}\ \emph {et~al.}(2015)\citenamefont
  {Bolotin}, \citenamefont {Kostenko}, \citenamefont {Lemets},\ and\
  \citenamefont {Yerokhin}}]{Bolotin_2015}%
  \BibitemOpen
  \bibfield  {author} {\bibinfo {author} {\bibfnamefont {Y.~L.}\ \bibnamefont
  {Bolotin}}, \bibinfo {author} {\bibfnamefont {A.}~\bibnamefont {Kostenko}},
  \bibinfo {author} {\bibfnamefont {O.~A.}\ \bibnamefont {Lemets}},\ and\
  \bibinfo {author} {\bibfnamefont {D.~A.}\ \bibnamefont {Yerokhin}},\ }\href
  {https://doi.org/10.1142/s0218271815300074} {\bibfield  {journal} {\bibinfo
  {journal} {International Journal of Modern Physics D}\ }\textbf {\bibinfo
  {volume} {24}},\ \bibinfo {pages} {1530007} (\bibinfo {year}
  {2015})}\BibitemShut {NoStop}%
\bibitem [{\citenamefont {Wang}\ \emph
  {et~al.}(2016{\natexlab{a}})\citenamefont {Wang}, \citenamefont {Abdalla},
  \citenamefont {Atrio-Barandela},\ and\ \citenamefont {Pavón}}]{Wang_2016}%
  \BibitemOpen
  \bibfield  {author} {\bibinfo {author} {\bibfnamefont {B.}~\bibnamefont
  {Wang}}, \bibinfo {author} {\bibfnamefont {E.}~\bibnamefont {Abdalla}},
  \bibinfo {author} {\bibfnamefont {F.}~\bibnamefont {Atrio-Barandela}},\ and\
  \bibinfo {author} {\bibfnamefont {D.}~\bibnamefont {Pavón}},\ }\href
  {https://doi.org/10.1088/0034-4885/79/9/096901} {\bibfield  {journal}
  {\bibinfo  {journal} {Reports on Progress in Physics}\ }\textbf {\bibinfo
  {volume} {79}},\ \bibinfo {pages} {096901} (\bibinfo {year}
  {2016}{\natexlab{a}})}\BibitemShut {NoStop}%
\bibitem [{\citenamefont {Ghedini}\ \emph {et~al.}(2024)\citenamefont
  {Ghedini}, \citenamefont {Hajjar},\ and\ \citenamefont
  {Mena}}]{Ghedini:2024mdu}%
  \BibitemOpen
  \bibfield  {author} {\bibinfo {author} {\bibfnamefont {P.}~\bibnamefont
  {Ghedini}}, \bibinfo {author} {\bibfnamefont {R.}~\bibnamefont {Hajjar}},\
  and\ \bibinfo {author} {\bibfnamefont {O.}~\bibnamefont {Mena}},\ }\href
  {https://doi.org/10.1016/j.dark.2024.101671} {\bibfield  {journal} {\bibinfo
  {journal} {Phys. Dark Univ.}\ }\textbf {\bibinfo {volume} {46}},\ \bibinfo
  {pages} {101671} (\bibinfo {year} {2024})},\ \Eprint
  {https://arxiv.org/abs/2409.02700} {arXiv:2409.02700 [astro-ph.CO]}
  \BibitemShut {NoStop}%
\bibitem [{\citenamefont {Amendola}(2000)}]{Amendola:1999er}%
  \BibitemOpen
  \bibfield  {author} {\bibinfo {author} {\bibfnamefont {L.}~\bibnamefont
  {Amendola}},\ }\href {https://doi.org/10.1103/PhysRevD.62.043511} {\bibfield
  {journal} {\bibinfo  {journal} {Phys. Rev. D}\ }\textbf {\bibinfo {volume}
  {62}},\ \bibinfo {pages} {043511} (\bibinfo {year} {2000})},\ \Eprint
  {https://arxiv.org/abs/astro-ph/9908023} {arXiv:astro-ph/9908023}
  \BibitemShut {NoStop}%
\bibitem [{\citenamefont {Amendola}\ and\ \citenamefont
  {Quercellini}(2003)}]{Amendola:2003eq}%
  \BibitemOpen
  \bibfield  {author} {\bibinfo {author} {\bibfnamefont {L.}~\bibnamefont
  {Amendola}}\ and\ \bibinfo {author} {\bibfnamefont {C.}~\bibnamefont
  {Quercellini}},\ }\href {https://doi.org/10.1103/PhysRevD.68.023514}
  {\bibfield  {journal} {\bibinfo  {journal} {Phys. Rev. D}\ }\textbf {\bibinfo
  {volume} {68}},\ \bibinfo {pages} {023514} (\bibinfo {year} {2003})},\
  \Eprint {https://arxiv.org/abs/astro-ph/0303228} {arXiv:astro-ph/0303228}
  \BibitemShut {NoStop}%
\bibitem [{\citenamefont {Cai}\ and\ \citenamefont {Wang}(2005)}]{Cai:2004dk}%
  \BibitemOpen
  \bibfield  {author} {\bibinfo {author} {\bibfnamefont {R.-G.}\ \bibnamefont
  {Cai}}\ and\ \bibinfo {author} {\bibfnamefont {A.}~\bibnamefont {Wang}},\
  }\href {https://doi.org/10.1088/1475-7516/2005/03/002} {\bibfield  {journal}
  {\bibinfo  {journal} {JCAP}\ }\textbf {\bibinfo {volume} {03}},\ \bibinfo
  {pages} {002}},\ \Eprint {https://arxiv.org/abs/hep-th/0411025}
  {arXiv:hep-th/0411025} \BibitemShut {NoStop}%
\bibitem [{\citenamefont {Huey}\ and\ \citenamefont
  {Wandelt}(2006)}]{Huey:2004qv}%
  \BibitemOpen
  \bibfield  {author} {\bibinfo {author} {\bibfnamefont {G.}~\bibnamefont
  {Huey}}\ and\ \bibinfo {author} {\bibfnamefont {B.~D.}\ \bibnamefont
  {Wandelt}},\ }\href {https://doi.org/10.1103/PhysRevD.74.023519} {\bibfield
  {journal} {\bibinfo  {journal} {Phys. Rev. D}\ }\textbf {\bibinfo {volume}
  {74}},\ \bibinfo {pages} {023519} (\bibinfo {year} {2006})},\ \Eprint
  {https://arxiv.org/abs/astro-ph/0407196} {arXiv:astro-ph/0407196}
  \BibitemShut {NoStop}%
\bibitem [{\citenamefont {Gumjudpai}\ \emph {et~al.}(2005)\citenamefont
  {Gumjudpai}, \citenamefont {Naskar}, \citenamefont {Sami},\ and\
  \citenamefont {Tsujikawa}}]{Gumjudpai:2005ry}%
  \BibitemOpen
  \bibfield  {author} {\bibinfo {author} {\bibfnamefont {B.}~\bibnamefont
  {Gumjudpai}}, \bibinfo {author} {\bibfnamefont {T.}~\bibnamefont {Naskar}},
  \bibinfo {author} {\bibfnamefont {M.}~\bibnamefont {Sami}},\ and\ \bibinfo
  {author} {\bibfnamefont {S.}~\bibnamefont {Tsujikawa}},\ }\href
  {https://doi.org/10.1088/1475-7516/2005/06/007} {\bibfield  {journal}
  {\bibinfo  {journal} {JCAP}\ }\textbf {\bibinfo {volume} {06}},\ \bibinfo
  {pages} {007}},\ \Eprint {https://arxiv.org/abs/hep-th/0502191}
  {arXiv:hep-th/0502191} \BibitemShut {NoStop}%
\bibitem [{\citenamefont {Brookfield}\ \emph
  {et~al.}(2006{\natexlab{a}})\citenamefont {Brookfield}, \citenamefont {van~de
  Bruck}, \citenamefont {Mota},\ and\ \citenamefont
  {Tocchini-Valentini}}]{Brookfield:2005td}%
  \BibitemOpen
  \bibfield  {author} {\bibinfo {author} {\bibfnamefont {A.~W.}\ \bibnamefont
  {Brookfield}}, \bibinfo {author} {\bibfnamefont {C.}~\bibnamefont {van~de
  Bruck}}, \bibinfo {author} {\bibfnamefont {D.~F.}\ \bibnamefont {Mota}},\
  and\ \bibinfo {author} {\bibfnamefont {D.}~\bibnamefont
  {Tocchini-Valentini}},\ }\href
  {https://doi.org/10.1103/PhysRevLett.96.061301} {\bibfield  {journal}
  {\bibinfo  {journal} {Phys. Rev. Lett.}\ }\textbf {\bibinfo {volume} {96}},\
  \bibinfo {pages} {061301} (\bibinfo {year} {2006}{\natexlab{a}})},\ \Eprint
  {https://arxiv.org/abs/astro-ph/0503349} {arXiv:astro-ph/0503349}
  \BibitemShut {NoStop}%
\bibitem [{\citenamefont {Berger}\ and\ \citenamefont
  {Shojaei}(2006)}]{Berger:2006db}%
  \BibitemOpen
  \bibfield  {author} {\bibinfo {author} {\bibfnamefont {M.~S.}\ \bibnamefont
  {Berger}}\ and\ \bibinfo {author} {\bibfnamefont {H.}~\bibnamefont
  {Shojaei}},\ }\href {https://doi.org/10.1103/PhysRevD.73.083528} {\bibfield
  {journal} {\bibinfo  {journal} {Phys. Rev. D}\ }\textbf {\bibinfo {volume}
  {73}},\ \bibinfo {pages} {083528} (\bibinfo {year} {2006})},\ \Eprint
  {https://arxiv.org/abs/gr-qc/0601086} {arXiv:gr-qc/0601086} \BibitemShut
  {NoStop}%
\bibitem [{\citenamefont {del Campo}\ \emph {et~al.}(2006)\citenamefont {del
  Campo}, \citenamefont {Herrera}, \citenamefont {Olivares},\ and\
  \citenamefont {Pavon}}]{delCampo:2006vv}%
  \BibitemOpen
  \bibfield  {author} {\bibinfo {author} {\bibfnamefont {S.}~\bibnamefont {del
  Campo}}, \bibinfo {author} {\bibfnamefont {R.}~\bibnamefont {Herrera}},
  \bibinfo {author} {\bibfnamefont {G.}~\bibnamefont {Olivares}},\ and\
  \bibinfo {author} {\bibfnamefont {D.}~\bibnamefont {Pavon}},\ }\href
  {https://doi.org/10.1103/PhysRevD.74.023501} {\bibfield  {journal} {\bibinfo
  {journal} {Phys. Rev. D}\ }\textbf {\bibinfo {volume} {74}},\ \bibinfo
  {pages} {023501} (\bibinfo {year} {2006})},\ \Eprint
  {https://arxiv.org/abs/astro-ph/0606520} {arXiv:astro-ph/0606520}
  \BibitemShut {NoStop}%
\bibitem [{\citenamefont {Barrow}\ and\ \citenamefont
  {Clifton}(2006)}]{Barrow:2006hia}%
  \BibitemOpen
  \bibfield  {author} {\bibinfo {author} {\bibfnamefont {J.~D.}\ \bibnamefont
  {Barrow}}\ and\ \bibinfo {author} {\bibfnamefont {T.}~\bibnamefont
  {Clifton}},\ }\href {https://doi.org/10.1103/PhysRevD.73.103520} {\bibfield
  {journal} {\bibinfo  {journal} {Phys. Rev. D}\ }\textbf {\bibinfo {volume}
  {73}},\ \bibinfo {pages} {103520} (\bibinfo {year} {2006})},\ \Eprint
  {https://arxiv.org/abs/gr-qc/0604063} {arXiv:gr-qc/0604063} \BibitemShut
  {NoStop}%
\bibitem [{\citenamefont {del Campo}\ \emph {et~al.}(2008)\citenamefont {del
  Campo}, \citenamefont {Herrera},\ and\ \citenamefont
  {Pavon}}]{delCampo:2008sr}%
  \BibitemOpen
  \bibfield  {author} {\bibinfo {author} {\bibfnamefont {S.}~\bibnamefont {del
  Campo}}, \bibinfo {author} {\bibfnamefont {R.}~\bibnamefont {Herrera}},\ and\
  \bibinfo {author} {\bibfnamefont {D.}~\bibnamefont {Pavon}},\ }\href
  {https://doi.org/10.1103/PhysRevD.78.021302} {\bibfield  {journal} {\bibinfo
  {journal} {Phys. Rev. D}\ }\textbf {\bibinfo {volume} {78}},\ \bibinfo
  {pages} {021302} (\bibinfo {year} {2008})},\ \Eprint
  {https://arxiv.org/abs/0806.2116} {arXiv:0806.2116 [astro-ph]} \BibitemShut
  {NoStop}%
\bibitem [{\citenamefont {Valiviita}\ \emph {et~al.}(2008)\citenamefont
  {Valiviita}, \citenamefont {Majerotto},\ and\ \citenamefont
  {Maartens}}]{Valiviita:2008iv}%
  \BibitemOpen
  \bibfield  {author} {\bibinfo {author} {\bibfnamefont {J.}~\bibnamefont
  {Valiviita}}, \bibinfo {author} {\bibfnamefont {E.}~\bibnamefont
  {Majerotto}},\ and\ \bibinfo {author} {\bibfnamefont {R.}~\bibnamefont
  {Maartens}},\ }\href {https://doi.org/10.1088/1475-7516/2008/07/020}
  {\bibfield  {journal} {\bibinfo  {journal} {JCAP}\ }\textbf {\bibinfo
  {volume} {07}},\ \bibinfo {pages} {020}},\ \Eprint
  {https://arxiv.org/abs/0804.0232} {arXiv:0804.0232 [astro-ph]} \BibitemShut
  {NoStop}%
\bibitem [{\citenamefont {Chongchitnan}(2009)}]{Chongchitnan:2008ry}%
  \BibitemOpen
  \bibfield  {author} {\bibinfo {author} {\bibfnamefont {S.}~\bibnamefont
  {Chongchitnan}},\ }\href {https://doi.org/10.1103/PhysRevD.79.043522}
  {\bibfield  {journal} {\bibinfo  {journal} {Phys. Rev. D}\ }\textbf {\bibinfo
  {volume} {79}},\ \bibinfo {pages} {043522} (\bibinfo {year} {2009})},\
  \Eprint {https://arxiv.org/abs/0810.5411} {arXiv:0810.5411 [astro-ph]}
  \BibitemShut {NoStop}%
\bibitem [{\citenamefont {Gavela}\ \emph {et~al.}(2009)\citenamefont {Gavela},
  \citenamefont {Hernandez}, \citenamefont {Lopez~Honorez}, \citenamefont
  {Mena},\ and\ \citenamefont {Rigolin}}]{Gavela:2009cy}%
  \BibitemOpen
  \bibfield  {author} {\bibinfo {author} {\bibfnamefont {M.~B.}\ \bibnamefont
  {Gavela}}, \bibinfo {author} {\bibfnamefont {D.}~\bibnamefont {Hernandez}},
  \bibinfo {author} {\bibfnamefont {L.}~\bibnamefont {Lopez~Honorez}}, \bibinfo
  {author} {\bibfnamefont {O.}~\bibnamefont {Mena}},\ and\ \bibinfo {author}
  {\bibfnamefont {S.}~\bibnamefont {Rigolin}},\ }\href
  {https://doi.org/10.1088/1475-7516/2009/07/034} {\bibfield  {journal}
  {\bibinfo  {journal} {JCAP}\ }\textbf {\bibinfo {volume} {07}},\ \bibinfo
  {pages} {034}},\ \bibinfo {note} {[Erratum: JCAP 05, E01 (2010)]},\ \Eprint
  {https://arxiv.org/abs/0901.1611} {arXiv:0901.1611 [astro-ph.CO]}
  \BibitemShut {NoStop}%
\bibitem [{\citenamefont {Xia}(2009)}]{Xia:2009zzb}%
  \BibitemOpen
  \bibfield  {author} {\bibinfo {author} {\bibfnamefont {J.-Q.}\ \bibnamefont
  {Xia}},\ }\href {https://doi.org/10.1103/PhysRevD.80.103514} {\bibfield
  {journal} {\bibinfo  {journal} {Phys. Rev. D}\ }\textbf {\bibinfo {volume}
  {80}},\ \bibinfo {pages} {103514} (\bibinfo {year} {2009})},\ \Eprint
  {https://arxiv.org/abs/0911.4820} {arXiv:0911.4820 [astro-ph.CO]}
  \BibitemShut {NoStop}%
\bibitem [{\citenamefont {Gavela}\ \emph {et~al.}(2010)\citenamefont {Gavela},
  \citenamefont {Lopez~Honorez}, \citenamefont {Mena},\ and\ \citenamefont
  {Rigolin}}]{Gavela:2010tm}%
  \BibitemOpen
  \bibfield  {author} {\bibinfo {author} {\bibfnamefont {M.~B.}\ \bibnamefont
  {Gavela}}, \bibinfo {author} {\bibfnamefont {L.}~\bibnamefont
  {Lopez~Honorez}}, \bibinfo {author} {\bibfnamefont {O.}~\bibnamefont
  {Mena}},\ and\ \bibinfo {author} {\bibfnamefont {S.}~\bibnamefont
  {Rigolin}},\ }\href {https://doi.org/10.1088/1475-7516/2010/11/044}
  {\bibfield  {journal} {\bibinfo  {journal} {JCAP}\ }\textbf {\bibinfo
  {volume} {11}},\ \bibinfo {pages} {044}},\ \Eprint
  {https://arxiv.org/abs/1005.0295} {arXiv:1005.0295 [astro-ph.CO]}
  \BibitemShut {NoStop}%
\bibitem [{\citenamefont {Mena}(2010)}]{Mena:2010zz}%
  \BibitemOpen
  \bibfield  {author} {\bibinfo {author} {\bibfnamefont {O.}~\bibnamefont
  {Mena}},\ }\href {https://doi.org/10.1088/1742-6596/259/1/012084} {\bibfield
  {journal} {\bibinfo  {journal} {J. Phys. Conf. Ser.}\ }\textbf {\bibinfo
  {volume} {259}},\ \bibinfo {pages} {012084} (\bibinfo {year}
  {2010})}\BibitemShut {NoStop}%
\bibitem [{\citenamefont {Baldi}\ and\ \citenamefont
  {Pettorino}(2011)}]{Baldi:2010td}%
  \BibitemOpen
  \bibfield  {author} {\bibinfo {author} {\bibfnamefont {M.}~\bibnamefont
  {Baldi}}\ and\ \bibinfo {author} {\bibfnamefont {V.}~\bibnamefont
  {Pettorino}},\ }\href {https://doi.org/10.1111/j.1745-3933.2010.00975.x}
  {\bibfield  {journal} {\bibinfo  {journal} {Mon. Not. Roy. Astron. Soc.}\
  }\textbf {\bibinfo {volume} {412}},\ \bibinfo {pages} {L1} (\bibinfo {year}
  {2011})},\ \Eprint {https://arxiv.org/abs/1006.3761} {arXiv:1006.3761
  [astro-ph.CO]} \BibitemShut {NoStop}%
\bibitem [{\citenamefont {Lopez~Honorez}\ \emph {et~al.}(2010)\citenamefont
  {Lopez~Honorez}, \citenamefont {Reid}, \citenamefont {Mena}, \citenamefont
  {Verde},\ and\ \citenamefont {Jimenez}}]{LopezHonorez:2010esq}%
  \BibitemOpen
  \bibfield  {author} {\bibinfo {author} {\bibfnamefont {L.}~\bibnamefont
  {Lopez~Honorez}}, \bibinfo {author} {\bibfnamefont {B.~A.}\ \bibnamefont
  {Reid}}, \bibinfo {author} {\bibfnamefont {O.}~\bibnamefont {Mena}}, \bibinfo
  {author} {\bibfnamefont {L.}~\bibnamefont {Verde}},\ and\ \bibinfo {author}
  {\bibfnamefont {R.}~\bibnamefont {Jimenez}},\ }\href
  {https://doi.org/10.1088/1475-7516/2010/09/029} {\bibfield  {journal}
  {\bibinfo  {journal} {JCAP}\ }\textbf {\bibinfo {volume} {09}},\ \bibinfo
  {pages} {029}},\ \Eprint {https://arxiv.org/abs/1006.0877} {arXiv:1006.0877
  [astro-ph.CO]} \BibitemShut {NoStop}%
\bibitem [{\citenamefont {Lee}\ and\ \citenamefont {Baldi}(2012)}]{Lee:2011tq}%
  \BibitemOpen
  \bibfield  {author} {\bibinfo {author} {\bibfnamefont {J.}~\bibnamefont
  {Lee}}\ and\ \bibinfo {author} {\bibfnamefont {M.}~\bibnamefont {Baldi}},\
  }\href {https://doi.org/10.1088/0004-637X/747/1/45} {\bibfield  {journal}
  {\bibinfo  {journal} {Astrophys. J.}\ }\textbf {\bibinfo {volume} {747}},\
  \bibinfo {pages} {45} (\bibinfo {year} {2012})},\ \Eprint
  {https://arxiv.org/abs/1110.0015} {arXiv:1110.0015 [astro-ph.CO]}
  \BibitemShut {NoStop}%
\bibitem [{\citenamefont {Beynon}\ \emph {et~al.}(2012)\citenamefont {Beynon},
  \citenamefont {Baldi}, \citenamefont {Bacon}, \citenamefont {Koyama},\ and\
  \citenamefont {Sabiu}}]{Beynon:2011hw}%
  \BibitemOpen
  \bibfield  {author} {\bibinfo {author} {\bibfnamefont {E.}~\bibnamefont
  {Beynon}}, \bibinfo {author} {\bibfnamefont {M.}~\bibnamefont {Baldi}},
  \bibinfo {author} {\bibfnamefont {D.~J.}\ \bibnamefont {Bacon}}, \bibinfo
  {author} {\bibfnamefont {K.}~\bibnamefont {Koyama}},\ and\ \bibinfo {author}
  {\bibfnamefont {C.}~\bibnamefont {Sabiu}},\ }\href
  {https://doi.org/10.1111/j.1365-2966.2012.20864.x} {\bibfield  {journal}
  {\bibinfo  {journal} {Mon. Not. Roy. Astron. Soc.}\ }\textbf {\bibinfo
  {volume} {422}},\ \bibinfo {pages} {3546} (\bibinfo {year} {2012})},\ \Eprint
  {https://arxiv.org/abs/1111.6974} {arXiv:1111.6974 [astro-ph.CO]}
  \BibitemShut {NoStop}%
\bibitem [{\citenamefont {Amendola}\ \emph {et~al.}(2012)\citenamefont
  {Amendola}, \citenamefont {Pettorino}, \citenamefont {Quercellini},\ and\
  \citenamefont {Vollmer}}]{Amendola:2011ie}%
  \BibitemOpen
  \bibfield  {author} {\bibinfo {author} {\bibfnamefont {L.}~\bibnamefont
  {Amendola}}, \bibinfo {author} {\bibfnamefont {V.}~\bibnamefont {Pettorino}},
  \bibinfo {author} {\bibfnamefont {C.}~\bibnamefont {Quercellini}},\ and\
  \bibinfo {author} {\bibfnamefont {A.}~\bibnamefont {Vollmer}},\ }\href
  {https://doi.org/10.1103/PhysRevD.85.103008} {\bibfield  {journal} {\bibinfo
  {journal} {Phys. Rev. D}\ }\textbf {\bibinfo {volume} {85}},\ \bibinfo
  {pages} {103008} (\bibinfo {year} {2012})},\ \Eprint
  {https://arxiv.org/abs/1111.1404} {arXiv:1111.1404 [astro-ph.CO]}
  \BibitemShut {NoStop}%
\bibitem [{\citenamefont {Farajollahi}\ \emph {et~al.}(2012)\citenamefont
  {Farajollahi}, \citenamefont {Ravanpak},\ and\ \citenamefont
  {Fadakar}}]{Farajollahi:2012zz}%
  \BibitemOpen
  \bibfield  {author} {\bibinfo {author} {\bibfnamefont {H.}~\bibnamefont
  {Farajollahi}}, \bibinfo {author} {\bibfnamefont {A.}~\bibnamefont
  {Ravanpak}},\ and\ \bibinfo {author} {\bibfnamefont {G.~F.}\ \bibnamefont
  {Fadakar}},\ }\href {https://doi.org/10.1016/j.physletb.2012.04.001}
  {\bibfield  {journal} {\bibinfo  {journal} {Phys. Lett. B}\ }\textbf
  {\bibinfo {volume} {711}},\ \bibinfo {pages} {225} (\bibinfo {year}
  {2012})},\ \Eprint {https://arxiv.org/abs/1206.5796} {arXiv:1206.5796
  [physics.gen-ph]} \BibitemShut {NoStop}%
\bibitem [{\citenamefont {Pettorino}\ \emph {et~al.}(2012)\citenamefont
  {Pettorino}, \citenamefont {Amendola}, \citenamefont {Baccigalupi},\ and\
  \citenamefont {Quercellini}}]{Pettorino:2012ts}%
  \BibitemOpen
  \bibfield  {author} {\bibinfo {author} {\bibfnamefont {V.}~\bibnamefont
  {Pettorino}}, \bibinfo {author} {\bibfnamefont {L.}~\bibnamefont {Amendola}},
  \bibinfo {author} {\bibfnamefont {C.}~\bibnamefont {Baccigalupi}},\ and\
  \bibinfo {author} {\bibfnamefont {C.}~\bibnamefont {Quercellini}},\ }\href
  {https://doi.org/10.1103/PhysRevD.86.103507} {\bibfield  {journal} {\bibinfo
  {journal} {Phys. Rev. D}\ }\textbf {\bibinfo {volume} {86}},\ \bibinfo
  {pages} {103507} (\bibinfo {year} {2012})},\ \Eprint
  {https://arxiv.org/abs/1207.3293} {arXiv:1207.3293 [astro-ph.CO]}
  \BibitemShut {NoStop}%
\bibitem [{\citenamefont {Chimento}\ and\ \citenamefont
  {Richarte}(2012)}]{Chimento:2012aea}%
  \BibitemOpen
  \bibfield  {author} {\bibinfo {author} {\bibfnamefont {L.~P.}\ \bibnamefont
  {Chimento}}\ and\ \bibinfo {author} {\bibfnamefont {M.~G.}\ \bibnamefont
  {Richarte}},\ }\href {https://doi.org/10.1103/PhysRevD.86.103501} {\bibfield
  {journal} {\bibinfo  {journal} {Phys. Rev. D}\ }\textbf {\bibinfo {volume}
  {86}},\ \bibinfo {pages} {103501} (\bibinfo {year} {2012})},\ \Eprint
  {https://arxiv.org/abs/1210.5505} {arXiv:1210.5505 [gr-qc]} \BibitemShut
  {NoStop}%
\bibitem [{\citenamefont {Salvatelli}\ \emph {et~al.}(2013)\citenamefont
  {Salvatelli}, \citenamefont {Marchini}, \citenamefont {Lopez-Honorez},\ and\
  \citenamefont {Mena}}]{Salvatelli:2013wra}%
  \BibitemOpen
  \bibfield  {author} {\bibinfo {author} {\bibfnamefont {V.}~\bibnamefont
  {Salvatelli}}, \bibinfo {author} {\bibfnamefont {A.}~\bibnamefont
  {Marchini}}, \bibinfo {author} {\bibfnamefont {L.}~\bibnamefont
  {Lopez-Honorez}},\ and\ \bibinfo {author} {\bibfnamefont {O.}~\bibnamefont
  {Mena}},\ }\href {https://doi.org/10.1103/PhysRevD.88.023531} {\bibfield
  {journal} {\bibinfo  {journal} {Phys. Rev. D}\ }\textbf {\bibinfo {volume}
  {88}},\ \bibinfo {pages} {023531} (\bibinfo {year} {2013})},\ \Eprint
  {https://arxiv.org/abs/1304.7119} {arXiv:1304.7119 [astro-ph.CO]}
  \BibitemShut {NoStop}%
\bibitem [{\citenamefont {Pettorino}(2013)}]{Pettorino:2013oxa}%
  \BibitemOpen
  \bibfield  {author} {\bibinfo {author} {\bibfnamefont {V.}~\bibnamefont
  {Pettorino}},\ }\href {https://doi.org/10.1103/PhysRevD.88.063519} {\bibfield
   {journal} {\bibinfo  {journal} {Phys. Rev. D}\ }\textbf {\bibinfo {volume}
  {88}},\ \bibinfo {pages} {063519} (\bibinfo {year} {2013})},\ \Eprint
  {https://arxiv.org/abs/1305.7457} {arXiv:1305.7457 [astro-ph.CO]}
  \BibitemShut {NoStop}%
\bibitem [{\citenamefont {Xia}(2013)}]{Xia:2013nua}%
  \BibitemOpen
  \bibfield  {author} {\bibinfo {author} {\bibfnamefont {J.-Q.}\ \bibnamefont
  {Xia}},\ }\href {https://doi.org/10.1088/1475-7516/2013/11/022} {\bibfield
  {journal} {\bibinfo  {journal} {JCAP}\ }\textbf {\bibinfo {volume} {11}},\
  \bibinfo {pages} {022}},\ \Eprint {https://arxiv.org/abs/1311.2131}
  {arXiv:1311.2131 [astro-ph.CO]} \BibitemShut {NoStop}%
\bibitem [{\citenamefont {Li}\ and\ \citenamefont {Zhang}(2014)}]{Li:2013bya}%
  \BibitemOpen
  \bibfield  {author} {\bibinfo {author} {\bibfnamefont {Y.-H.}\ \bibnamefont
  {Li}}\ and\ \bibinfo {author} {\bibfnamefont {X.}~\bibnamefont {Zhang}},\
  }\href {https://doi.org/10.1103/PhysRevD.89.083009} {\bibfield  {journal}
  {\bibinfo  {journal} {Phys. Rev. D}\ }\textbf {\bibinfo {volume} {89}},\
  \bibinfo {pages} {083009} (\bibinfo {year} {2014})},\ \Eprint
  {https://arxiv.org/abs/1312.6328} {arXiv:1312.6328 [astro-ph.CO]}
  \BibitemShut {NoStop}%
\bibitem [{\citenamefont {Yang}\ and\ \citenamefont
  {Xu}(2014{\natexlab{a}})}]{Yang:2014okp}%
  \BibitemOpen
  \bibfield  {author} {\bibinfo {author} {\bibfnamefont {W.}~\bibnamefont
  {Yang}}\ and\ \bibinfo {author} {\bibfnamefont {L.}~\bibnamefont {Xu}},\
  }\href {https://doi.org/10.1088/1475-7516/2014/08/034} {\bibfield  {journal}
  {\bibinfo  {journal} {JCAP}\ }\textbf {\bibinfo {volume} {08}},\ \bibinfo
  {pages} {034}},\ \Eprint {https://arxiv.org/abs/1401.5177} {arXiv:1401.5177
  [astro-ph.CO]} \BibitemShut {NoStop}%
\bibitem [{\citenamefont {Yang}\ and\ \citenamefont
  {Xu}(2014{\natexlab{b}})}]{Yang:2014gza}%
  \BibitemOpen
  \bibfield  {author} {\bibinfo {author} {\bibfnamefont {W.}~\bibnamefont
  {Yang}}\ and\ \bibinfo {author} {\bibfnamefont {L.}~\bibnamefont {Xu}},\
  }\href {https://doi.org/10.1103/PhysRevD.89.083517} {\bibfield  {journal}
  {\bibinfo  {journal} {Phys. Rev. D}\ }\textbf {\bibinfo {volume} {89}},\
  \bibinfo {pages} {083517} (\bibinfo {year} {2014}{\natexlab{b}})},\ \Eprint
  {https://arxiv.org/abs/1401.1286} {arXiv:1401.1286 [astro-ph.CO]}
  \BibitemShut {NoStop}%
\bibitem [{\citenamefont {Salvatelli}\ \emph {et~al.}(2014)\citenamefont
  {Salvatelli}, \citenamefont {Said}, \citenamefont {Bruni}, \citenamefont
  {Melchiorri},\ and\ \citenamefont {Wands}}]{Salvatelli:2014zta}%
  \BibitemOpen
  \bibfield  {author} {\bibinfo {author} {\bibfnamefont {V.}~\bibnamefont
  {Salvatelli}}, \bibinfo {author} {\bibfnamefont {N.}~\bibnamefont {Said}},
  \bibinfo {author} {\bibfnamefont {M.}~\bibnamefont {Bruni}}, \bibinfo
  {author} {\bibfnamefont {A.}~\bibnamefont {Melchiorri}},\ and\ \bibinfo
  {author} {\bibfnamefont {D.}~\bibnamefont {Wands}},\ }\href
  {https://doi.org/10.1103/PhysRevLett.113.181301} {\bibfield  {journal}
  {\bibinfo  {journal} {Phys. Rev. Lett.}\ }\textbf {\bibinfo {volume} {113}},\
  \bibinfo {pages} {181301} (\bibinfo {year} {2014})},\ \Eprint
  {https://arxiv.org/abs/1406.7297} {arXiv:1406.7297 [astro-ph.CO]}
  \BibitemShut {NoStop}%
\bibitem [{\citenamefont {Wang}\ \emph {et~al.}(2014)\citenamefont {Wang},
  \citenamefont {Wands}, \citenamefont {Zhao},\ and\ \citenamefont
  {Xu}}]{Wang:2014xca}%
  \BibitemOpen
  \bibfield  {author} {\bibinfo {author} {\bibfnamefont {Y.}~\bibnamefont
  {Wang}}, \bibinfo {author} {\bibfnamefont {D.}~\bibnamefont {Wands}},
  \bibinfo {author} {\bibfnamefont {G.-B.}\ \bibnamefont {Zhao}},\ and\
  \bibinfo {author} {\bibfnamefont {L.}~\bibnamefont {Xu}},\ }\href
  {https://doi.org/10.1103/PhysRevD.90.023502} {\bibfield  {journal} {\bibinfo
  {journal} {Phys. Rev. D}\ }\textbf {\bibinfo {volume} {90}},\ \bibinfo
  {pages} {023502} (\bibinfo {year} {2014})},\ \Eprint
  {https://arxiv.org/abs/1404.5706} {arXiv:1404.5706 [astro-ph.CO]}
  \BibitemShut {NoStop}%
\bibitem [{\citenamefont {Li}\ \emph {et~al.}(2014)\citenamefont {Li},
  \citenamefont {Zhang},\ and\ \citenamefont {Geng}}]{Li:2014eba}%
  \BibitemOpen
  \bibfield  {author} {\bibinfo {author} {\bibfnamefont {E.-K.}\ \bibnamefont
  {Li}}, \bibinfo {author} {\bibfnamefont {Y.}~\bibnamefont {Zhang}},\ and\
  \bibinfo {author} {\bibfnamefont {J.-L.}\ \bibnamefont {Geng}},\ }\href
  {https://doi.org/10.1103/PhysRevD.90.083534} {\bibfield  {journal} {\bibinfo
  {journal} {Phys. Rev. D}\ }\textbf {\bibinfo {volume} {90}},\ \bibinfo
  {pages} {083534} (\bibinfo {year} {2014})},\ \Eprint
  {https://arxiv.org/abs/1412.5482} {arXiv:1412.5482 [gr-qc]} \BibitemShut
  {NoStop}%
\bibitem [{\citenamefont {Pan}\ \emph {et~al.}(2015)\citenamefont {Pan},
  \citenamefont {Bhattacharya},\ and\ \citenamefont
  {Chakraborty}}]{Pan:2012ki}%
  \BibitemOpen
  \bibfield  {author} {\bibinfo {author} {\bibfnamefont {S.}~\bibnamefont
  {Pan}}, \bibinfo {author} {\bibfnamefont {S.}~\bibnamefont {Bhattacharya}},\
  and\ \bibinfo {author} {\bibfnamefont {S.}~\bibnamefont {Chakraborty}},\
  }\href {https://doi.org/10.1093/mnras/stv1495} {\bibfield  {journal}
  {\bibinfo  {journal} {Mon. Not. Roy. Astron. Soc.}\ }\textbf {\bibinfo
  {volume} {452}},\ \bibinfo {pages} {3038} (\bibinfo {year} {2015})},\ \Eprint
  {https://arxiv.org/abs/1210.0396} {arXiv:1210.0396 [gr-qc]} \BibitemShut
  {NoStop}%
\bibitem [{\citenamefont {Li}\ \emph {et~al.}(2016)\citenamefont {Li},
  \citenamefont {Zhang},\ and\ \citenamefont {Zhang}}]{Li:2015vla}%
  \BibitemOpen
  \bibfield  {author} {\bibinfo {author} {\bibfnamefont {Y.-H.}\ \bibnamefont
  {Li}}, \bibinfo {author} {\bibfnamefont {J.-F.}\ \bibnamefont {Zhang}},\ and\
  \bibinfo {author} {\bibfnamefont {X.}~\bibnamefont {Zhang}},\ }\href
  {https://doi.org/10.1103/PhysRevD.93.023002} {\bibfield  {journal} {\bibinfo
  {journal} {Phys. Rev. D}\ }\textbf {\bibinfo {volume} {93}},\ \bibinfo
  {pages} {023002} (\bibinfo {year} {2016})},\ \Eprint
  {https://arxiv.org/abs/1506.06349} {arXiv:1506.06349 [astro-ph.CO]}
  \BibitemShut {NoStop}%
\bibitem [{\citenamefont {Cui}\ \emph {et~al.}(2015)\citenamefont {Cui},
  \citenamefont {Yin}, \citenamefont {Wang}, \citenamefont {Li},\ and\
  \citenamefont {Zhang}}]{Cui:2015ueu}%
  \BibitemOpen
  \bibfield  {author} {\bibinfo {author} {\bibfnamefont {J.-L.}\ \bibnamefont
  {Cui}}, \bibinfo {author} {\bibfnamefont {L.}~\bibnamefont {Yin}}, \bibinfo
  {author} {\bibfnamefont {L.-F.}\ \bibnamefont {Wang}}, \bibinfo {author}
  {\bibfnamefont {Y.-H.}\ \bibnamefont {Li}},\ and\ \bibinfo {author}
  {\bibfnamefont {X.}~\bibnamefont {Zhang}},\ }\href
  {https://doi.org/10.1088/1475-7516/2015/09/024} {\bibfield  {journal}
  {\bibinfo  {journal} {JCAP}\ }\textbf {\bibinfo {volume} {09}},\ \bibinfo
  {pages} {024}},\ \Eprint {https://arxiv.org/abs/1503.08948} {arXiv:1503.08948
  [astro-ph.CO]} \BibitemShut {NoStop}%
\bibitem [{\citenamefont {Landim}(2016)}]{Landim:2015uda}%
  \BibitemOpen
  \bibfield  {author} {\bibinfo {author} {\bibfnamefont {R.~C.~G.}\
  \bibnamefont {Landim}},\ }\href
  {https://doi.org/10.1140/epjc/s10052-016-3894-2} {\bibfield  {journal}
  {\bibinfo  {journal} {Eur. Phys. J. C}\ }\textbf {\bibinfo {volume} {76}},\
  \bibinfo {pages} {31} (\bibinfo {year} {2016})},\ \Eprint
  {https://arxiv.org/abs/1507.00902} {arXiv:1507.00902 [gr-qc]} \BibitemShut
  {NoStop}%
\bibitem [{\citenamefont {Yang}\ \emph {et~al.}(2016)\citenamefont {Yang},
  \citenamefont {Li}, \citenamefont {Wu},\ and\ \citenamefont
  {Lu}}]{Yang:2016evp}%
  \BibitemOpen
  \bibfield  {author} {\bibinfo {author} {\bibfnamefont {W.}~\bibnamefont
  {Yang}}, \bibinfo {author} {\bibfnamefont {H.}~\bibnamefont {Li}}, \bibinfo
  {author} {\bibfnamefont {Y.}~\bibnamefont {Wu}},\ and\ \bibinfo {author}
  {\bibfnamefont {J.}~\bibnamefont {Lu}},\ }\href
  {https://doi.org/10.1088/1475-7516/2016/10/007} {\bibfield  {journal}
  {\bibinfo  {journal} {JCAP}\ }\textbf {\bibinfo {volume} {10}},\ \bibinfo
  {pages} {007}},\ \Eprint {https://arxiv.org/abs/1608.07039} {arXiv:1608.07039
  [astro-ph.CO]} \BibitemShut {NoStop}%
\bibitem [{\citenamefont {Wang}\ \emph
  {et~al.}(2016{\natexlab{b}})\citenamefont {Wang}, \citenamefont {Abdalla},
  \citenamefont {Atrio-Barandela},\ and\ \citenamefont {Pavon}}]{Wang:2016lxa}%
  \BibitemOpen
  \bibfield  {author} {\bibinfo {author} {\bibfnamefont {B.}~\bibnamefont
  {Wang}}, \bibinfo {author} {\bibfnamefont {E.}~\bibnamefont {Abdalla}},
  \bibinfo {author} {\bibfnamefont {F.}~\bibnamefont {Atrio-Barandela}},\ and\
  \bibinfo {author} {\bibfnamefont {D.}~\bibnamefont {Pavon}},\ }\href
  {https://doi.org/10.1088/0034-4885/79/9/096901} {\bibfield  {journal}
  {\bibinfo  {journal} {Rept. Prog. Phys.}\ }\textbf {\bibinfo {volume} {79}},\
  \bibinfo {pages} {096901} (\bibinfo {year} {2016}{\natexlab{b}})},\ \Eprint
  {https://arxiv.org/abs/1603.08299} {arXiv:1603.08299 [astro-ph.CO]}
  \BibitemShut {NoStop}%
\bibitem [{\citenamefont {Nunes}\ \emph {et~al.}(2016)\citenamefont {Nunes},
  \citenamefont {Pan},\ and\ \citenamefont {Saridakis}}]{Nunes:2016dlj}%
  \BibitemOpen
  \bibfield  {author} {\bibinfo {author} {\bibfnamefont {R.~C.}\ \bibnamefont
  {Nunes}}, \bibinfo {author} {\bibfnamefont {S.}~\bibnamefont {Pan}},\ and\
  \bibinfo {author} {\bibfnamefont {E.~N.}\ \bibnamefont {Saridakis}},\ }\href
  {https://doi.org/10.1103/PhysRevD.94.023508} {\bibfield  {journal} {\bibinfo
  {journal} {Phys. Rev. D}\ }\textbf {\bibinfo {volume} {94}},\ \bibinfo
  {pages} {023508} (\bibinfo {year} {2016})},\ \Eprint
  {https://arxiv.org/abs/1605.01712} {arXiv:1605.01712 [astro-ph.CO]}
  \BibitemShut {NoStop}%
\bibitem [{\citenamefont {van~de Bruck}\ \emph {et~al.}(2017)\citenamefont
  {van~de Bruck}, \citenamefont {Mifsud},\ and\ \citenamefont
  {Morrice}}]{vandeBruck:2016hpz}%
  \BibitemOpen
  \bibfield  {author} {\bibinfo {author} {\bibfnamefont {C.}~\bibnamefont
  {van~de Bruck}}, \bibinfo {author} {\bibfnamefont {J.}~\bibnamefont
  {Mifsud}},\ and\ \bibinfo {author} {\bibfnamefont {J.}~\bibnamefont
  {Morrice}},\ }\href {https://doi.org/10.1103/PhysRevD.95.043513} {\bibfield
  {journal} {\bibinfo  {journal} {Phys. Rev. D}\ }\textbf {\bibinfo {volume}
  {95}},\ \bibinfo {pages} {043513} (\bibinfo {year} {2017})},\ \Eprint
  {https://arxiv.org/abs/1609.09855} {arXiv:1609.09855 [astro-ph.CO]}
  \BibitemShut {NoStop}%
\bibitem [{\citenamefont {Xia}\ and\ \citenamefont {Wang}(2016)}]{Xia:2016vnp}%
  \BibitemOpen
  \bibfield  {author} {\bibinfo {author} {\bibfnamefont {D.-M.}\ \bibnamefont
  {Xia}}\ and\ \bibinfo {author} {\bibfnamefont {S.}~\bibnamefont {Wang}},\
  }\href {https://doi.org/10.1093/mnras/stw2073} {\bibfield  {journal}
  {\bibinfo  {journal} {Mon. Not. Roy. Astron. Soc.}\ }\textbf {\bibinfo
  {volume} {463}},\ \bibinfo {pages} {952} (\bibinfo {year} {2016})},\ \Eprint
  {https://arxiv.org/abs/1608.04545} {arXiv:1608.04545 [astro-ph.CO]}
  \BibitemShut {NoStop}%
\bibitem [{\citenamefont {Pan}\ and\ \citenamefont
  {Sharov}(2017)}]{Pan:2016ngu}%
  \BibitemOpen
  \bibfield  {author} {\bibinfo {author} {\bibfnamefont {S.}~\bibnamefont
  {Pan}}\ and\ \bibinfo {author} {\bibfnamefont {G.}~\bibnamefont {Sharov}},\
  }\href {https://doi.org/10.1093/mnras/stx2278} {\bibfield  {journal}
  {\bibinfo  {journal} {Mon. Not. Roy. Astron. Soc.}\ }\textbf {\bibinfo
  {volume} {472}},\ \bibinfo {pages} {4736} (\bibinfo {year} {2017})},\ \Eprint
  {https://arxiv.org/abs/1609.02287} {arXiv:1609.02287 [gr-qc]} \BibitemShut
  {NoStop}%
\bibitem [{\citenamefont {Fay}(2016)}]{Fay:2016yow}%
  \BibitemOpen
  \bibfield  {author} {\bibinfo {author} {\bibfnamefont {S.}~\bibnamefont
  {Fay}},\ }\href {https://doi.org/10.1093/mnras/stw1087} {\bibfield  {journal}
  {\bibinfo  {journal} {Mon. Not. Roy. Astron. Soc.}\ }\textbf {\bibinfo
  {volume} {460}},\ \bibinfo {pages} {1863} (\bibinfo {year} {2016})},\ \Eprint
  {https://arxiv.org/abs/1605.01644} {arXiv:1605.01644 [astro-ph.CO]}
  \BibitemShut {NoStop}%
\bibitem [{\citenamefont {Kumar}\ and\ \citenamefont
  {Nunes}(2017)}]{Kumar:2017dnp}%
  \BibitemOpen
  \bibfield  {author} {\bibinfo {author} {\bibfnamefont {S.}~\bibnamefont
  {Kumar}}\ and\ \bibinfo {author} {\bibfnamefont {R.~C.}\ \bibnamefont
  {Nunes}},\ }\href {https://doi.org/10.1103/PhysRevD.96.103511} {\bibfield
  {journal} {\bibinfo  {journal} {Phys. Rev.}\ }\textbf {\bibinfo {volume}
  {D96}},\ \bibinfo {pages} {103511} (\bibinfo {year} {2017})},\ \Eprint
  {https://arxiv.org/abs/1702.02143} {arXiv:1702.02143 [astro-ph.CO]}
  \BibitemShut {NoStop}%
\bibitem [{\citenamefont {Di~Valentino}\ \emph {et~al.}(2017)\citenamefont
  {Di~Valentino}, \citenamefont {Melchiorri},\ and\ \citenamefont
  {Mena}}]{DiValentino:2017iww}%
  \BibitemOpen
  \bibfield  {author} {\bibinfo {author} {\bibfnamefont {E.}~\bibnamefont
  {Di~Valentino}}, \bibinfo {author} {\bibfnamefont {A.}~\bibnamefont
  {Melchiorri}},\ and\ \bibinfo {author} {\bibfnamefont {O.}~\bibnamefont
  {Mena}},\ }\href {https://doi.org/10.1103/PhysRevD.96.043503} {\bibfield
  {journal} {\bibinfo  {journal} {Phys. Rev. D}\ }\textbf {\bibinfo {volume}
  {96}},\ \bibinfo {pages} {043503} (\bibinfo {year} {2017})},\ \Eprint
  {https://arxiv.org/abs/1704.08342} {arXiv:1704.08342 [astro-ph.CO]}
  \BibitemShut {NoStop}%
\bibitem [{\citenamefont {Sharov}\ \emph {et~al.}(2017)\citenamefont {Sharov},
  \citenamefont {Bhattacharya}, \citenamefont {Pan}, \citenamefont {Nunes},\
  and\ \citenamefont {Chakraborty}}]{Sharov:2017iue}%
  \BibitemOpen
  \bibfield  {author} {\bibinfo {author} {\bibfnamefont {G.~S.}\ \bibnamefont
  {Sharov}}, \bibinfo {author} {\bibfnamefont {S.}~\bibnamefont
  {Bhattacharya}}, \bibinfo {author} {\bibfnamefont {S.}~\bibnamefont {Pan}},
  \bibinfo {author} {\bibfnamefont {R.~C.}\ \bibnamefont {Nunes}},\ and\
  \bibinfo {author} {\bibfnamefont {S.}~\bibnamefont {Chakraborty}},\ }\href
  {https://doi.org/10.1093/mnras/stw3358} {\bibfield  {journal} {\bibinfo
  {journal} {Mon. Not. Roy. Astron. Soc.}\ }\textbf {\bibinfo {volume} {466}},\
  \bibinfo {pages} {3497} (\bibinfo {year} {2017})},\ \Eprint
  {https://arxiv.org/abs/1701.00780} {arXiv:1701.00780 [gr-qc]} \BibitemShut
  {NoStop}%
\bibitem [{\citenamefont {Yang}\ \emph
  {et~al.}(2017{\natexlab{a}})\citenamefont {Yang}, \citenamefont {Banerjee},\
  and\ \citenamefont {Pan}}]{Yang:2017yme}%
  \BibitemOpen
  \bibfield  {author} {\bibinfo {author} {\bibfnamefont {W.}~\bibnamefont
  {Yang}}, \bibinfo {author} {\bibfnamefont {N.}~\bibnamefont {Banerjee}},\
  and\ \bibinfo {author} {\bibfnamefont {S.}~\bibnamefont {Pan}},\ }\href
  {https://doi.org/10.1103/PhysRevD.95.123527} {\bibfield  {journal} {\bibinfo
  {journal} {Phys. Rev. D}\ }\textbf {\bibinfo {volume} {95}},\ \bibinfo
  {pages} {123527} (\bibinfo {year} {2017}{\natexlab{a}})},\ \Eprint
  {https://arxiv.org/abs/1705.09278} {arXiv:1705.09278 [astro-ph.CO]}
  \BibitemShut {NoStop}%
\bibitem [{\citenamefont {Yang}\ \emph
  {et~al.}(2017{\natexlab{b}})\citenamefont {Yang}, \citenamefont {Pan},\ and\
  \citenamefont {Mota}}]{Yang:2017ccc}%
  \BibitemOpen
  \bibfield  {author} {\bibinfo {author} {\bibfnamefont {W.}~\bibnamefont
  {Yang}}, \bibinfo {author} {\bibfnamefont {S.}~\bibnamefont {Pan}},\ and\
  \bibinfo {author} {\bibfnamefont {D.~F.}\ \bibnamefont {Mota}},\ }\href
  {https://doi.org/10.1103/PhysRevD.96.123508} {\bibfield  {journal} {\bibinfo
  {journal} {Phys. Rev. D}\ }\textbf {\bibinfo {volume} {96}},\ \bibinfo
  {pages} {123508} (\bibinfo {year} {2017}{\natexlab{b}})},\ \Eprint
  {https://arxiv.org/abs/1709.00006} {arXiv:1709.00006 [astro-ph.CO]}
  \BibitemShut {NoStop}%
\bibitem [{\citenamefont {Pan}\ \emph {et~al.}(2018)\citenamefont {Pan},
  \citenamefont {Mukherjee},\ and\ \citenamefont {Banerjee}}]{Pan:2017ent}%
  \BibitemOpen
  \bibfield  {author} {\bibinfo {author} {\bibfnamefont {S.}~\bibnamefont
  {Pan}}, \bibinfo {author} {\bibfnamefont {A.}~\bibnamefont {Mukherjee}},\
  and\ \bibinfo {author} {\bibfnamefont {N.}~\bibnamefont {Banerjee}},\ }\href
  {https://doi.org/10.1093/mnras/sty755} {\bibfield  {journal} {\bibinfo
  {journal} {Mon. Not. Roy. Astron. Soc.}\ }\textbf {\bibinfo {volume} {477}},\
  \bibinfo {pages} {1189} (\bibinfo {year} {2018})},\ \Eprint
  {https://arxiv.org/abs/1710.03725} {arXiv:1710.03725 [astro-ph.CO]}
  \BibitemShut {NoStop}%
\bibitem [{\citenamefont {Mifsud}\ and\ \citenamefont {Van
  De~Bruck}(2017)}]{Mifsud:2017fsy}%
  \BibitemOpen
  \bibfield  {author} {\bibinfo {author} {\bibfnamefont {J.}~\bibnamefont
  {Mifsud}}\ and\ \bibinfo {author} {\bibfnamefont {C.}~\bibnamefont {Van
  De~Bruck}},\ }\href {https://doi.org/10.1088/1475-7516/2017/11/001}
  {\bibfield  {journal} {\bibinfo  {journal} {JCAP}\ }\textbf {\bibinfo
  {volume} {11}},\ \bibinfo {pages} {001}},\ \Eprint
  {https://arxiv.org/abs/1707.07667} {arXiv:1707.07667 [astro-ph.CO]}
  \BibitemShut {NoStop}%
\bibitem [{\citenamefont {Van De~Bruck}\ and\ \citenamefont
  {Mifsud}(2018)}]{VanDeBruck:2017mua}%
  \BibitemOpen
  \bibfield  {author} {\bibinfo {author} {\bibfnamefont {C.}~\bibnamefont {Van
  De~Bruck}}\ and\ \bibinfo {author} {\bibfnamefont {J.}~\bibnamefont
  {Mifsud}},\ }\href {https://doi.org/10.1103/PhysRevD.97.023506} {\bibfield
  {journal} {\bibinfo  {journal} {Phys. Rev. D}\ }\textbf {\bibinfo {volume}
  {97}},\ \bibinfo {pages} {023506} (\bibinfo {year} {2018})},\ \Eprint
  {https://arxiv.org/abs/1709.04882} {arXiv:1709.04882 [astro-ph.CO]}
  \BibitemShut {NoStop}%
\bibitem [{\citenamefont {Yang}\ \emph
  {et~al.}(2018{\natexlab{a}})\citenamefont {Yang}, \citenamefont {Pan},\ and\
  \citenamefont {Barrow}}]{Yang:2017zjs}%
  \BibitemOpen
  \bibfield  {author} {\bibinfo {author} {\bibfnamefont {W.}~\bibnamefont
  {Yang}}, \bibinfo {author} {\bibfnamefont {S.}~\bibnamefont {Pan}},\ and\
  \bibinfo {author} {\bibfnamefont {J.~D.}\ \bibnamefont {Barrow}},\ }\href
  {https://doi.org/10.1103/PhysRevD.97.043529} {\bibfield  {journal} {\bibinfo
  {journal} {Phys. Rev. D}\ }\textbf {\bibinfo {volume} {97}},\ \bibinfo
  {pages} {043529} (\bibinfo {year} {2018}{\natexlab{a}})},\ \Eprint
  {https://arxiv.org/abs/1706.04953} {arXiv:1706.04953 [astro-ph.CO]}
  \BibitemShut {NoStop}%
\bibitem [{\citenamefont {Yang}\ \emph
  {et~al.}(2018{\natexlab{b}})\citenamefont {Yang}, \citenamefont {Pan},
  \citenamefont {Di~Valentino}, \citenamefont {Nunes}, \citenamefont
  {Vagnozzi},\ and\ \citenamefont {Mota}}]{Yang:2018euj}%
  \BibitemOpen
  \bibfield  {author} {\bibinfo {author} {\bibfnamefont {W.}~\bibnamefont
  {Yang}}, \bibinfo {author} {\bibfnamefont {S.}~\bibnamefont {Pan}}, \bibinfo
  {author} {\bibfnamefont {E.}~\bibnamefont {Di~Valentino}}, \bibinfo {author}
  {\bibfnamefont {R.~C.}\ \bibnamefont {Nunes}}, \bibinfo {author}
  {\bibfnamefont {S.}~\bibnamefont {Vagnozzi}},\ and\ \bibinfo {author}
  {\bibfnamefont {D.~F.}\ \bibnamefont {Mota}},\ }\href
  {https://doi.org/10.1088/1475-7516/2018/09/019} {\bibfield  {journal}
  {\bibinfo  {journal} {JCAP}\ }\textbf {\bibinfo {volume} {09}},\ \bibinfo
  {pages} {019}},\ \Eprint {https://arxiv.org/abs/1805.08252} {arXiv:1805.08252
  [astro-ph.CO]} \BibitemShut {NoStop}%
\bibitem [{\citenamefont {Yang}\ \emph
  {et~al.}(2018{\natexlab{c}})\citenamefont {Yang}, \citenamefont {Pan},
  \citenamefont {Herrera},\ and\ \citenamefont {Chakraborty}}]{Yang:2018xlt}%
  \BibitemOpen
  \bibfield  {author} {\bibinfo {author} {\bibfnamefont {W.}~\bibnamefont
  {Yang}}, \bibinfo {author} {\bibfnamefont {S.}~\bibnamefont {Pan}}, \bibinfo
  {author} {\bibfnamefont {R.}~\bibnamefont {Herrera}},\ and\ \bibinfo {author}
  {\bibfnamefont {S.}~\bibnamefont {Chakraborty}},\ }\href
  {https://doi.org/10.1103/PhysRevD.98.043517} {\bibfield  {journal} {\bibinfo
  {journal} {Phys. Rev. D}\ }\textbf {\bibinfo {volume} {98}},\ \bibinfo
  {pages} {043517} (\bibinfo {year} {2018}{\natexlab{c}})},\ \Eprint
  {https://arxiv.org/abs/1808.01669} {arXiv:1808.01669 [gr-qc]} \BibitemShut
  {NoStop}%
\bibitem [{\citenamefont {Li}\ \emph {et~al.}(2018)\citenamefont {Li},
  \citenamefont {Yang}, \citenamefont {Wu},\ and\ \citenamefont
  {Jiang}}]{Li:2018jiu}%
  \BibitemOpen
  \bibfield  {author} {\bibinfo {author} {\bibfnamefont {H.}~\bibnamefont
  {Li}}, \bibinfo {author} {\bibfnamefont {W.}~\bibnamefont {Yang}}, \bibinfo
  {author} {\bibfnamefont {Y.}~\bibnamefont {Wu}},\ and\ \bibinfo {author}
  {\bibfnamefont {Y.}~\bibnamefont {Jiang}},\ }\href
  {https://doi.org/10.1016/j.dark.2018.04.001} {\bibfield  {journal} {\bibinfo
  {journal} {Phys. Dark Univ.}\ }\textbf {\bibinfo {volume} {20}},\ \bibinfo
  {pages} {78} (\bibinfo {year} {2018})}\BibitemShut {NoStop}%
\bibitem [{\citenamefont {Barros}(2019)}]{Barros:2019rdv}%
  \BibitemOpen
  \bibfield  {author} {\bibinfo {author} {\bibfnamefont {B.~J.}\ \bibnamefont
  {Barros}},\ }\href {https://doi.org/10.1103/PhysRevD.99.064051} {\bibfield
  {journal} {\bibinfo  {journal} {Phys. Rev. D}\ }\textbf {\bibinfo {volume}
  {99}},\ \bibinfo {pages} {064051} (\bibinfo {year} {2019})},\ \Eprint
  {https://arxiv.org/abs/1901.03972} {arXiv:1901.03972 [gr-qc]} \BibitemShut
  {NoStop}%
\bibitem [{\citenamefont {Teixeira}\ \emph {et~al.}(2019)\citenamefont
  {Teixeira}, \citenamefont {Nunes},\ and\ \citenamefont
  {Nunes}}]{Teixeira:2019tfi}%
  \BibitemOpen
  \bibfield  {author} {\bibinfo {author} {\bibfnamefont {E.~M.}\ \bibnamefont
  {Teixeira}}, \bibinfo {author} {\bibfnamefont {A.}~\bibnamefont {Nunes}},\
  and\ \bibinfo {author} {\bibfnamefont {N.~J.}\ \bibnamefont {Nunes}},\ }\href
  {https://doi.org/10.1103/PhysRevD.100.043539} {\bibfield  {journal} {\bibinfo
   {journal} {Phys. Rev. D}\ }\textbf {\bibinfo {volume} {100}},\ \bibinfo
  {pages} {043539} (\bibinfo {year} {2019})},\ \Eprint
  {https://arxiv.org/abs/1903.06028} {arXiv:1903.06028 [gr-qc]} \BibitemShut
  {NoStop}%
\bibitem [{\citenamefont {Di~Valentino}\ \emph
  {et~al.}(2020{\natexlab{a}})\citenamefont {Di~Valentino}, \citenamefont
  {Melchiorri}, \citenamefont {Mena},\ and\ \citenamefont
  {Vagnozzi}}]{DiValentino:2019jae}%
  \BibitemOpen
  \bibfield  {author} {\bibinfo {author} {\bibfnamefont {E.}~\bibnamefont
  {Di~Valentino}}, \bibinfo {author} {\bibfnamefont {A.}~\bibnamefont
  {Melchiorri}}, \bibinfo {author} {\bibfnamefont {O.}~\bibnamefont {Mena}},\
  and\ \bibinfo {author} {\bibfnamefont {S.}~\bibnamefont {Vagnozzi}},\ }\href
  {https://doi.org/10.1103/PhysRevD.101.063502} {\bibfield  {journal} {\bibinfo
   {journal} {Phys. Rev. D}\ }\textbf {\bibinfo {volume} {101}},\ \bibinfo
  {pages} {063502} (\bibinfo {year} {2020}{\natexlab{a}})},\ \Eprint
  {https://arxiv.org/abs/1910.09853} {arXiv:1910.09853 [astro-ph.CO]}
  \BibitemShut {NoStop}%
\bibitem [{\citenamefont {Paliathanasis}\ \emph {et~al.}(2019)\citenamefont
  {Paliathanasis}, \citenamefont {Pan},\ and\ \citenamefont
  {Yang}}]{Paliathanasis:2019hbi}%
  \BibitemOpen
  \bibfield  {author} {\bibinfo {author} {\bibfnamefont {A.}~\bibnamefont
  {Paliathanasis}}, \bibinfo {author} {\bibfnamefont {S.}~\bibnamefont {Pan}},\
  and\ \bibinfo {author} {\bibfnamefont {W.}~\bibnamefont {Yang}},\ }\href
  {https://doi.org/10.1142/S021827181950161X} {\bibfield  {journal} {\bibinfo
  {journal} {Int. J. Mod. Phys. D}\ }\textbf {\bibinfo {volume} {28}},\
  \bibinfo {pages} {1950161} (\bibinfo {year} {2019})},\ \Eprint
  {https://arxiv.org/abs/1903.02370} {arXiv:1903.02370 [gr-qc]} \BibitemShut
  {NoStop}%
\bibitem [{\citenamefont {Yang}\ \emph
  {et~al.}(2020{\natexlab{a}})\citenamefont {Yang}, \citenamefont {Pan},
  \citenamefont {Nunes},\ and\ \citenamefont {Mota}}]{Yang:2019uog}%
  \BibitemOpen
  \bibfield  {author} {\bibinfo {author} {\bibfnamefont {W.}~\bibnamefont
  {Yang}}, \bibinfo {author} {\bibfnamefont {S.}~\bibnamefont {Pan}}, \bibinfo
  {author} {\bibfnamefont {R.~C.}\ \bibnamefont {Nunes}},\ and\ \bibinfo
  {author} {\bibfnamefont {D.~F.}\ \bibnamefont {Mota}},\ }\href
  {https://doi.org/10.1088/1475-7516/2020/04/008} {\bibfield  {journal}
  {\bibinfo  {journal} {JCAP}\ }\textbf {\bibinfo {volume} {04}},\ \bibinfo
  {pages} {008}},\ \Eprint {https://arxiv.org/abs/1910.08821} {arXiv:1910.08821
  [astro-ph.CO]} \BibitemShut {NoStop}%
\bibitem [{\citenamefont {Pan}\ \emph {et~al.}(2019{\natexlab{a}})\citenamefont
  {Pan}, \citenamefont {Yang}, \citenamefont {Di~Valentino}, \citenamefont
  {Saridakis},\ and\ \citenamefont {Chakraborty}}]{Pan:2019gop}%
  \BibitemOpen
  \bibfield  {author} {\bibinfo {author} {\bibfnamefont {S.}~\bibnamefont
  {Pan}}, \bibinfo {author} {\bibfnamefont {W.}~\bibnamefont {Yang}}, \bibinfo
  {author} {\bibfnamefont {E.}~\bibnamefont {Di~Valentino}}, \bibinfo {author}
  {\bibfnamefont {E.~N.}\ \bibnamefont {Saridakis}},\ and\ \bibinfo {author}
  {\bibfnamefont {S.}~\bibnamefont {Chakraborty}},\ }\href
  {https://doi.org/10.1103/PhysRevD.100.103520} {\bibfield  {journal} {\bibinfo
   {journal} {Phys. Rev. D}\ }\textbf {\bibinfo {volume} {100}},\ \bibinfo
  {pages} {103520} (\bibinfo {year} {2019}{\natexlab{a}})},\ \Eprint
  {https://arxiv.org/abs/1907.07540} {arXiv:1907.07540 [astro-ph.CO]}
  \BibitemShut {NoStop}%
\bibitem [{\citenamefont {Nakamura}\ \emph {et~al.}(2019)\citenamefont
  {Nakamura}, \citenamefont {Kase},\ and\ \citenamefont
  {Tsujikawa}}]{Nakamura:2019phn}%
  \BibitemOpen
  \bibfield  {author} {\bibinfo {author} {\bibfnamefont {S.}~\bibnamefont
  {Nakamura}}, \bibinfo {author} {\bibfnamefont {R.}~\bibnamefont {Kase}},\
  and\ \bibinfo {author} {\bibfnamefont {S.}~\bibnamefont {Tsujikawa}},\ }\href
  {https://doi.org/10.1088/1475-7516/2019/12/032} {\bibfield  {journal}
  {\bibinfo  {journal} {JCAP}\ }\textbf {\bibinfo {volume} {12}},\ \bibinfo
  {pages} {032}},\ \Eprint {https://arxiv.org/abs/1907.12216} {arXiv:1907.12216
  [gr-qc]} \BibitemShut {NoStop}%
\bibitem [{\citenamefont {Di~Valentino}\ \emph
  {et~al.}(2020{\natexlab{b}})\citenamefont {Di~Valentino}, \citenamefont
  {Melchiorri}, \citenamefont {Mena},\ and\ \citenamefont
  {Vagnozzi}}]{DiValentino:2019ffd}%
  \BibitemOpen
  \bibfield  {author} {\bibinfo {author} {\bibfnamefont {E.}~\bibnamefont
  {Di~Valentino}}, \bibinfo {author} {\bibfnamefont {A.}~\bibnamefont
  {Melchiorri}}, \bibinfo {author} {\bibfnamefont {O.}~\bibnamefont {Mena}},\
  and\ \bibinfo {author} {\bibfnamefont {S.}~\bibnamefont {Vagnozzi}},\ }\href
  {https://doi.org/10.1016/j.dark.2020.100666} {\bibfield  {journal} {\bibinfo
  {journal} {Phys. Dark Univ.}\ }\textbf {\bibinfo {volume} {30}},\ \bibinfo
  {pages} {100666} (\bibinfo {year} {2020}{\natexlab{b}})},\ \Eprint
  {https://arxiv.org/abs/1908.04281} {arXiv:1908.04281 [astro-ph.CO]}
  \BibitemShut {NoStop}%
\bibitem [{\citenamefont {Yang}\ \emph
  {et~al.}(2020{\natexlab{b}})\citenamefont {Yang}, \citenamefont {Pan},
  \citenamefont {Di~Valentino}, \citenamefont {Wang},\ and\ \citenamefont
  {Wang}}]{Yang:2019bpr}%
  \BibitemOpen
  \bibfield  {author} {\bibinfo {author} {\bibfnamefont {W.}~\bibnamefont
  {Yang}}, \bibinfo {author} {\bibfnamefont {S.}~\bibnamefont {Pan}}, \bibinfo
  {author} {\bibfnamefont {E.}~\bibnamefont {Di~Valentino}}, \bibinfo {author}
  {\bibfnamefont {B.}~\bibnamefont {Wang}},\ and\ \bibinfo {author}
  {\bibfnamefont {A.}~\bibnamefont {Wang}},\ }\href
  {https://doi.org/10.1088/1475-7516/2020/05/050} {\bibfield  {journal}
  {\bibinfo  {journal} {JCAP}\ }\textbf {\bibinfo {volume} {05}},\ \bibinfo
  {pages} {050}},\ \Eprint {https://arxiv.org/abs/1904.11980} {arXiv:1904.11980
  [astro-ph.CO]} \BibitemShut {NoStop}%
\bibitem [{\citenamefont {Kumar}\ \emph {et~al.}(2019)\citenamefont {Kumar},
  \citenamefont {Nunes},\ and\ \citenamefont {Yadav}}]{Kumar:2019wfs}%
  \BibitemOpen
  \bibfield  {author} {\bibinfo {author} {\bibfnamefont {S.}~\bibnamefont
  {Kumar}}, \bibinfo {author} {\bibfnamefont {R.~C.}\ \bibnamefont {Nunes}},\
  and\ \bibinfo {author} {\bibfnamefont {S.~K.}\ \bibnamefont {Yadav}},\ }\href
  {https://doi.org/10.1140/epjc/s10052-019-7087-7} {\bibfield  {journal}
  {\bibinfo  {journal} {Eur. Phys. J.}\ }\textbf {\bibinfo {volume} {C79}},\
  \bibinfo {pages} {576} (\bibinfo {year} {2019})},\ \Eprint
  {https://arxiv.org/abs/1903.04865} {arXiv:1903.04865 [astro-ph.CO]}
  \BibitemShut {NoStop}%
\bibitem [{\citenamefont {Martinelli}\ \emph {et~al.}(2019)\citenamefont
  {Martinelli}, \citenamefont {Hogg}, \citenamefont {Peirone}, \citenamefont
  {Bruni},\ and\ \citenamefont {Wands}}]{Martinelli:2019dau}%
  \BibitemOpen
  \bibfield  {author} {\bibinfo {author} {\bibfnamefont {M.}~\bibnamefont
  {Martinelli}}, \bibinfo {author} {\bibfnamefont {N.~B.}\ \bibnamefont
  {Hogg}}, \bibinfo {author} {\bibfnamefont {S.}~\bibnamefont {Peirone}},
  \bibinfo {author} {\bibfnamefont {M.}~\bibnamefont {Bruni}},\ and\ \bibinfo
  {author} {\bibfnamefont {D.}~\bibnamefont {Wands}},\ }\href
  {https://doi.org/10.1093/mnras/stz1915} {\bibfield  {journal} {\bibinfo
  {journal} {Mon. Not. Roy. Astron. Soc.}\ }\textbf {\bibinfo {volume} {488}},\
  \bibinfo {pages} {3423} (\bibinfo {year} {2019})},\ \Eprint
  {https://arxiv.org/abs/1902.10694} {arXiv:1902.10694 [astro-ph.CO]}
  \BibitemShut {NoStop}%
\bibitem [{\citenamefont {Costa}\ \emph {et~al.}(2019)\citenamefont {Costa}
  \emph {et~al.}}]{Costa:2019uvk}%
  \BibitemOpen
  \bibfield  {author} {\bibinfo {author} {\bibfnamefont {A.~A.}\ \bibnamefont
  {Costa}} \emph {et~al.},\ }\href {https://doi.org/10.1093/mnras/stz1675}
  {\bibfield  {journal} {\bibinfo  {journal} {Mon. Not. Roy. Astron. Soc.}\
  }\textbf {\bibinfo {volume} {488}},\ \bibinfo {pages} {78} (\bibinfo {year}
  {2019})},\ \Eprint {https://arxiv.org/abs/1901.02540} {arXiv:1901.02540
  [astro-ph.CO]} \BibitemShut {NoStop}%
\bibitem [{\citenamefont {Arevalo}\ \emph {et~al.}(2019)\citenamefont
  {Arevalo}, \citenamefont {Cid}, \citenamefont {Chimento},\ and\ \citenamefont
  {Mella}}]{Arevalo:2019axj}%
  \BibitemOpen
  \bibfield  {author} {\bibinfo {author} {\bibfnamefont {F.}~\bibnamefont
  {Arevalo}}, \bibinfo {author} {\bibfnamefont {A.}~\bibnamefont {Cid}},
  \bibinfo {author} {\bibfnamefont {L.~P.}\ \bibnamefont {Chimento}},\ and\
  \bibinfo {author} {\bibfnamefont {P.}~\bibnamefont {Mella}},\ }\href
  {https://doi.org/10.1140/epjc/s10052-019-6872-7} {\bibfield  {journal}
  {\bibinfo  {journal} {Eur. Phys. J. C}\ }\textbf {\bibinfo {volume} {79}},\
  \bibinfo {pages} {355} (\bibinfo {year} {2019})},\ \Eprint
  {https://arxiv.org/abs/1901.04300} {arXiv:1901.04300 [gr-qc]} \BibitemShut
  {NoStop}%
\bibitem [{\citenamefont {Pan}\ \emph {et~al.}(2019{\natexlab{b}})\citenamefont
  {Pan}, \citenamefont {Yang}, \citenamefont {Singha},\ and\ \citenamefont
  {Saridakis}}]{Pan:2019jqh}%
  \BibitemOpen
  \bibfield  {author} {\bibinfo {author} {\bibfnamefont {S.}~\bibnamefont
  {Pan}}, \bibinfo {author} {\bibfnamefont {W.}~\bibnamefont {Yang}}, \bibinfo
  {author} {\bibfnamefont {C.}~\bibnamefont {Singha}},\ and\ \bibinfo {author}
  {\bibfnamefont {E.~N.}\ \bibnamefont {Saridakis}},\ }\href
  {https://doi.org/10.1103/PhysRevD.100.083539} {\bibfield  {journal} {\bibinfo
   {journal} {Phys. Rev. D}\ }\textbf {\bibinfo {volume} {100}},\ \bibinfo
  {pages} {083539} (\bibinfo {year} {2019}{\natexlab{b}})},\ \Eprint
  {https://arxiv.org/abs/1903.10969} {arXiv:1903.10969 [astro-ph.CO]}
  \BibitemShut {NoStop}%
\bibitem [{\citenamefont {Cheng}\ \emph {et~al.}(2020)\citenamefont {Cheng},
  \citenamefont {Ma}, \citenamefont {Wu}, \citenamefont {Zhang},\ and\
  \citenamefont {Chen}}]{Cheng:2019bkh}%
  \BibitemOpen
  \bibfield  {author} {\bibinfo {author} {\bibfnamefont {G.}~\bibnamefont
  {Cheng}}, \bibinfo {author} {\bibfnamefont {Y.-Z.}\ \bibnamefont {Ma}},
  \bibinfo {author} {\bibfnamefont {F.}~\bibnamefont {Wu}}, \bibinfo {author}
  {\bibfnamefont {J.}~\bibnamefont {Zhang}},\ and\ \bibinfo {author}
  {\bibfnamefont {X.}~\bibnamefont {Chen}},\ }\href
  {https://doi.org/10.1103/PhysRevD.102.043517} {\bibfield  {journal} {\bibinfo
   {journal} {Phys. Rev. D}\ }\textbf {\bibinfo {volume} {102}},\ \bibinfo
  {pages} {043517} (\bibinfo {year} {2020})},\ \Eprint
  {https://arxiv.org/abs/1911.04520} {arXiv:1911.04520 [astro-ph.CO]}
  \BibitemShut {NoStop}%
\bibitem [{\citenamefont {Oikonomou}(2019)}]{Oikonomou:2019nmm}%
  \BibitemOpen
  \bibfield  {author} {\bibinfo {author} {\bibfnamefont {V.~K.}\ \bibnamefont
  {Oikonomou}},\ }\href {https://doi.org/10.1016/j.aop.2019.167934} {\bibfield
  {journal} {\bibinfo  {journal} {Annals Phys.}\ }\textbf {\bibinfo {volume}
  {409}},\ \bibinfo {pages} {167934} (\bibinfo {year} {2019})},\ \Eprint
  {https://arxiv.org/abs/1907.02600} {arXiv:1907.02600 [gr-qc]} \BibitemShut
  {NoStop}%
\bibitem [{\citenamefont {Kase}\ and\ \citenamefont
  {Tsujikawa}(2020)}]{Kase:2019veo}%
  \BibitemOpen
  \bibfield  {author} {\bibinfo {author} {\bibfnamefont {R.}~\bibnamefont
  {Kase}}\ and\ \bibinfo {author} {\bibfnamefont {S.}~\bibnamefont
  {Tsujikawa}},\ }\href {https://doi.org/10.1103/PhysRevD.101.063511}
  {\bibfield  {journal} {\bibinfo  {journal} {Phys. Rev. D}\ }\textbf {\bibinfo
  {volume} {101}},\ \bibinfo {pages} {063511} (\bibinfo {year} {2020})},\
  \Eprint {https://arxiv.org/abs/1910.02699} {arXiv:1910.02699 [gr-qc]}
  \BibitemShut {NoStop}%
\bibitem [{\citenamefont {Pan}\ \emph {et~al.}(2020{\natexlab{a}})\citenamefont
  {Pan}, \citenamefont {Sharov},\ and\ \citenamefont {Yang}}]{Pan:2020zza}%
  \BibitemOpen
  \bibfield  {author} {\bibinfo {author} {\bibfnamefont {S.}~\bibnamefont
  {Pan}}, \bibinfo {author} {\bibfnamefont {G.~S.}\ \bibnamefont {Sharov}},\
  and\ \bibinfo {author} {\bibfnamefont {W.}~\bibnamefont {Yang}},\ }\href
  {https://doi.org/10.1103/PhysRevD.101.103533} {\bibfield  {journal} {\bibinfo
   {journal} {Phys. Rev. D}\ }\textbf {\bibinfo {volume} {101}},\ \bibinfo
  {pages} {103533} (\bibinfo {year} {2020}{\natexlab{a}})},\ \Eprint
  {https://arxiv.org/abs/2001.03120} {arXiv:2001.03120 [astro-ph.CO]}
  \BibitemShut {NoStop}%
\bibitem [{\citenamefont {G{\'o}mez-Valent}\ \emph {et~al.}(2020)\citenamefont
  {G{\'o}mez-Valent}, \citenamefont {Pettorino},\ and\ \citenamefont
  {Amendola}}]{Gomez-Valent:2020mqn}%
  \BibitemOpen
  \bibfield  {author} {\bibinfo {author} {\bibfnamefont {A.}~\bibnamefont
  {G{\'o}mez-Valent}}, \bibinfo {author} {\bibfnamefont {V.}~\bibnamefont
  {Pettorino}},\ and\ \bibinfo {author} {\bibfnamefont {L.}~\bibnamefont
  {Amendola}},\ }\href {https://doi.org/10.1103/PhysRevD.101.123513} {\bibfield
   {journal} {\bibinfo  {journal} {Phys. Rev. D}\ }\textbf {\bibinfo {volume}
  {101}},\ \bibinfo {pages} {123513} (\bibinfo {year} {2020})},\ \Eprint
  {https://arxiv.org/abs/2004.00610} {arXiv:2004.00610 [astro-ph.CO]}
  \BibitemShut {NoStop}%
\bibitem [{\citenamefont {Di~Valentino}\ and\ \citenamefont
  {Mena}(2020)}]{DiValentino:2020leo}%
  \BibitemOpen
  \bibfield  {author} {\bibinfo {author} {\bibfnamefont {E.}~\bibnamefont
  {Di~Valentino}}\ and\ \bibinfo {author} {\bibfnamefont {O.}~\bibnamefont
  {Mena}},\ }\href {https://doi.org/10.1093/mnrasl/slaa175} {\bibfield
  {journal} {\bibinfo  {journal} {Mon. Not. Roy. Astron. Soc.}\ }\textbf
  {\bibinfo {volume} {500}},\ \bibinfo {pages} {L22} (\bibinfo {year}
  {2020})},\ \Eprint {https://arxiv.org/abs/2009.12620} {arXiv:2009.12620
  [astro-ph.CO]} \BibitemShut {NoStop}%
\bibitem [{\citenamefont {Yang}\ \emph
  {et~al.}(2020{\natexlab{c}})\citenamefont {Yang}, \citenamefont
  {Di~Valentino}, \citenamefont {Mena}, \citenamefont {Pan},\ and\
  \citenamefont {Nunes}}]{Yang:2020uga}%
  \BibitemOpen
  \bibfield  {author} {\bibinfo {author} {\bibfnamefont {W.}~\bibnamefont
  {Yang}}, \bibinfo {author} {\bibfnamefont {E.}~\bibnamefont {Di~Valentino}},
  \bibinfo {author} {\bibfnamefont {O.}~\bibnamefont {Mena}}, \bibinfo {author}
  {\bibfnamefont {S.}~\bibnamefont {Pan}},\ and\ \bibinfo {author}
  {\bibfnamefont {R.~C.}\ \bibnamefont {Nunes}},\ }\href
  {https://doi.org/10.1103/PhysRevD.101.083509} {\bibfield  {journal} {\bibinfo
   {journal} {Phys. Rev. D}\ }\textbf {\bibinfo {volume} {101}},\ \bibinfo
  {pages} {083509} (\bibinfo {year} {2020}{\natexlab{c}})},\ \Eprint
  {https://arxiv.org/abs/2001.10852} {arXiv:2001.10852 [astro-ph.CO]}
  \BibitemShut {NoStop}%
\bibitem [{\citenamefont {Lucca}\ and\ \citenamefont
  {Hooper}(2020)}]{Lucca:2020zjb}%
  \BibitemOpen
  \bibfield  {author} {\bibinfo {author} {\bibfnamefont {M.}~\bibnamefont
  {Lucca}}\ and\ \bibinfo {author} {\bibfnamefont {D.~C.}\ \bibnamefont
  {Hooper}},\ }\href {https://doi.org/10.1103/PhysRevD.102.123502} {\bibfield
  {journal} {\bibinfo  {journal} {Phys. Rev. D}\ }\textbf {\bibinfo {volume}
  {102}},\ \bibinfo {pages} {123502} (\bibinfo {year} {2020})},\ \Eprint
  {https://arxiv.org/abs/2002.06127} {arXiv:2002.06127 [astro-ph.CO]}
  \BibitemShut {NoStop}%
\bibitem [{\citenamefont {Di~Valentino}\ \emph {et~al.}(2021)\citenamefont
  {Di~Valentino}, \citenamefont {Melchiorri}, \citenamefont {Mena},
  \citenamefont {Pan},\ and\ \citenamefont {Yang}}]{DiValentino:2020kpf}%
  \BibitemOpen
  \bibfield  {author} {\bibinfo {author} {\bibfnamefont {E.}~\bibnamefont
  {Di~Valentino}}, \bibinfo {author} {\bibfnamefont {A.}~\bibnamefont
  {Melchiorri}}, \bibinfo {author} {\bibfnamefont {O.}~\bibnamefont {Mena}},
  \bibinfo {author} {\bibfnamefont {S.}~\bibnamefont {Pan}},\ and\ \bibinfo
  {author} {\bibfnamefont {W.}~\bibnamefont {Yang}},\ }\href
  {https://doi.org/10.1093/mnrasl/slaa207} {\bibfield  {journal} {\bibinfo
  {journal} {Mon. Not. Roy. Astron. Soc.}\ }\textbf {\bibinfo {volume} {502}},\
  \bibinfo {pages} {L23} (\bibinfo {year} {2021})},\ \Eprint
  {https://arxiv.org/abs/2011.00283} {arXiv:2011.00283 [astro-ph.CO]}
  \BibitemShut {NoStop}%
\bibitem [{\citenamefont {Hogg}\ \emph {et~al.}(2020)\citenamefont {Hogg},
  \citenamefont {Bruni}, \citenamefont {Crittenden}, \citenamefont
  {Martinelli},\ and\ \citenamefont {Peirone}}]{Hogg:2020rdp}%
  \BibitemOpen
  \bibfield  {author} {\bibinfo {author} {\bibfnamefont {N.~B.}\ \bibnamefont
  {Hogg}}, \bibinfo {author} {\bibfnamefont {M.}~\bibnamefont {Bruni}},
  \bibinfo {author} {\bibfnamefont {R.}~\bibnamefont {Crittenden}}, \bibinfo
  {author} {\bibfnamefont {M.}~\bibnamefont {Martinelli}},\ and\ \bibinfo
  {author} {\bibfnamefont {S.}~\bibnamefont {Peirone}},\ }\href
  {https://doi.org/10.1016/j.dark.2020.100583} {\bibfield  {journal} {\bibinfo
  {journal} {Phys. Dark Univ.}\ }\textbf {\bibinfo {volume} {29}},\ \bibinfo
  {pages} {100583} (\bibinfo {year} {2020})},\ \Eprint
  {https://arxiv.org/abs/2002.10449} {arXiv:2002.10449 [astro-ph.CO]}
  \BibitemShut {NoStop}%
\bibitem [{\citenamefont {Yao}\ and\ \citenamefont {Meng}(2020)}]{Yao:2020hkw}%
  \BibitemOpen
  \bibfield  {author} {\bibinfo {author} {\bibfnamefont {Y.-H.}\ \bibnamefont
  {Yao}}\ and\ \bibinfo {author} {\bibfnamefont {X.-H.}\ \bibnamefont {Meng}},\
  }\href {https://doi.org/10.1016/j.dark.2020.100729} {\bibfield  {journal}
  {\bibinfo  {journal} {Phys. Dark Univ.}\ }\textbf {\bibinfo {volume} {30}},\
  \bibinfo {pages} {100729} (\bibinfo {year} {2020})},\ \Eprint
  {https://arxiv.org/abs/2205.14928} {arXiv:2205.14928 [astro-ph.CO]}
  \BibitemShut {NoStop}%
\bibitem [{\citenamefont {Beltr\'an~Jim\'enez}\ \emph
  {et~al.}(2021)\citenamefont {Beltr\'an~Jim\'enez}, \citenamefont {Bettoni},
  \citenamefont {Figueruelo}, \citenamefont {Teppa~Pannia},\ and\ \citenamefont
  {Tsujikawa}}]{BeltranJimenez:2020qdu}%
  \BibitemOpen
  \bibfield  {author} {\bibinfo {author} {\bibfnamefont {J.}~\bibnamefont
  {Beltr\'an~Jim\'enez}}, \bibinfo {author} {\bibfnamefont {D.}~\bibnamefont
  {Bettoni}}, \bibinfo {author} {\bibfnamefont {D.}~\bibnamefont {Figueruelo}},
  \bibinfo {author} {\bibfnamefont {F.~A.}\ \bibnamefont {Teppa~Pannia}},\ and\
  \bibinfo {author} {\bibfnamefont {S.}~\bibnamefont {Tsujikawa}},\ }\href
  {https://doi.org/10.1088/1475-7516/2021/03/085} {\bibfield  {journal}
  {\bibinfo  {journal} {JCAP}\ }\textbf {\bibinfo {volume} {03}},\ \bibinfo
  {pages} {085}},\ \Eprint {https://arxiv.org/abs/2012.12204} {arXiv:2012.12204
  [astro-ph.CO]} \BibitemShut {NoStop}%
\bibitem [{\citenamefont {Yao}\ and\ \citenamefont {Meng}(2021)}]{Yao:2020pji}%
  \BibitemOpen
  \bibfield  {author} {\bibinfo {author} {\bibfnamefont {Y.}~\bibnamefont
  {Yao}}\ and\ \bibinfo {author} {\bibfnamefont {X.-H.}\ \bibnamefont {Meng}},\
  }\href {https://doi.org/10.1016/j.dark.2021.100852} {\bibfield  {journal}
  {\bibinfo  {journal} {Phys. Dark Univ.}\ }\textbf {\bibinfo {volume} {33}},\
  \bibinfo {pages} {100852} (\bibinfo {year} {2021})},\ \Eprint
  {https://arxiv.org/abs/2011.09160} {arXiv:2011.09160 [astro-ph.CO]}
  \BibitemShut {NoStop}%
\bibitem [{\citenamefont {Lucca}(2021{\natexlab{a}})}]{Lucca:2021dxo}%
  \BibitemOpen
  \bibfield  {author} {\bibinfo {author} {\bibfnamefont {M.}~\bibnamefont
  {Lucca}},\ }\href {https://doi.org/10.1016/j.dark.2021.100899} {\bibfield
  {journal} {\bibinfo  {journal} {Phys. Dark Univ.}\ }\textbf {\bibinfo
  {volume} {34}},\ \bibinfo {pages} {100899} (\bibinfo {year}
  {2021}{\natexlab{a}})},\ \Eprint {https://arxiv.org/abs/2105.09249}
  {arXiv:2105.09249 [astro-ph.CO]} \BibitemShut {NoStop}%
\bibitem [{\citenamefont {Lucca}(2021{\natexlab{b}})}]{Lucca:2021eqy}%
  \BibitemOpen
  \bibfield  {author} {\bibinfo {author} {\bibfnamefont {M.}~\bibnamefont
  {Lucca}},\ }\href {https://doi.org/10.1103/PhysRevD.104.083510} {\bibfield
  {journal} {\bibinfo  {journal} {Phys. Rev. D}\ }\textbf {\bibinfo {volume}
  {104}},\ \bibinfo {pages} {083510} (\bibinfo {year} {2021}{\natexlab{b}})},\
  \Eprint {https://arxiv.org/abs/2106.15196} {arXiv:2106.15196 [astro-ph.CO]}
  \BibitemShut {NoStop}%
\bibitem [{\citenamefont {Nunes}\ and\ \citenamefont
  {Di~Valentino}(2021)}]{Nunes:2021zzi}%
  \BibitemOpen
  \bibfield  {author} {\bibinfo {author} {\bibfnamefont {R.~C.}\ \bibnamefont
  {Nunes}}\ and\ \bibinfo {author} {\bibfnamefont {E.}~\bibnamefont
  {Di~Valentino}},\ }\href {https://doi.org/10.1103/PhysRevD.104.063529}
  {\bibfield  {journal} {\bibinfo  {journal} {Phys. Rev. D}\ }\textbf {\bibinfo
  {volume} {104}},\ \bibinfo {pages} {063529} (\bibinfo {year} {2021})},\
  \Eprint {https://arxiv.org/abs/2107.09151} {arXiv:2107.09151 [astro-ph.CO]}
  \BibitemShut {NoStop}%
\bibitem [{\citenamefont {Gariazzo}\ \emph {et~al.}(2022)\citenamefont
  {Gariazzo}, \citenamefont {Di~Valentino}, \citenamefont {Mena},\ and\
  \citenamefont {Nunes}}]{Gariazzo:2021qtg}%
  \BibitemOpen
  \bibfield  {author} {\bibinfo {author} {\bibfnamefont {S.}~\bibnamefont
  {Gariazzo}}, \bibinfo {author} {\bibfnamefont {E.}~\bibnamefont
  {Di~Valentino}}, \bibinfo {author} {\bibfnamefont {O.}~\bibnamefont {Mena}},\
  and\ \bibinfo {author} {\bibfnamefont {R.~C.}\ \bibnamefont {Nunes}},\ }\href
  {https://doi.org/10.1103/PhysRevD.106.023530} {\bibfield  {journal} {\bibinfo
   {journal} {Phys. Rev. D}\ }\textbf {\bibinfo {volume} {106}},\ \bibinfo
  {pages} {023530} (\bibinfo {year} {2022})},\ \Eprint
  {https://arxiv.org/abs/2111.03152} {arXiv:2111.03152 [astro-ph.CO]}
  \BibitemShut {NoStop}%
\bibitem [{\citenamefont {Potting}\ and\ \citenamefont
  {S\'a}(2022)}]{Potting:2021bje}%
  \BibitemOpen
  \bibfield  {author} {\bibinfo {author} {\bibfnamefont {R.}~\bibnamefont
  {Potting}}\ and\ \bibinfo {author} {\bibfnamefont {P.~M.}\ \bibnamefont
  {S\'a}},\ }\href {https://doi.org/10.1142/S0218271822501115} {\bibfield
  {journal} {\bibinfo  {journal} {Int. J. Mod. Phys. D}\ }\textbf {\bibinfo
  {volume} {31}},\ \bibinfo {pages} {2250111} (\bibinfo {year} {2022})},\
  \Eprint {https://arxiv.org/abs/2112.07608} {arXiv:2112.07608 [gr-qc]}
  \BibitemShut {NoStop}%
\bibitem [{\citenamefont {S\'a}(2021)}]{Sa:2021eft}%
  \BibitemOpen
  \bibfield  {author} {\bibinfo {author} {\bibfnamefont {P.~M.}\ \bibnamefont
  {S\'a}},\ }\href {https://doi.org/10.1103/PhysRevD.103.123517} {\bibfield
  {journal} {\bibinfo  {journal} {Phys. Rev. D}\ }\textbf {\bibinfo {volume}
  {103}},\ \bibinfo {pages} {123517} (\bibinfo {year} {2021})},\ \Eprint
  {https://arxiv.org/abs/2103.01693} {arXiv:2103.01693 [gr-qc]} \BibitemShut
  {NoStop}%
\bibitem [{\citenamefont {da~Fonseca}\ \emph {et~al.}(2022)\citenamefont
  {da~Fonseca}, \citenamefont {Barreiro},\ and\ \citenamefont
  {Nunes}}]{daFonseca:2021imp}%
  \BibitemOpen
  \bibfield  {author} {\bibinfo {author} {\bibfnamefont {V.}~\bibnamefont
  {da~Fonseca}}, \bibinfo {author} {\bibfnamefont {T.}~\bibnamefont
  {Barreiro}},\ and\ \bibinfo {author} {\bibfnamefont {N.~J.}\ \bibnamefont
  {Nunes}},\ }\href {https://doi.org/10.1016/j.dark.2021.100940} {\bibfield
  {journal} {\bibinfo  {journal} {Phys. Dark Univ.}\ }\textbf {\bibinfo
  {volume} {35}},\ \bibinfo {pages} {100940} (\bibinfo {year} {2022})},\
  \Eprint {https://arxiv.org/abs/2104.14889} {arXiv:2104.14889 [astro-ph.CO]}
  \BibitemShut {NoStop}%
\bibitem [{\citenamefont {Gao}\ \emph {et~al.}(2021)\citenamefont {Gao},
  \citenamefont {Zhao}, \citenamefont {Xue},\ and\ \citenamefont
  {Zhang}}]{Gao:2021xnk}%
  \BibitemOpen
  \bibfield  {author} {\bibinfo {author} {\bibfnamefont {L.-Y.}\ \bibnamefont
  {Gao}}, \bibinfo {author} {\bibfnamefont {Z.-W.}\ \bibnamefont {Zhao}},
  \bibinfo {author} {\bibfnamefont {S.-S.}\ \bibnamefont {Xue}},\ and\ \bibinfo
  {author} {\bibfnamefont {X.}~\bibnamefont {Zhang}},\ }\href
  {https://doi.org/10.1088/1475-7516/2021/07/005} {\bibfield  {journal}
  {\bibinfo  {journal} {JCAP}\ }\textbf {\bibinfo {volume} {07}},\ \bibinfo
  {pages} {005}},\ \Eprint {https://arxiv.org/abs/2101.10714} {arXiv:2101.10714
  [astro-ph.CO]} \BibitemShut {NoStop}%
\bibitem [{\citenamefont {Thipaksorn}\ \emph {et~al.}(2022)\citenamefont
  {Thipaksorn}, \citenamefont {Sapa},\ and\ \citenamefont
  {Karwan}}]{Thipaksorn:2022yul}%
  \BibitemOpen
  \bibfield  {author} {\bibinfo {author} {\bibfnamefont {W.}~\bibnamefont
  {Thipaksorn}}, \bibinfo {author} {\bibfnamefont {S.}~\bibnamefont {Sapa}},\
  and\ \bibinfo {author} {\bibfnamefont {K.}~\bibnamefont {Karwan}},\ }\href
  {https://doi.org/10.1103/PhysRevD.105.063527} {\bibfield  {journal} {\bibinfo
   {journal} {Phys. Rev. D}\ }\textbf {\bibinfo {volume} {105}},\ \bibinfo
  {pages} {063527} (\bibinfo {year} {2022})},\ \Eprint
  {https://arxiv.org/abs/2201.03261} {arXiv:2201.03261 [gr-qc]} \BibitemShut
  {NoStop}%
\bibitem [{\citenamefont {G{\'o}mez-Valent}\ \emph {et~al.}(2022)\citenamefont
  {G{\'o}mez-Valent}, \citenamefont {Zheng}, \citenamefont {Amendola},
  \citenamefont {Wetterich},\ and\ \citenamefont
  {Pettorino}}]{Gomez-Valent:2022bku}%
  \BibitemOpen
  \bibfield  {author} {\bibinfo {author} {\bibfnamefont {A.}~\bibnamefont
  {G{\'o}mez-Valent}}, \bibinfo {author} {\bibfnamefont {Z.}~\bibnamefont
  {Zheng}}, \bibinfo {author} {\bibfnamefont {L.}~\bibnamefont {Amendola}},
  \bibinfo {author} {\bibfnamefont {C.}~\bibnamefont {Wetterich}},\ and\
  \bibinfo {author} {\bibfnamefont {V.}~\bibnamefont {Pettorino}},\ }\href
  {https://doi.org/10.1103/PhysRevD.106.103522} {\bibfield  {journal} {\bibinfo
   {journal} {Phys. Rev. D}\ }\textbf {\bibinfo {volume} {106}},\ \bibinfo
  {pages} {103522} (\bibinfo {year} {2022})},\ \Eprint
  {https://arxiv.org/abs/2207.14487} {arXiv:2207.14487 [astro-ph.CO]}
  \BibitemShut {NoStop}%
\bibitem [{\citenamefont {Harko}\ \emph {et~al.}(2022)\citenamefont {Harko},
  \citenamefont {Asadi}, \citenamefont {Moshafi},\ and\ \citenamefont
  {Sheikhahmadi}}]{Harko:2022unn}%
  \BibitemOpen
  \bibfield  {author} {\bibinfo {author} {\bibfnamefont {T.}~\bibnamefont
  {Harko}}, \bibinfo {author} {\bibfnamefont {K.}~\bibnamefont {Asadi}},
  \bibinfo {author} {\bibfnamefont {H.}~\bibnamefont {Moshafi}},\ and\ \bibinfo
  {author} {\bibfnamefont {H.}~\bibnamefont {Sheikhahmadi}},\ }\href
  {https://doi.org/10.1016/j.dark.2022.101131} {\bibfield  {journal} {\bibinfo
  {journal} {Phys. Dark Univ.}\ }\textbf {\bibinfo {volume} {38}},\ \bibinfo
  {pages} {101131} (\bibinfo {year} {2022})},\ \Eprint
  {https://arxiv.org/abs/2203.08907} {arXiv:2203.08907 [gr-qc]} \BibitemShut
  {NoStop}%
\bibitem [{\citenamefont {Yengejeh}\ \emph {et~al.}(2023)\citenamefont
  {Yengejeh}, \citenamefont {Fakhry}, \citenamefont {Firouzjaee},\ and\
  \citenamefont {Fathi}}]{Yengejeh:2022tpa}%
  \BibitemOpen
  \bibfield  {author} {\bibinfo {author} {\bibfnamefont {M.~G.}\ \bibnamefont
  {Yengejeh}}, \bibinfo {author} {\bibfnamefont {S.}~\bibnamefont {Fakhry}},
  \bibinfo {author} {\bibfnamefont {J.~T.}\ \bibnamefont {Firouzjaee}},\ and\
  \bibinfo {author} {\bibfnamefont {H.}~\bibnamefont {Fathi}},\ }\href
  {https://doi.org/10.1016/j.dark.2022.101144} {\bibfield  {journal} {\bibinfo
  {journal} {Phys. Dark Univ.}\ }\textbf {\bibinfo {volume} {39}},\ \bibinfo
  {pages} {101144} (\bibinfo {year} {2023})},\ \Eprint
  {https://arxiv.org/abs/2206.01030} {arXiv:2206.01030 [astro-ph.CO]}
  \BibitemShut {NoStop}%
\bibitem [{\citenamefont {Nunes}\ \emph {et~al.}(2022)\citenamefont {Nunes},
  \citenamefont {Vagnozzi}, \citenamefont {Kumar}, \citenamefont
  {Di~Valentino},\ and\ \citenamefont {Mena}}]{Nunes:2022bhn}%
  \BibitemOpen
  \bibfield  {author} {\bibinfo {author} {\bibfnamefont {R.~C.}\ \bibnamefont
  {Nunes}}, \bibinfo {author} {\bibfnamefont {S.}~\bibnamefont {Vagnozzi}},
  \bibinfo {author} {\bibfnamefont {S.}~\bibnamefont {Kumar}}, \bibinfo
  {author} {\bibfnamefont {E.}~\bibnamefont {Di~Valentino}},\ and\ \bibinfo
  {author} {\bibfnamefont {O.}~\bibnamefont {Mena}},\ }\href
  {https://doi.org/10.1103/PhysRevD.105.123506} {\bibfield  {journal} {\bibinfo
   {journal} {Phys. Rev. D}\ }\textbf {\bibinfo {volume} {105}},\ \bibinfo
  {pages} {123506} (\bibinfo {year} {2022})},\ \Eprint
  {https://arxiv.org/abs/2203.08093} {arXiv:2203.08093 [astro-ph.CO]}
  \BibitemShut {NoStop}%
\bibitem [{\citenamefont {Landim}(2022)}]{Landim:2022jgr}%
  \BibitemOpen
  \bibfield  {author} {\bibinfo {author} {\bibfnamefont {R.~G.}\ \bibnamefont
  {Landim}},\ }\href {https://doi.org/10.1103/PhysRevD.106.043527} {\bibfield
  {journal} {\bibinfo  {journal} {Phys. Rev. D}\ }\textbf {\bibinfo {volume}
  {106}},\ \bibinfo {pages} {043527} (\bibinfo {year} {2022})},\ \Eprint
  {https://arxiv.org/abs/2206.10205} {arXiv:2206.10205 [astro-ph.CO]}
  \BibitemShut {NoStop}%
\bibitem [{\citenamefont {Yao}\ and\ \citenamefont {Meng}(2023)}]{Yao:2022kub}%
  \BibitemOpen
  \bibfield  {author} {\bibinfo {author} {\bibfnamefont {Y.-H.}\ \bibnamefont
  {Yao}}\ and\ \bibinfo {author} {\bibfnamefont {X.-H.}\ \bibnamefont {Meng}},\
  }\href {https://doi.org/10.1016/j.dark.2022.101165} {\bibfield  {journal}
  {\bibinfo  {journal} {Phys. Dark Univ.}\ }\textbf {\bibinfo {volume} {39}},\
  \bibinfo {pages} {101165} (\bibinfo {year} {2023})},\ \Eprint
  {https://arxiv.org/abs/2207.05955} {arXiv:2207.05955 [astro-ph.CO]}
  \BibitemShut {NoStop}%
\bibitem [{\citenamefont {Li}\ \emph {et~al.}(2024{\natexlab{a}})\citenamefont
  {Li}, \citenamefont {Jin}, \citenamefont {Li}, \citenamefont {Zhang},\ and\
  \citenamefont {Zhang}}]{Li:2023gtu}%
  \BibitemOpen
  \bibfield  {author} {\bibinfo {author} {\bibfnamefont {T.-N.}\ \bibnamefont
  {Li}}, \bibinfo {author} {\bibfnamefont {S.-J.}\ \bibnamefont {Jin}},
  \bibinfo {author} {\bibfnamefont {H.-L.}\ \bibnamefont {Li}}, \bibinfo
  {author} {\bibfnamefont {J.-F.}\ \bibnamefont {Zhang}},\ and\ \bibinfo
  {author} {\bibfnamefont {X.}~\bibnamefont {Zhang}},\ }\href
  {https://doi.org/10.3847/1538-4357/ad1bc9} {\bibfield  {journal} {\bibinfo
  {journal} {Astrophys. J.}\ }\textbf {\bibinfo {volume} {963}},\ \bibinfo
  {pages} {52} (\bibinfo {year} {2024}{\natexlab{a}})},\ \Eprint
  {https://arxiv.org/abs/2310.15879} {arXiv:2310.15879 [astro-ph.CO]}
  \BibitemShut {NoStop}%
\bibitem [{\citenamefont {Giar{\`e}}\ \emph
  {et~al.}(2024{\natexlab{a}})\citenamefont {Giar{\`e}}, \citenamefont {Zhai},
  \citenamefont {Pan}, \citenamefont {Di~Valentino}, \citenamefont {Nunes},\
  and\ \citenamefont {van~de Bruck}}]{Giare:2024ytc}%
  \BibitemOpen
  \bibfield  {author} {\bibinfo {author} {\bibfnamefont {W.}~\bibnamefont
  {Giar{\`e}}}, \bibinfo {author} {\bibfnamefont {Y.}~\bibnamefont {Zhai}},
  \bibinfo {author} {\bibfnamefont {S.}~\bibnamefont {Pan}}, \bibinfo {author}
  {\bibfnamefont {E.}~\bibnamefont {Di~Valentino}}, \bibinfo {author}
  {\bibfnamefont {R.~C.}\ \bibnamefont {Nunes}},\ and\ \bibinfo {author}
  {\bibfnamefont {C.}~\bibnamefont {van~de Bruck}},\ }\href
  {https://doi.org/10.1103/PhysRevD.110.063527} {\bibfield  {journal} {\bibinfo
   {journal} {Phys. Rev. D}\ }\textbf {\bibinfo {volume} {110}},\ \bibinfo
  {pages} {063527} (\bibinfo {year} {2024}{\natexlab{a}})},\ \Eprint
  {https://arxiv.org/abs/2404.02110} {arXiv:2404.02110 [astro-ph.CO]}
  \BibitemShut {NoStop}%
\bibitem [{\citenamefont {Kritpetch}\ \emph {et~al.}(2025)\citenamefont
  {Kritpetch}, \citenamefont {Roy},\ and\ \citenamefont
  {Banerjee}}]{Kritpetch:2024rgi}%
  \BibitemOpen
  \bibfield  {author} {\bibinfo {author} {\bibfnamefont {C.}~\bibnamefont
  {Kritpetch}}, \bibinfo {author} {\bibfnamefont {N.}~\bibnamefont {Roy}},\
  and\ \bibinfo {author} {\bibfnamefont {N.}~\bibnamefont {Banerjee}},\ }\href
  {https://doi.org/10.1103/PhysRevD.111.103501} {\bibfield  {journal} {\bibinfo
   {journal} {Phys. Rev. D}\ }\textbf {\bibinfo {volume} {111}},\ \bibinfo
  {pages} {103501} (\bibinfo {year} {2025})},\ \Eprint
  {https://arxiv.org/abs/2405.10604} {arXiv:2405.10604 [gr-qc]} \BibitemShut
  {NoStop}%
\bibitem [{\citenamefont {Giar{\`e}}\ \emph
  {et~al.}(2024{\natexlab{b}})\citenamefont {Giar{\`e}}, \citenamefont
  {Sabogal}, \citenamefont {Nunes},\ and\ \citenamefont
  {Di~Valentino}}]{Giare:2024smz}%
  \BibitemOpen
  \bibfield  {author} {\bibinfo {author} {\bibfnamefont {W.}~\bibnamefont
  {Giar{\`e}}}, \bibinfo {author} {\bibfnamefont {M.~A.}\ \bibnamefont
  {Sabogal}}, \bibinfo {author} {\bibfnamefont {R.~C.}\ \bibnamefont {Nunes}},\
  and\ \bibinfo {author} {\bibfnamefont {E.}~\bibnamefont {Di~Valentino}},\
  }\href {https://doi.org/10.1103/PhysRevLett.133.251003} {\bibfield  {journal}
  {\bibinfo  {journal} {Phys. Rev. Lett.}\ }\textbf {\bibinfo {volume} {133}},\
  \bibinfo {pages} {251003} (\bibinfo {year} {2024}{\natexlab{b}})},\ \Eprint
  {https://arxiv.org/abs/2404.15232} {arXiv:2404.15232 [astro-ph.CO]}
  \BibitemShut {NoStop}%
\bibitem [{\citenamefont {Li}\ \emph {et~al.}(2024{\natexlab{b}})\citenamefont
  {Li}, \citenamefont {Wu}, \citenamefont {Du}, \citenamefont {Jin},
  \citenamefont {Li}, \citenamefont {Zhang},\ and\ \citenamefont
  {Zhang}}]{Li:2024qso}%
  \BibitemOpen
  \bibfield  {author} {\bibinfo {author} {\bibfnamefont {T.-N.}\ \bibnamefont
  {Li}}, \bibinfo {author} {\bibfnamefont {P.-J.}\ \bibnamefont {Wu}}, \bibinfo
  {author} {\bibfnamefont {G.-H.}\ \bibnamefont {Du}}, \bibinfo {author}
  {\bibfnamefont {S.-J.}\ \bibnamefont {Jin}}, \bibinfo {author} {\bibfnamefont
  {H.-L.}\ \bibnamefont {Li}}, \bibinfo {author} {\bibfnamefont {J.-F.}\
  \bibnamefont {Zhang}},\ and\ \bibinfo {author} {\bibfnamefont
  {X.}~\bibnamefont {Zhang}},\ }\href
  {https://doi.org/10.3847/1538-4357/ad87f0} {\bibfield  {journal} {\bibinfo
  {journal} {Astrophys. J.}\ }\textbf {\bibinfo {volume} {976}},\ \bibinfo
  {pages} {1} (\bibinfo {year} {2024}{\natexlab{b}})},\ \Eprint
  {https://arxiv.org/abs/2407.14934} {arXiv:2407.14934 [astro-ph.CO]}
  \BibitemShut {NoStop}%
\bibitem [{\citenamefont {Halder}\ \emph {et~al.}(2024)\citenamefont {Halder},
  \citenamefont {Pan}, \citenamefont {S{\'a}},\ and\ \citenamefont
  {Saha}}]{Halder:2024gag}%
  \BibitemOpen
  \bibfield  {author} {\bibinfo {author} {\bibfnamefont {S.}~\bibnamefont
  {Halder}}, \bibinfo {author} {\bibfnamefont {S.}~\bibnamefont {Pan}},
  \bibinfo {author} {\bibfnamefont {P.~M.}\ \bibnamefont {S{\'a}}},\ and\
  \bibinfo {author} {\bibfnamefont {T.}~\bibnamefont {Saha}},\ }\href
  {https://doi.org/10.1103/PhysRevD.110.063529} {\bibfield  {journal} {\bibinfo
   {journal} {Phys. Rev. D}\ }\textbf {\bibinfo {volume} {110}},\ \bibinfo
  {pages} {063529} (\bibinfo {year} {2024})},\ \Eprint
  {https://arxiv.org/abs/2407.15804} {arXiv:2407.15804 [gr-qc]} \BibitemShut
  {NoStop}%
\bibitem [{\citenamefont {Carrion}\ \emph {et~al.}(2025)\citenamefont
  {Carrion}, \citenamefont {Spurio~Mancini}, \citenamefont {Piras},\ and\
  \citenamefont {Hidalgo}}]{Carrion:2024jur}%
  \BibitemOpen
  \bibfield  {author} {\bibinfo {author} {\bibfnamefont {K.}~\bibnamefont
  {Carrion}}, \bibinfo {author} {\bibfnamefont {A.}~\bibnamefont
  {Spurio~Mancini}}, \bibinfo {author} {\bibfnamefont {D.}~\bibnamefont
  {Piras}},\ and\ \bibinfo {author} {\bibfnamefont {J.~C.}\ \bibnamefont
  {Hidalgo}},\ }\href {https://doi.org/10.1093/mnras/staf663} {\bibfield
  {journal} {\bibinfo  {journal} {Mon. Not. Roy. Astron. Soc.}\ }\textbf
  {\bibinfo {volume} {539}},\ \bibinfo {pages} {3220} (\bibinfo {year}
  {2025})},\ \Eprint {https://arxiv.org/abs/2410.10603} {arXiv:2410.10603
  [astro-ph.CO]} \BibitemShut {NoStop}%
\bibitem [{\citenamefont {Giani}\ \emph {et~al.}(2025)\citenamefont {Giani},
  \citenamefont {Von~Marttens},\ and\ \citenamefont
  {Camilleri}}]{Giani:2024nnv}%
  \BibitemOpen
  \bibfield  {author} {\bibinfo {author} {\bibfnamefont {L.}~\bibnamefont
  {Giani}}, \bibinfo {author} {\bibfnamefont {R.}~\bibnamefont
  {Von~Marttens}},\ and\ \bibinfo {author} {\bibfnamefont {R.}~\bibnamefont
  {Camilleri}},\ }\href {https://doi.org/10.1103/zr92-m7py} {\bibfield
  {journal} {\bibinfo  {journal} {Phys. Rev. Lett.}\ }\textbf {\bibinfo
  {volume} {135}},\ \bibinfo {pages} {071004} (\bibinfo {year} {2025})},\
  \Eprint {https://arxiv.org/abs/2410.15295} {arXiv:2410.15295 [astro-ph.CO]}
  \BibitemShut {NoStop}%
\bibitem [{\citenamefont {Tsedrik}\ \emph {et~al.}(2025)\citenamefont {Tsedrik}
  \emph {et~al.}}]{Tsedrik:2025cwc}%
  \BibitemOpen
  \bibfield  {author} {\bibinfo {author} {\bibfnamefont {M.}~\bibnamefont
  {Tsedrik}} \emph {et~al.},\ }\href@noop {} {\bibinfo {title} {{Interacting
  dark energy constraints from the full-shape analyses of BOSS DR12 and DES
  Year 3 measurements}}} (\bibinfo {year} {2025}),\ \Eprint
  {https://arxiv.org/abs/2502.03390} {arXiv:2502.03390 [astro-ph.CO]}
  \BibitemShut {NoStop}%
\bibitem [{\citenamefont {Liu}\ \emph {et~al.}(2025{\natexlab{a}})\citenamefont
  {Liu}, \citenamefont {Fu}, \citenamefont {Xu}, \citenamefont {Ding},
  \citenamefont {Huang},\ and\ \citenamefont {Qing}}]{Liu:2025pxy}%
  \BibitemOpen
  \bibfield  {author} {\bibinfo {author} {\bibfnamefont {K.}~\bibnamefont
  {Liu}}, \bibinfo {author} {\bibfnamefont {X.}~\bibnamefont {Fu}}, \bibinfo
  {author} {\bibfnamefont {B.}~\bibnamefont {Xu}}, \bibinfo {author}
  {\bibfnamefont {C.}~\bibnamefont {Ding}}, \bibinfo {author} {\bibfnamefont
  {Y.}~\bibnamefont {Huang}},\ and\ \bibinfo {author} {\bibfnamefont
  {X.}~\bibnamefont {Qing}},\ }\href@noop {} {\bibinfo {title} {{The growth of
  linear perturbations in the interacting dark energy models and observational
  constraints}}} (\bibinfo {year} {2025}{\natexlab{a}}),\ \Eprint
  {https://arxiv.org/abs/2503.05208} {arXiv:2503.05208 [astro-ph.CO]}
  \BibitemShut {NoStop}%
\bibitem [{\citenamefont {Zhai}\ \emph {et~al.}(2025)\citenamefont {Zhai},
  \citenamefont {de~Cesare}, \citenamefont {van~de Bruck}, \citenamefont
  {Di~Valentino},\ and\ \citenamefont {Wilson-Ewing}}]{Zhai:2025hfi}%
  \BibitemOpen
  \bibfield  {author} {\bibinfo {author} {\bibfnamefont {Y.}~\bibnamefont
  {Zhai}}, \bibinfo {author} {\bibfnamefont {M.}~\bibnamefont {de~Cesare}},
  \bibinfo {author} {\bibfnamefont {C.}~\bibnamefont {van~de Bruck}}, \bibinfo
  {author} {\bibfnamefont {E.}~\bibnamefont {Di~Valentino}},\ and\ \bibinfo
  {author} {\bibfnamefont {E.}~\bibnamefont {Wilson-Ewing}},\ }\href@noop {}
  {\bibinfo {title} {{A low-redshift preference for an interacting dark energy
  model}}} (\bibinfo {year} {2025}),\ \Eprint
  {https://arxiv.org/abs/2503.15659} {arXiv:2503.15659 [astro-ph.CO]}
  \BibitemShut {NoStop}%
\bibitem [{\citenamefont {Li}\ \emph {et~al.}(2025{\natexlab{a}})\citenamefont
  {Li}, \citenamefont {Du}, \citenamefont {Li}, \citenamefont {Wu},
  \citenamefont {Jin}, \citenamefont {Zhang},\ and\ \citenamefont
  {Zhang}}]{Li:2025owk}%
  \BibitemOpen
  \bibfield  {author} {\bibinfo {author} {\bibfnamefont {T.-N.}\ \bibnamefont
  {Li}}, \bibinfo {author} {\bibfnamefont {G.-H.}\ \bibnamefont {Du}}, \bibinfo
  {author} {\bibfnamefont {Y.-H.}\ \bibnamefont {Li}}, \bibinfo {author}
  {\bibfnamefont {P.-J.}\ \bibnamefont {Wu}}, \bibinfo {author} {\bibfnamefont
  {S.-J.}\ \bibnamefont {Jin}}, \bibinfo {author} {\bibfnamefont {J.-F.}\
  \bibnamefont {Zhang}},\ and\ \bibinfo {author} {\bibfnamefont
  {X.}~\bibnamefont {Zhang}},\ }\href@noop {} {\bibinfo {title} {{Probing the
  sign-changeable interaction between dark energy and dark matter with DESI
  baryon acoustic oscillations and DES supernovae data}}} (\bibinfo {year}
  {2025}{\natexlab{a}}),\ \Eprint {https://arxiv.org/abs/2501.07361}
  {arXiv:2501.07361 [astro-ph.CO]} \BibitemShut {NoStop}%
\bibitem [{\citenamefont {Silva}\ \emph {et~al.}(2025)\citenamefont {Silva},
  \citenamefont {Sabogal}, \citenamefont {Scherer}, \citenamefont {Nunes},
  \citenamefont {Di~Valentino},\ and\ \citenamefont {Kumar}}]{Silva:2025hxw}%
  \BibitemOpen
  \bibfield  {author} {\bibinfo {author} {\bibfnamefont {E.}~\bibnamefont
  {Silva}}, \bibinfo {author} {\bibfnamefont {M.~A.}\ \bibnamefont {Sabogal}},
  \bibinfo {author} {\bibfnamefont {M.}~\bibnamefont {Scherer}}, \bibinfo
  {author} {\bibfnamefont {R.~C.}\ \bibnamefont {Nunes}}, \bibinfo {author}
  {\bibfnamefont {E.}~\bibnamefont {Di~Valentino}},\ and\ \bibinfo {author}
  {\bibfnamefont {S.}~\bibnamefont {Kumar}},\ }\href
  {https://doi.org/10.1103/qqc6-76z4} {\bibfield  {journal} {\bibinfo
  {journal} {Phys. Rev. D}\ }\textbf {\bibinfo {volume} {111}},\ \bibinfo
  {pages} {123511} (\bibinfo {year} {2025})},\ \Eprint
  {https://arxiv.org/abs/2503.23225} {arXiv:2503.23225 [astro-ph.CO]}
  \BibitemShut {NoStop}%
\bibitem [{\citenamefont {Chakraborty}\ \emph {et~al.}(2025)\citenamefont
  {Chakraborty}, \citenamefont {Chanda}, \citenamefont {Das},\ and\
  \citenamefont {Dutta}}]{Chakraborty:2025syu}%
  \BibitemOpen
  \bibfield  {author} {\bibinfo {author} {\bibfnamefont {A.}~\bibnamefont
  {Chakraborty}}, \bibinfo {author} {\bibfnamefont {P.~K.}\ \bibnamefont
  {Chanda}}, \bibinfo {author} {\bibfnamefont {S.}~\bibnamefont {Das}},\ and\
  \bibinfo {author} {\bibfnamefont {K.}~\bibnamefont {Dutta}},\ }\href@noop {}
  {\bibinfo {title} {{DESI results: Hint towards coupled dark matter and dark
  energy}}} (\bibinfo {year} {2025}),\ \Eprint
  {https://arxiv.org/abs/2503.10806} {arXiv:2503.10806 [astro-ph.CO]}
  \BibitemShut {NoStop}%
\bibitem [{\citenamefont {You}\ \emph {et~al.}(2025)\citenamefont {You},
  \citenamefont {Wang},\ and\ \citenamefont {Yang}}]{You:2025uon}%
  \BibitemOpen
  \bibfield  {author} {\bibinfo {author} {\bibfnamefont {C.}~\bibnamefont
  {You}}, \bibinfo {author} {\bibfnamefont {D.}~\bibnamefont {Wang}},\ and\
  \bibinfo {author} {\bibfnamefont {T.}~\bibnamefont {Yang}},\ }\href
  {https://doi.org/10.1103/f6v7-n9fr} {\bibfield  {journal} {\bibinfo
  {journal} {Phys. Rev. D}\ }\textbf {\bibinfo {volume} {112}},\ \bibinfo
  {pages} {043503} (\bibinfo {year} {2025})},\ \Eprint
  {https://arxiv.org/abs/2504.00985} {arXiv:2504.00985 [astro-ph.CO]}
  \BibitemShut {NoStop}%
\bibitem [{\citenamefont {Aoki}\ \emph {et~al.}(2025)\citenamefont {Aoki},
  \citenamefont {Beltr{\'a}n~Jim{\'e}nez}, \citenamefont {Pookkillath},\ and\
  \citenamefont {Tsujikawa}}]{Aoki:2025bmj}%
  \BibitemOpen
  \bibfield  {author} {\bibinfo {author} {\bibfnamefont {K.}~\bibnamefont
  {Aoki}}, \bibinfo {author} {\bibfnamefont {J.}~\bibnamefont
  {Beltr{\'a}n~Jim{\'e}nez}}, \bibinfo {author} {\bibfnamefont {M.~C.}\
  \bibnamefont {Pookkillath}},\ and\ \bibinfo {author} {\bibfnamefont
  {S.}~\bibnamefont {Tsujikawa}},\ }\href@noop {} {\bibinfo {title} {{Effective
  field theory of coupled dark energy and dark matter}}} (\bibinfo {year}
  {2025}),\ \Eprint {https://arxiv.org/abs/2504.17293} {arXiv:2504.17293
  [astro-ph.CO]} \BibitemShut {NoStop}%
\bibitem [{\citenamefont {Yang}\ \emph
  {et~al.}(2025{\natexlab{a}})\citenamefont {Yang}, \citenamefont {Pan},
  \citenamefont {Di~Valentino}, \citenamefont {Mena}, \citenamefont {Mota},\
  and\ \citenamefont {Chakraborty}}]{Yang:2025ume}%
  \BibitemOpen
  \bibfield  {author} {\bibinfo {author} {\bibfnamefont {W.}~\bibnamefont
  {Yang}}, \bibinfo {author} {\bibfnamefont {S.}~\bibnamefont {Pan}}, \bibinfo
  {author} {\bibfnamefont {E.}~\bibnamefont {Di~Valentino}}, \bibinfo {author}
  {\bibfnamefont {O.}~\bibnamefont {Mena}}, \bibinfo {author} {\bibfnamefont
  {D.~F.}\ \bibnamefont {Mota}},\ and\ \bibinfo {author} {\bibfnamefont
  {S.}~\bibnamefont {Chakraborty}},\ }\href
  {https://doi.org/10.1103/PhysRevD.111.103509} {\bibfield  {journal} {\bibinfo
   {journal} {Phys. Rev. D}\ }\textbf {\bibinfo {volume} {111}},\ \bibinfo
  {pages} {103509} (\bibinfo {year} {2025}{\natexlab{a}})},\ \Eprint
  {https://arxiv.org/abs/2504.11973} {arXiv:2504.11973 [astro-ph.CO]}
  \BibitemShut {NoStop}%
\bibitem [{\citenamefont {Abedin}\ \emph {et~al.}(2025)\citenamefont {Abedin},
  \citenamefont {Wang}, \citenamefont {Ma},\ and\ \citenamefont
  {Pan}}]{Abedin:2025yru}%
  \BibitemOpen
  \bibfield  {author} {\bibinfo {author} {\bibfnamefont {M.}~\bibnamefont
  {Abedin}}, \bibinfo {author} {\bibfnamefont {G.-J.}\ \bibnamefont {Wang}},
  \bibinfo {author} {\bibfnamefont {Y.-Z.}\ \bibnamefont {Ma}},\ and\ \bibinfo
  {author} {\bibfnamefont {S.}~\bibnamefont {Pan}},\ }\href
  {https://doi.org/10.1093/mnras/staf762} {\bibfield  {journal} {\bibinfo
  {journal} {Mon. Not. Roy. Astron. Soc.}\ }\textbf {\bibinfo {volume} {540}},\
  \bibinfo {pages} {2253} (\bibinfo {year} {2025})},\ \Eprint
  {https://arxiv.org/abs/2505.04336} {arXiv:2505.04336 [astro-ph.CO]}
  \BibitemShut {NoStop}%
\bibitem [{\citenamefont {Pan}\ \emph {et~al.}(2025)\citenamefont {Pan},
  \citenamefont {Paul}, \citenamefont {Saridakis},\ and\ \citenamefont
  {Yang}}]{Pan:2025qwy}%
  \BibitemOpen
  \bibfield  {author} {\bibinfo {author} {\bibfnamefont {S.}~\bibnamefont
  {Pan}}, \bibinfo {author} {\bibfnamefont {S.}~\bibnamefont {Paul}}, \bibinfo
  {author} {\bibfnamefont {E.~N.}\ \bibnamefont {Saridakis}},\ and\ \bibinfo
  {author} {\bibfnamefont {W.}~\bibnamefont {Yang}},\ }\href@noop {} {\bibinfo
  {title} {{Interacting dark energy after DESI DR2: a challenge for
  $\Lambda$CDM paradigm?}}} (\bibinfo {year} {2025}),\ \Eprint
  {https://arxiv.org/abs/2504.00994} {arXiv:2504.00994 [astro-ph.CO]}
  \BibitemShut {NoStop}%
\bibitem [{\citenamefont {van~der Westhuizen}\ \emph
  {et~al.}(2025)\citenamefont {van~der Westhuizen}, \citenamefont {Figueruelo},
  \citenamefont {Thubisi}, \citenamefont {Sahlu}, \citenamefont {Abebe},\ and\
  \citenamefont {Paliathanasis}}]{vanderWesthuizen:2025iam}%
  \BibitemOpen
  \bibfield  {author} {\bibinfo {author} {\bibfnamefont {M.}~\bibnamefont
  {van~der Westhuizen}}, \bibinfo {author} {\bibfnamefont {D.}~\bibnamefont
  {Figueruelo}}, \bibinfo {author} {\bibfnamefont {R.}~\bibnamefont {Thubisi}},
  \bibinfo {author} {\bibfnamefont {S.}~\bibnamefont {Sahlu}}, \bibinfo
  {author} {\bibfnamefont {A.}~\bibnamefont {Abebe}},\ and\ \bibinfo {author}
  {\bibfnamefont {A.}~\bibnamefont {Paliathanasis}},\ }\href@noop {} {\bibinfo
  {title} {{Compartmentalization in the Dark Sector of the Universe after DESI
  DR2 BAO data}}} (\bibinfo {year} {2025}),\ \Eprint
  {https://arxiv.org/abs/2505.23306} {arXiv:2505.23306 [astro-ph.CO]}
  \BibitemShut {NoStop}%
\bibitem [{\citenamefont {Liu}\ \emph {et~al.}(2025{\natexlab{b}})\citenamefont
  {Liu}, \citenamefont {Wu}, \citenamefont {Pan},\ and\ \citenamefont
  {Yang}}]{Liu:2025vda}%
  \BibitemOpen
  \bibfield  {author} {\bibinfo {author} {\bibfnamefont {W.}~\bibnamefont
  {Liu}}, \bibinfo {author} {\bibfnamefont {Y.}~\bibnamefont {Wu}}, \bibinfo
  {author} {\bibfnamefont {S.}~\bibnamefont {Pan}},\ and\ \bibinfo {author}
  {\bibfnamefont {W.}~\bibnamefont {Yang}},\ }\href
  {https://doi.org/10.1016/j.jheap.2025.100403} {\bibfield  {journal} {\bibinfo
   {journal} {JHEAp}\ }\textbf {\bibinfo {volume} {47}},\ \bibinfo {pages}
  {100403} (\bibinfo {year} {2025}{\natexlab{b}})}\BibitemShut {NoStop}%
\bibitem [{\citenamefont {Yang}\ \emph
  {et~al.}(2025{\natexlab{b}})\citenamefont {Yang}, \citenamefont {Dai},\ and\
  \citenamefont {Wang}}]{Yang:2025boq}%
  \BibitemOpen
  \bibfield  {author} {\bibinfo {author} {\bibfnamefont {Y.}~\bibnamefont
  {Yang}}, \bibinfo {author} {\bibfnamefont {X.}~\bibnamefont {Dai}},\ and\
  \bibinfo {author} {\bibfnamefont {Y.}~\bibnamefont {Wang}},\ }\href
  {https://doi.org/10.1103/8ync-vrtz} {\bibfield  {journal} {\bibinfo
  {journal} {Phys. Rev. D}\ }\textbf {\bibinfo {volume} {111}},\ \bibinfo
  {pages} {103534} (\bibinfo {year} {2025}{\natexlab{b}})},\ \Eprint
  {https://arxiv.org/abs/2505.09879} {arXiv:2505.09879 [astro-ph.CO]}
  \BibitemShut {NoStop}%
\bibitem [{\citenamefont {Paliathanasis}(2025)}]{Paliathanasis:2025xxm}%
  \BibitemOpen
  \bibfield  {author} {\bibinfo {author} {\bibfnamefont {A.}~\bibnamefont
  {Paliathanasis}},\ }\href@noop {} {\bibinfo {title} {{Observational
  Constraints on Scalar Field--Matter Interaction in Weyl Integrable
  Spacetime}}} (\bibinfo {year} {2025}),\ \Eprint
  {https://arxiv.org/abs/2506.16223} {arXiv:2506.16223 [gr-qc]} \BibitemShut
  {NoStop}%
\bibitem [{\citenamefont {Li}\ and\ \citenamefont {Zhang}(2025)}]{Li:2025ula}%
  \BibitemOpen
  \bibfield  {author} {\bibinfo {author} {\bibfnamefont {Y.-H.}\ \bibnamefont
  {Li}}\ and\ \bibinfo {author} {\bibfnamefont {X.}~\bibnamefont {Zhang}},\
  }\href@noop {} {\bibinfo {title} {{Cosmic Sign-Reversal: Non-Parametric
  Reconstruction of Interacting Dark Energy with DESI DR2}}} (\bibinfo {year}
  {2025}),\ \Eprint {https://arxiv.org/abs/2506.18477} {arXiv:2506.18477
  [astro-ph.CO]} \BibitemShut {NoStop}%
\bibitem [{\citenamefont {Yan}\ \emph {et~al.}(2025)\citenamefont {Yan},
  \citenamefont {Pan}, \citenamefont {Wang}, \citenamefont {Xu},\ and\
  \citenamefont {Peng}}]{Yan:2025iga}%
  \BibitemOpen
  \bibfield  {author} {\bibinfo {author} {\bibfnamefont {H.}~\bibnamefont
  {Yan}}, \bibinfo {author} {\bibfnamefont {Y.}~\bibnamefont {Pan}}, \bibinfo
  {author} {\bibfnamefont {J.-X.}\ \bibnamefont {Wang}}, \bibinfo {author}
  {\bibfnamefont {W.-X.}\ \bibnamefont {Xu}},\ and\ \bibinfo {author}
  {\bibfnamefont {Z.-H.}\ \bibnamefont {Peng}},\ }\href@noop {} {\bibinfo
  {title} {{Investigating Interacting Dark Energy Models Using Fast Radio Burst
  Observations}}} (\bibinfo {year} {2025}),\ \Eprint
  {https://arxiv.org/abs/2507.16308} {arXiv:2507.16308 [astro-ph.CO]}
  \BibitemShut {NoStop}%
\bibitem [{\citenamefont {Wang}\ \emph {et~al.}(2025)\citenamefont {Wang},
  \citenamefont {Cai}, \citenamefont {Guo},\ and\ \citenamefont
  {Wang}}]{Wang:2025znm}%
  \BibitemOpen
  \bibfield  {author} {\bibinfo {author} {\bibfnamefont {J.-Q.}\ \bibnamefont
  {Wang}}, \bibinfo {author} {\bibfnamefont {R.-G.}\ \bibnamefont {Cai}},
  \bibinfo {author} {\bibfnamefont {Z.-K.}\ \bibnamefont {Guo}},\ and\ \bibinfo
  {author} {\bibfnamefont {S.-J.}\ \bibnamefont {Wang}},\ }\href@noop {}
  {\bibinfo {title} {{Resolving the Planck-DESI tension by non-minimally
  coupled quintessence}}} (\bibinfo {year} {2025}),\ \Eprint
  {https://arxiv.org/abs/2508.01759} {arXiv:2508.01759 [astro-ph.CO]}
  \BibitemShut {NoStop}%
\bibitem [{\citenamefont {Ghosh}\ and\ \citenamefont
  {Devi}(2024)}]{Ghosh:2024ojq}%
  \BibitemOpen
  \bibfield  {author} {\bibinfo {author} {\bibfnamefont {S.}~\bibnamefont
  {Ghosh}}\ and\ \bibinfo {author} {\bibfnamefont {B.}~\bibnamefont {Devi}},\
  }\href {https://doi.org/10.1088/1742-6596/2919/1/012017} {\bibfield
  {journal} {\bibinfo  {journal} {J. Phys. Conf. Ser.}\ }\textbf {\bibinfo
  {volume} {2919}},\ \bibinfo {pages} {012017} (\bibinfo {year}
  {2024})}\BibitemShut {NoStop}%
\bibitem [{\citenamefont {Adame}\ \emph {et~al.}(2025)\citenamefont {Adame}
  \emph {et~al.}}]{DESI:2024mwx}%
  \BibitemOpen
  \bibfield  {author} {\bibinfo {author} {\bibfnamefont {A.~G.}\ \bibnamefont
  {Adame}} \emph {et~al.} (\bibinfo {collaboration} {DESI}),\ }\href
  {https://doi.org/10.1088/1475-7516/2025/02/021} {\bibfield  {journal}
  {\bibinfo  {journal} {JCAP}\ }\textbf {\bibinfo {volume} {02}},\ \bibinfo
  {pages} {021}},\ \Eprint {https://arxiv.org/abs/2404.03002} {arXiv:2404.03002
  [astro-ph.CO]} \BibitemShut {NoStop}%
\bibitem [{\citenamefont {Abdul~Karim}\ \emph {et~al.}(2025)\citenamefont
  {Abdul~Karim} \emph {et~al.}}]{DESI:2025zgx}%
  \BibitemOpen
  \bibfield  {author} {\bibinfo {author} {\bibfnamefont {M.}~\bibnamefont
  {Abdul~Karim}} \emph {et~al.} (\bibinfo {collaboration} {DESI}),\ }\href@noop
  {} {\bibinfo {title} {{DESI DR2 Results II: Measurements of Baryon Acoustic
  Oscillations and Cosmological Constraints}}} (\bibinfo {year} {2025}),\
  \Eprint {https://arxiv.org/abs/2503.14738} {arXiv:2503.14738 [astro-ph.CO]}
  \BibitemShut {NoStop}%
\bibitem [{\citenamefont {Gu}\ \emph {et~al.}(2025)\citenamefont {Gu} \emph
  {et~al.}}]{DESI:2025wyn}%
  \BibitemOpen
  \bibfield  {author} {\bibinfo {author} {\bibfnamefont {G.}~\bibnamefont {Gu}}
  \emph {et~al.} (\bibinfo {collaboration} {DESI}),\ }\href@noop {} {\
  (\bibinfo {year} {2025})},\ \Eprint {https://arxiv.org/abs/2504.06118}
  {arXiv:2504.06118 [astro-ph.CO]} \BibitemShut {NoStop}%
\bibitem [{\citenamefont {Lodha}\ \emph {et~al.}(2025)\citenamefont {Lodha}
  \emph {et~al.}}]{DESI:2025fii}%
  \BibitemOpen
  \bibfield  {author} {\bibinfo {author} {\bibfnamefont {K.}~\bibnamefont
  {Lodha}} \emph {et~al.} (\bibinfo {collaboration} {DESI}),\ }\href
  {https://doi.org/10.1103/w4c6-1r5j} {\bibfield  {journal} {\bibinfo
  {journal} {Phys. Rev. D}\ }\textbf {\bibinfo {volume} {112}},\ \bibinfo
  {pages} {083511} (\bibinfo {year} {2025})},\ \Eprint
  {https://arxiv.org/abs/2503.14743} {arXiv:2503.14743 [astro-ph.CO]}
  \BibitemShut {NoStop}%
\bibitem [{\citenamefont {Efstathiou}(1999)}]{Efstathiou:1999tm}%
  \BibitemOpen
  \bibfield  {author} {\bibinfo {author} {\bibfnamefont {G.}~\bibnamefont
  {Efstathiou}},\ }\href {https://doi.org/10.1046/j.1365-8711.1999.02997.x}
  {\bibfield  {journal} {\bibinfo  {journal} {Mon. Not. Roy. Astron. Soc.}\
  }\textbf {\bibinfo {volume} {310}},\ \bibinfo {pages} {842} (\bibinfo {year}
  {1999})},\ \Eprint {https://arxiv.org/abs/astro-ph/9904356}
  {arXiv:astro-ph/9904356} \BibitemShut {NoStop}%
\bibitem [{\citenamefont {Barboza}\ and\ \citenamefont
  {Alcaniz}(2008)}]{Barboza:2008rh}%
  \BibitemOpen
  \bibfield  {author} {\bibinfo {author} {\bibfnamefont {E.~M.}\ \bibnamefont
  {Barboza}, \bibfnamefont {Jr.}}\ and\ \bibinfo {author} {\bibfnamefont
  {J.~S.}\ \bibnamefont {Alcaniz}},\ }\href
  {https://doi.org/10.1016/j.physletb.2008.08.012} {\bibfield  {journal}
  {\bibinfo  {journal} {Phys. Lett. B}\ }\textbf {\bibinfo {volume} {666}},\
  \bibinfo {pages} {415} (\bibinfo {year} {2008})},\ \Eprint
  {https://arxiv.org/abs/0805.1713} {arXiv:0805.1713 [astro-ph]} \BibitemShut
  {NoStop}%
\bibitem [{\citenamefont {Dimakis}\ \emph {et~al.}(2016)\citenamefont
  {Dimakis}, \citenamefont {Karagiorgos}, \citenamefont {Zampeli},
  \citenamefont {Paliathanasis}, \citenamefont {Christodoulakis},\ and\
  \citenamefont {Terzis}}]{Dimakis:2016mip}%
  \BibitemOpen
  \bibfield  {author} {\bibinfo {author} {\bibfnamefont {N.}~\bibnamefont
  {Dimakis}}, \bibinfo {author} {\bibfnamefont {A.}~\bibnamefont
  {Karagiorgos}}, \bibinfo {author} {\bibfnamefont {A.}~\bibnamefont
  {Zampeli}}, \bibinfo {author} {\bibfnamefont {A.}~\bibnamefont
  {Paliathanasis}}, \bibinfo {author} {\bibfnamefont {T.}~\bibnamefont
  {Christodoulakis}},\ and\ \bibinfo {author} {\bibfnamefont {P.~A.}\
  \bibnamefont {Terzis}},\ }\href {https://doi.org/10.1103/PhysRevD.93.123518}
  {\bibfield  {journal} {\bibinfo  {journal} {Phys. Rev. D}\ }\textbf {\bibinfo
  {volume} {93}},\ \bibinfo {pages} {123518} (\bibinfo {year} {2016})},\
  \Eprint {https://arxiv.org/abs/1604.05168} {arXiv:1604.05168 [gr-qc]}
  \BibitemShut {NoStop}%
\bibitem [{\citenamefont {Pan}\ \emph {et~al.}(2020{\natexlab{b}})\citenamefont
  {Pan}, \citenamefont {Yang},\ and\ \citenamefont
  {Paliathanasis}}]{Pan:2019brc}%
  \BibitemOpen
  \bibfield  {author} {\bibinfo {author} {\bibfnamefont {S.}~\bibnamefont
  {Pan}}, \bibinfo {author} {\bibfnamefont {W.}~\bibnamefont {Yang}},\ and\
  \bibinfo {author} {\bibfnamefont {A.}~\bibnamefont {Paliathanasis}},\ }\href
  {https://doi.org/10.1140/epjc/s10052-020-7832-y} {\bibfield  {journal}
  {\bibinfo  {journal} {Eur. Phys. J. C}\ }\textbf {\bibinfo {volume} {80}},\
  \bibinfo {pages} {274} (\bibinfo {year} {2020}{\natexlab{b}})},\ \Eprint
  {https://arxiv.org/abs/1902.07108} {arXiv:1902.07108 [astro-ph.CO]}
  \BibitemShut {NoStop}%
\bibitem [{\citenamefont {Jassal}\ \emph {et~al.}(2005)\citenamefont {Jassal},
  \citenamefont {Bagla},\ and\ \citenamefont {Padmanabhan}}]{Jassal:2005qc}%
  \BibitemOpen
  \bibfield  {author} {\bibinfo {author} {\bibfnamefont {H.~K.}\ \bibnamefont
  {Jassal}}, \bibinfo {author} {\bibfnamefont {J.~S.}\ \bibnamefont {Bagla}},\
  and\ \bibinfo {author} {\bibfnamefont {T.}~\bibnamefont {Padmanabhan}},\
  }\href {https://doi.org/10.1103/PhysRevD.72.103503} {\bibfield  {journal}
  {\bibinfo  {journal} {Phys. Rev. D}\ }\textbf {\bibinfo {volume} {72}},\
  \bibinfo {pages} {103503} (\bibinfo {year} {2005})},\ \Eprint
  {https://arxiv.org/abs/astro-ph/0506748} {arXiv:astro-ph/0506748}
  \BibitemShut {NoStop}%
\bibitem [{\citenamefont {Cheng}\ \emph {et~al.}(2025)\citenamefont {Cheng},
  \citenamefont {Pan},\ and\ \citenamefont {Di~Valentino}}]{Cheng:2025yue}%
  \BibitemOpen
  \bibfield  {author} {\bibinfo {author} {\bibfnamefont {H.}~\bibnamefont
  {Cheng}}, \bibinfo {author} {\bibfnamefont {S.}~\bibnamefont {Pan}},\ and\
  \bibinfo {author} {\bibfnamefont {E.}~\bibnamefont {Di~Valentino}},\
  }\href@noop {} {\  (\bibinfo {year} {2025})},\ \Eprint
  {https://arxiv.org/abs/2512.09866} {arXiv:2512.09866 [astro-ph.CO]}
  \BibitemShut {NoStop}%
\bibitem [{\citenamefont {Alam}\ and\ \citenamefont
  {Hossain}(2025)}]{Alam:2025epg}%
  \BibitemOpen
  \bibfield  {author} {\bibinfo {author} {\bibfnamefont {S.}~\bibnamefont
  {Alam}}\ and\ \bibinfo {author} {\bibfnamefont {M.~W.}\ \bibnamefont
  {Hossain}},\ }\href@noop {} {\  (\bibinfo {year} {2025})},\ \Eprint
  {https://arxiv.org/abs/2510.03779} {arXiv:2510.03779 [astro-ph.CO]}
  \BibitemShut {NoStop}%
\bibitem [{\citenamefont {Lee}\ \emph {et~al.}(2025)\citenamefont {Lee},
  \citenamefont {Yang}, \citenamefont {Di~Valentino}, \citenamefont {Pan},\
  and\ \citenamefont {van~de Bruck}}]{Lee:2025pzo}%
  \BibitemOpen
  \bibfield  {author} {\bibinfo {author} {\bibfnamefont {D.~H.}\ \bibnamefont
  {Lee}}, \bibinfo {author} {\bibfnamefont {W.}~\bibnamefont {Yang}}, \bibinfo
  {author} {\bibfnamefont {E.}~\bibnamefont {Di~Valentino}}, \bibinfo {author}
  {\bibfnamefont {S.}~\bibnamefont {Pan}},\ and\ \bibinfo {author}
  {\bibfnamefont {C.}~\bibnamefont {van~de Bruck}},\ }\href@noop {} {\
  (\bibinfo {year} {2025})},\ \Eprint {https://arxiv.org/abs/2507.11432}
  {arXiv:2507.11432 [astro-ph.CO]} \BibitemShut {NoStop}%
\bibitem [{\citenamefont {Chevallier}\ and\ \citenamefont
  {Polarski}(2001)}]{Chevallier:2000qy}%
  \BibitemOpen
  \bibfield  {author} {\bibinfo {author} {\bibfnamefont {M.}~\bibnamefont
  {Chevallier}}\ and\ \bibinfo {author} {\bibfnamefont {D.}~\bibnamefont
  {Polarski}},\ }\href {https://doi.org/10.1142/S0218271801000822} {\bibfield
  {journal} {\bibinfo  {journal} {Int. J. Mod. Phys. D}\ }\textbf {\bibinfo
  {volume} {10}},\ \bibinfo {pages} {213} (\bibinfo {year} {2001})},\ \Eprint
  {https://arxiv.org/abs/gr-qc/0009008} {arXiv:gr-qc/0009008} \BibitemShut
  {NoStop}%
\bibitem [{\citenamefont {Linder}(2003)}]{Linder:2002et}%
  \BibitemOpen
  \bibfield  {author} {\bibinfo {author} {\bibfnamefont {E.~V.}\ \bibnamefont
  {Linder}},\ }\href {https://doi.org/10.1103/PhysRevLett.90.091301} {\bibfield
   {journal} {\bibinfo  {journal} {Phys. Rev. Lett.}\ }\textbf {\bibinfo
  {volume} {90}},\ \bibinfo {pages} {091301} (\bibinfo {year} {2003})},\
  \Eprint {https://arxiv.org/abs/astro-ph/0208512} {arXiv:astro-ph/0208512}
  \BibitemShut {NoStop}%
\bibitem [{\citenamefont {Gonz{\'a}lez-Fuentes}\ and\ \citenamefont
  {G{\'o}mez-Valent}(2025)}]{Gonzalez-Fuentes:2025lei}%
  \BibitemOpen
  \bibfield  {author} {\bibinfo {author} {\bibfnamefont {A.}~\bibnamefont
  {Gonz{\'a}lez-Fuentes}}\ and\ \bibinfo {author} {\bibfnamefont
  {A.}~\bibnamefont {G{\'o}mez-Valent}},\ }\href@noop {} {\  (\bibinfo {year}
  {2025})},\ \Eprint {https://arxiv.org/abs/2506.11758} {arXiv:2506.11758
  [astro-ph.CO]} \BibitemShut {NoStop}%
\bibitem [{\citenamefont {Wang}\ and\ \citenamefont
  {Freese}(2025)}]{Wang:2025vfb}%
  \BibitemOpen
  \bibfield  {author} {\bibinfo {author} {\bibfnamefont {Y.}~\bibnamefont
  {Wang}}\ and\ \bibinfo {author} {\bibfnamefont {K.}~\bibnamefont {Freese}},\
  }\href@noop {} {\  (\bibinfo {year} {2025})},\ \Eprint
  {https://arxiv.org/abs/2505.17415} {arXiv:2505.17415 [astro-ph.CO]}
  \BibitemShut {NoStop}%
\bibitem [{\citenamefont {Calderon}\ \emph
  {et~al.}(2024{\natexlab{a}})\citenamefont {Calderon} \emph
  {et~al.}}]{DESI:2024aqx}%
  \BibitemOpen
  \bibfield  {author} {\bibinfo {author} {\bibfnamefont {R.}~\bibnamefont
  {Calderon}} \emph {et~al.} (\bibinfo {collaboration} {DESI}),\ }\href
  {https://doi.org/10.1088/1475-7516/2024/10/048} {\bibfield  {journal}
  {\bibinfo  {journal} {JCAP}\ }\textbf {\bibinfo {volume} {10}},\ \bibinfo
  {pages} {048}},\ \Eprint {https://arxiv.org/abs/2405.04216} {arXiv:2405.04216
  [astro-ph.CO]} \BibitemShut {NoStop}%
\bibitem [{\citenamefont {Ormondroyd}\ \emph {et~al.}(2025)\citenamefont
  {Ormondroyd}, \citenamefont {Handley}, \citenamefont {Hobson},\ and\
  \citenamefont {Lasenby}}]{Ormondroyd:2025exu}%
  \BibitemOpen
  \bibfield  {author} {\bibinfo {author} {\bibfnamefont {A.~N.}\ \bibnamefont
  {Ormondroyd}}, \bibinfo {author} {\bibfnamefont {W.~J.}\ \bibnamefont
  {Handley}}, \bibinfo {author} {\bibfnamefont {M.~P.}\ \bibnamefont
  {Hobson}},\ and\ \bibinfo {author} {\bibfnamefont {A.~N.}\ \bibnamefont
  {Lasenby}},\ }\href {https://doi.org/10.1093/mnras/staf1144} {\bibfield
  {journal} {\bibinfo  {journal} {Mon. Not. Roy. Astron. Soc.}\ }\textbf
  {\bibinfo {volume} {3388}},\ \bibinfo {pages} {3400} (\bibinfo {year}
  {2025})},\ \Eprint {https://arxiv.org/abs/2503.08658} {arXiv:2503.08658
  [astro-ph.CO]} \BibitemShut {NoStop}%
\bibitem [{\citenamefont {Dinda}\ and\ \citenamefont
  {Maartens}(2025)}]{Dinda:2024ktd}%
  \BibitemOpen
  \bibfield  {author} {\bibinfo {author} {\bibfnamefont {B.~R.}\ \bibnamefont
  {Dinda}}\ and\ \bibinfo {author} {\bibfnamefont {R.}~\bibnamefont
  {Maartens}},\ }\href {https://doi.org/10.1088/1475-7516/2025/01/120}
  {\bibfield  {journal} {\bibinfo  {journal} {JCAP}\ }\textbf {\bibinfo
  {volume} {01}},\ \bibinfo {pages} {120}},\ \Eprint
  {https://arxiv.org/abs/2407.17252} {arXiv:2407.17252 [astro-ph.CO]}
  \BibitemShut {NoStop}%
\bibitem [{\citenamefont {de~Souza}\ \emph {et~al.}(2025)\citenamefont
  {de~Souza}, \citenamefont {Sousa-Neto}, \citenamefont {Gonz{\'a}lez},\ and\
  \citenamefont {Alcaniz}}]{deSouza:2025vdv}%
  \BibitemOpen
  \bibfield  {author} {\bibinfo {author} {\bibfnamefont {R.}~\bibnamefont
  {de~Souza}}, \bibinfo {author} {\bibfnamefont {A.}~\bibnamefont
  {Sousa-Neto}}, \bibinfo {author} {\bibfnamefont {J.~E.}\ \bibnamefont
  {Gonz{\'a}lez}},\ and\ \bibinfo {author} {\bibfnamefont {J.}~\bibnamefont
  {Alcaniz}},\ }\href@noop {} {\  (\bibinfo {year} {2025})},\ \Eprint
  {https://arxiv.org/abs/2511.13666} {arXiv:2511.13666 [astro-ph.CO]}
  \BibitemShut {NoStop}%
\bibitem [{\citenamefont {Berti}\ \emph {et~al.}(2025)\citenamefont {Berti},
  \citenamefont {Bellini}, \citenamefont {Bonvin}, \citenamefont {Kunz},
  \citenamefont {Viel},\ and\ \citenamefont {Zumalacarregui}}]{Berti:2025phi}%
  \BibitemOpen
  \bibfield  {author} {\bibinfo {author} {\bibfnamefont {M.}~\bibnamefont
  {Berti}}, \bibinfo {author} {\bibfnamefont {E.}~\bibnamefont {Bellini}},
  \bibinfo {author} {\bibfnamefont {C.}~\bibnamefont {Bonvin}}, \bibinfo
  {author} {\bibfnamefont {M.}~\bibnamefont {Kunz}}, \bibinfo {author}
  {\bibfnamefont {M.}~\bibnamefont {Viel}},\ and\ \bibinfo {author}
  {\bibfnamefont {M.}~\bibnamefont {Zumalacarregui}},\ }\href
  {https://doi.org/10.1103/dj3k-84v4} {\bibfield  {journal} {\bibinfo
  {journal} {Phys. Rev. D}\ }\textbf {\bibinfo {volume} {112}},\ \bibinfo
  {pages} {023518} (\bibinfo {year} {2025})},\ \Eprint
  {https://arxiv.org/abs/2503.13198} {arXiv:2503.13198 [astro-ph.CO]}
  \BibitemShut {NoStop}%
\bibitem [{\citenamefont {Calderon}\ \emph
  {et~al.}(2024{\natexlab{b}})\citenamefont {Calderon}, \citenamefont {Lodha},
  \citenamefont {Shafieloo}, \citenamefont {Linder}, \citenamefont {Sohn},
  \citenamefont {de~Mattia}, \citenamefont {Cervantes-Cota}, \citenamefont
  {Crittenden}, \citenamefont {Davis}, \citenamefont {Ishak}, \citenamefont
  {Kim}, \citenamefont {Matthewson}, \citenamefont {Niz}, \citenamefont {Park},
  \citenamefont {Aguilar}, \citenamefont {Ahlen}, \citenamefont {Allen},
  \citenamefont {Brooks}, \citenamefont {Claybaugh}, \citenamefont {de~la
  Macorra}, \citenamefont {Dey}, \citenamefont {Dey}, \citenamefont {Doel},
  \citenamefont {Forero-Romero}, \citenamefont {Gaztañaga}, \citenamefont
  {Gontcho}, \citenamefont {Honscheid}, \citenamefont {Howlett}, \citenamefont
  {Juneau}, \citenamefont {Kremin}, \citenamefont {Landriau}, \citenamefont
  {Le~Guillou}, \citenamefont {Levi}, \citenamefont {Manera}, \citenamefont
  {Miquel}, \citenamefont {Moustakas}, \citenamefont {Newman}, \citenamefont
  {Palanque-Delabrouille}, \citenamefont {Percival}, \citenamefont {Poppett},
  \citenamefont {Prada}, \citenamefont {Rezaie}, \citenamefont {Rossi},
  \citenamefont {Ruhlmann-Kleider}, \citenamefont {Sanchez}, \citenamefont
  {Schlegel}, \citenamefont {Schubnell}, \citenamefont {Seo}, \citenamefont
  {Sprayberry}, \citenamefont {Tarlé}, \citenamefont {Taylor}, \citenamefont
  {Vargas-Magaña}, \citenamefont {Weaver}, \citenamefont {Zarrouk},\ and\
  \citenamefont {Zou}}]{Calderon_2024}%
  \BibitemOpen
  \bibfield  {author} {\bibinfo {author} {\bibfnamefont {R.}~\bibnamefont
  {Calderon}}, \bibinfo {author} {\bibfnamefont {K.}~\bibnamefont {Lodha}},
  \bibinfo {author} {\bibfnamefont {A.}~\bibnamefont {Shafieloo}}, \bibinfo
  {author} {\bibfnamefont {E.}~\bibnamefont {Linder}}, \bibinfo {author}
  {\bibfnamefont {W.}~\bibnamefont {Sohn}}, \bibinfo {author} {\bibfnamefont
  {A.}~\bibnamefont {de~Mattia}}, \bibinfo {author} {\bibfnamefont
  {J.}~\bibnamefont {Cervantes-Cota}}, \bibinfo {author} {\bibfnamefont
  {R.}~\bibnamefont {Crittenden}}, \bibinfo {author} {\bibfnamefont
  {T.}~\bibnamefont {Davis}}, \bibinfo {author} {\bibfnamefont
  {M.}~\bibnamefont {Ishak}}, \bibinfo {author} {\bibfnamefont
  {A.}~\bibnamefont {Kim}}, \bibinfo {author} {\bibfnamefont {W.}~\bibnamefont
  {Matthewson}}, \bibinfo {author} {\bibfnamefont {G.}~\bibnamefont {Niz}},
  \bibinfo {author} {\bibfnamefont {S.}~\bibnamefont {Park}}, \bibinfo {author}
  {\bibfnamefont {J.}~\bibnamefont {Aguilar}}, \bibinfo {author} {\bibfnamefont
  {S.}~\bibnamefont {Ahlen}}, \bibinfo {author} {\bibfnamefont
  {S.}~\bibnamefont {Allen}}, \bibinfo {author} {\bibfnamefont
  {D.}~\bibnamefont {Brooks}}, \bibinfo {author} {\bibfnamefont
  {T.}~\bibnamefont {Claybaugh}}, \bibinfo {author} {\bibfnamefont
  {A.}~\bibnamefont {de~la Macorra}}, \bibinfo {author} {\bibfnamefont
  {A.}~\bibnamefont {Dey}}, \bibinfo {author} {\bibfnamefont {B.}~\bibnamefont
  {Dey}}, \bibinfo {author} {\bibfnamefont {P.}~\bibnamefont {Doel}}, \bibinfo
  {author} {\bibfnamefont {J.}~\bibnamefont {Forero-Romero}}, \bibinfo {author}
  {\bibfnamefont {E.}~\bibnamefont {Gaztañaga}}, \bibinfo {author}
  {\bibfnamefont {S.~A.}\ \bibnamefont {Gontcho}}, \bibinfo {author}
  {\bibfnamefont {K.}~\bibnamefont {Honscheid}}, \bibinfo {author}
  {\bibfnamefont {C.}~\bibnamefont {Howlett}}, \bibinfo {author} {\bibfnamefont
  {S.}~\bibnamefont {Juneau}}, \bibinfo {author} {\bibfnamefont
  {A.}~\bibnamefont {Kremin}}, \bibinfo {author} {\bibfnamefont
  {M.}~\bibnamefont {Landriau}}, \bibinfo {author} {\bibfnamefont
  {L.}~\bibnamefont {Le~Guillou}}, \bibinfo {author} {\bibfnamefont
  {M.}~\bibnamefont {Levi}}, \bibinfo {author} {\bibfnamefont {M.}~\bibnamefont
  {Manera}}, \bibinfo {author} {\bibfnamefont {R.}~\bibnamefont {Miquel}},
  \bibinfo {author} {\bibfnamefont {J.}~\bibnamefont {Moustakas}}, \bibinfo
  {author} {\bibfnamefont {J.}~\bibnamefont {Newman}}, \bibinfo {author}
  {\bibfnamefont {N.}~\bibnamefont {Palanque-Delabrouille}}, \bibinfo {author}
  {\bibfnamefont {W.}~\bibnamefont {Percival}}, \bibinfo {author}
  {\bibfnamefont {C.}~\bibnamefont {Poppett}}, \bibinfo {author} {\bibfnamefont
  {F.}~\bibnamefont {Prada}}, \bibinfo {author} {\bibfnamefont
  {M.}~\bibnamefont {Rezaie}}, \bibinfo {author} {\bibfnamefont
  {G.}~\bibnamefont {Rossi}}, \bibinfo {author} {\bibfnamefont
  {V.}~\bibnamefont {Ruhlmann-Kleider}}, \bibinfo {author} {\bibfnamefont
  {E.}~\bibnamefont {Sanchez}}, \bibinfo {author} {\bibfnamefont
  {D.}~\bibnamefont {Schlegel}}, \bibinfo {author} {\bibfnamefont
  {M.}~\bibnamefont {Schubnell}}, \bibinfo {author} {\bibfnamefont
  {H.}~\bibnamefont {Seo}}, \bibinfo {author} {\bibfnamefont {D.}~\bibnamefont
  {Sprayberry}}, \bibinfo {author} {\bibfnamefont {G.}~\bibnamefont {Tarlé}},
  \bibinfo {author} {\bibfnamefont {P.}~\bibnamefont {Taylor}}, \bibinfo
  {author} {\bibfnamefont {M.}~\bibnamefont {Vargas-Magaña}}, \bibinfo
  {author} {\bibfnamefont {B.}~\bibnamefont {Weaver}}, \bibinfo {author}
  {\bibfnamefont {P.}~\bibnamefont {Zarrouk}},\ and\ \bibinfo {author}
  {\bibfnamefont {H.}~\bibnamefont {Zou}},\ }\href
  {https://doi.org/10.1088/1475-7516/2024/10/048} {\bibfield  {journal}
  {\bibinfo  {journal} {Journal of Cosmology and Astroparticle Physics}\
  }\textbf {\bibinfo {volume} {2024}}\bibinfo  {number} { (10)},\ \bibinfo
  {pages} {048}}\BibitemShut {NoStop}%
\bibitem [{\citenamefont {Nojiri}\ \emph {et~al.}(2025)\citenamefont {Nojiri},
  \citenamefont {Odintsov},\ and\ \citenamefont {Oikonomou}}]{Nojiri:2025low}%
  \BibitemOpen
\bibfield  {number} {  }\bibfield  {author} {\bibinfo {author} {\bibfnamefont
  {S.}~\bibnamefont {Nojiri}}, \bibinfo {author} {\bibfnamefont {S.~D.}\
  \bibnamefont {Odintsov}},\ and\ \bibinfo {author} {\bibfnamefont {V.~K.}\
  \bibnamefont {Oikonomou}},\ }\href {https://doi.org/10.1103/16yg-966k}
  {\bibfield  {journal} {\bibinfo  {journal} {Phys. Rev. D}\ }\textbf {\bibinfo
  {volume} {112}},\ \bibinfo {pages} {104035} (\bibinfo {year} {2025})},\
  \Eprint {https://arxiv.org/abs/2506.21010} {arXiv:2506.21010 [gr-qc]}
  \BibitemShut {NoStop}%
\bibitem [{\citenamefont {Liu}\ \emph {et~al.}(2019)\citenamefont {Liu},
  \citenamefont {Qin}, \citenamefont {Zhang}, \citenamefont {Zhang},\ and\
  \citenamefont {Yu}}]{Liu:2015mkm}%
  \BibitemOpen
  \bibfield  {author} {\bibinfo {author} {\bibfnamefont {Z.-E.}\ \bibnamefont
  {Liu}}, \bibinfo {author} {\bibfnamefont {H.-F.}\ \bibnamefont {Qin}},
  \bibinfo {author} {\bibfnamefont {J.}~\bibnamefont {Zhang}}, \bibinfo
  {author} {\bibfnamefont {T.-J.}\ \bibnamefont {Zhang}},\ and\ \bibinfo
  {author} {\bibfnamefont {H.-R.}\ \bibnamefont {Yu}},\ }\href
  {https://doi.org/10.1016/j.dark.2019.100379} {\bibfield  {journal} {\bibinfo
  {journal} {Phys. Dark Univ.}\ }\textbf {\bibinfo {volume} {26}},\ \bibinfo
  {pages} {100379} (\bibinfo {year} {2019})},\ \Eprint
  {https://arxiv.org/abs/1501.02971} {arXiv:1501.02971 [astro-ph.CO]}
  \BibitemShut {NoStop}%
\bibitem [{\citenamefont {Rasmussen}(2004)}]{Rasmussen2004}%
  \BibitemOpen
  \bibfield  {author} {\bibinfo {author} {\bibfnamefont {C.~E.}\ \bibnamefont
  {Rasmussen}},\ }\bibinfo {title} {Gaussian processes in machine learning},\
  in\ \href {https://doi.org/10.1007/978-3-540-28650-9_4} {\emph {\bibinfo
  {booktitle} {Advanced Lectures on Machine Learning: ML Summer Schools 2003,
  Canberra, Australia, February 2 - 14, 2003, T{\"u}bingen, Germany, August 4 -
  16, 2003, Revised Lectures}}},\ \bibinfo {editor} {edited by\ \bibinfo
  {editor} {\bibfnamefont {O.}~\bibnamefont {Bousquet}}, \bibinfo {editor}
  {\bibfnamefont {U.}~\bibnamefont {von Luxburg}},\ and\ \bibinfo {editor}
  {\bibfnamefont {G.}~\bibnamefont {R{\"a}tsch}}}\ (\bibinfo  {publisher}
  {Springer Berlin Heidelberg},\ \bibinfo {address} {Berlin, Heidelberg},\
  \bibinfo {year} {2004})\ pp.\ \bibinfo {pages} {63--71}\BibitemShut {NoStop}%
\bibitem [{\citenamefont {G\'omez-Valent}\ and\ \citenamefont
  {Amendola}(2018)}]{Gomez-Valent:2018hwc}%
  \BibitemOpen
  \bibfield  {author} {\bibinfo {author} {\bibfnamefont {A.}~\bibnamefont
  {G\'omez-Valent}}\ and\ \bibinfo {author} {\bibfnamefont {L.}~\bibnamefont
  {Amendola}},\ }\href {https://doi.org/10.1088/1475-7516/2018/04/051}
  {\bibfield  {journal} {\bibinfo  {journal} {JCAP}\ }\textbf {\bibinfo
  {volume} {04}},\ \bibinfo {pages} {051}},\ \Eprint
  {https://arxiv.org/abs/1802.01505} {arXiv:1802.01505 [astro-ph.CO]}
  \BibitemShut {NoStop}%
\bibitem [{\citenamefont {G{\'o}mez-Valent}(2019)}]{Gomez-Valent:2018gvm}%
  \BibitemOpen
  \bibfield  {author} {\bibinfo {author} {\bibfnamefont {A.}~\bibnamefont
  {G{\'o}mez-Valent}},\ }\href {https://doi.org/10.1088/1475-7516/2019/05/026}
  {\bibfield  {journal} {\bibinfo  {journal} {JCAP}\ }\textbf {\bibinfo
  {volume} {05}},\ \bibinfo {pages} {026}},\ \Eprint
  {https://arxiv.org/abs/1810.02278} {arXiv:1810.02278 [astro-ph.CO]}
  \BibitemShut {NoStop}%
\bibitem [{\citenamefont {Fritsch}\ and\ \citenamefont
  {Butland}(1984)}]{doi:10.1137/0905021}%
  \BibitemOpen
  \bibfield  {author} {\bibinfo {author} {\bibfnamefont {F.~N.}\ \bibnamefont
  {Fritsch}}\ and\ \bibinfo {author} {\bibfnamefont {J.}~\bibnamefont
  {Butland}},\ }\href {https://doi.org/10.1137/0905021} {\bibfield  {journal}
  {\bibinfo  {journal} {SIAM Journal on Scientific and Statistical Computing}\
  }\textbf {\bibinfo {volume} {5}},\ \bibinfo {pages} {300} (\bibinfo {year}
  {1984})},\ \Eprint {https://arxiv.org/abs/https://doi.org/10.1137/0905021}
  {https://doi.org/10.1137/0905021} \BibitemShut {NoStop}%
\bibitem [{\citenamefont {Fritsch}\ and\ \citenamefont
  {Carlson}(1980)}]{doi:10.1137/0717021}%
  \BibitemOpen
  \bibfield  {author} {\bibinfo {author} {\bibfnamefont {F.~N.}\ \bibnamefont
  {Fritsch}}\ and\ \bibinfo {author} {\bibfnamefont {R.~E.}\ \bibnamefont
  {Carlson}},\ }\href {https://doi.org/10.1137/0717021} {\bibfield  {journal}
  {\bibinfo  {journal} {SIAM Journal on Numerical Analysis}\ }\textbf {\bibinfo
  {volume} {17}},\ \bibinfo {pages} {238} (\bibinfo {year} {1980})},\ \Eprint
  {https://arxiv.org/abs/https://doi.org/10.1137/0717021}
  {https://doi.org/10.1137/0717021} \BibitemShut {NoStop}%
\bibitem [{\citenamefont {Gariazzo}\ \emph {et~al.}(2023)\citenamefont
  {Gariazzo}, \citenamefont {Mena},\ and\ \citenamefont
  {Schwetz}}]{Gariazzo:2023joe}%
  \BibitemOpen
  \bibfield  {author} {\bibinfo {author} {\bibfnamefont {S.}~\bibnamefont
  {Gariazzo}}, \bibinfo {author} {\bibfnamefont {O.}~\bibnamefont {Mena}},\
  and\ \bibinfo {author} {\bibfnamefont {T.}~\bibnamefont {Schwetz}},\ }\href
  {https://doi.org/10.1016/j.dark.2023.101226} {\bibfield  {journal} {\bibinfo
  {journal} {Phys. Dark Univ.}\ }\textbf {\bibinfo {volume} {40}},\ \bibinfo
  {pages} {101226} (\bibinfo {year} {2023})},\ \Eprint
  {https://arxiv.org/abs/2302.14159} {arXiv:2302.14159 [hep-ph]} \BibitemShut
  {NoStop}%
\bibitem [{\citenamefont {Bert{\'o}lez-Mart{\'\i}nez}\ \emph
  {et~al.}(2025{\natexlab{a}})\citenamefont {Bert{\'o}lez-Mart{\'\i}nez},
  \citenamefont {Esteban}, \citenamefont {Hajjar}, \citenamefont {Mena},\ and\
  \citenamefont {Salvado}}]{Bertolez-Martinez:2024wez}%
  \BibitemOpen
  \bibfield  {author} {\bibinfo {author} {\bibfnamefont {T.}~\bibnamefont
  {Bert{\'o}lez-Mart{\'\i}nez}}, \bibinfo {author} {\bibfnamefont
  {I.}~\bibnamefont {Esteban}}, \bibinfo {author} {\bibfnamefont
  {R.}~\bibnamefont {Hajjar}}, \bibinfo {author} {\bibfnamefont
  {O.}~\bibnamefont {Mena}},\ and\ \bibinfo {author} {\bibfnamefont
  {J.}~\bibnamefont {Salvado}},\ }\href
  {https://doi.org/10.1088/1475-7516/2025/06/058} {\bibfield  {journal}
  {\bibinfo  {journal} {JCAP}\ }\textbf {\bibinfo {volume} {06}},\ \bibinfo
  {pages} {058}},\ \Eprint {https://arxiv.org/abs/2411.14524} {arXiv:2411.14524
  [astro-ph.CO]} \BibitemShut {NoStop}%
\bibitem [{\citenamefont {Naredo-Tuero}\ \emph {et~al.}(2024)\citenamefont
  {Naredo-Tuero}, \citenamefont {Escudero}, \citenamefont
  {Fern\'andez-Mart\'\i{}nez}, \citenamefont {Marcano},\ and\ \citenamefont
  {Poulin}}]{Naredo-Tuero:2024sgf}%
  \BibitemOpen
  \bibfield  {author} {\bibinfo {author} {\bibfnamefont {D.}~\bibnamefont
  {Naredo-Tuero}}, \bibinfo {author} {\bibfnamefont {M.}~\bibnamefont
  {Escudero}}, \bibinfo {author} {\bibfnamefont {E.}~\bibnamefont
  {Fern\'andez-Mart\'\i{}nez}}, \bibinfo {author} {\bibfnamefont
  {X.}~\bibnamefont {Marcano}},\ and\ \bibinfo {author} {\bibfnamefont
  {V.}~\bibnamefont {Poulin}},\ }\href@noop {} {\  (\bibinfo {year} {2024})},\
  \Eprint {https://arxiv.org/abs/2407.13831} {arXiv:2407.13831 [astro-ph.CO]}
  \BibitemShut {NoStop}%
\bibitem [{\citenamefont {Craig}\ \emph {et~al.}(2024)\citenamefont {Craig},
  \citenamefont {Green}, \citenamefont {Meyers},\ and\ \citenamefont
  {Rajendran}}]{Craig:2024tky}%
  \BibitemOpen
  \bibfield  {author} {\bibinfo {author} {\bibfnamefont {N.}~\bibnamefont
  {Craig}}, \bibinfo {author} {\bibfnamefont {D.}~\bibnamefont {Green}},
  \bibinfo {author} {\bibfnamefont {J.}~\bibnamefont {Meyers}},\ and\ \bibinfo
  {author} {\bibfnamefont {S.}~\bibnamefont {Rajendran}},\ }\href
  {https://doi.org/10.1007/JHEP09(2024)097} {\bibfield  {journal} {\bibinfo
  {journal} {JHEP}\ }\textbf {\bibinfo {volume} {09}},\ \bibinfo {pages}
  {097}},\ \Eprint {https://arxiv.org/abs/2405.00836} {arXiv:2405.00836
  [astro-ph.CO]} \BibitemShut {NoStop}%
\bibitem [{\citenamefont {Jiang}\ \emph {et~al.}(2025)\citenamefont {Jiang},
  \citenamefont {Giar{\`e}}, \citenamefont {Gariazzo}, \citenamefont
  {Dainotti}, \citenamefont {Di~Valentino}, \citenamefont {Mena}, \citenamefont
  {Pedrotti}, \citenamefont {da~Costa},\ and\ \citenamefont
  {Vagnozzi}}]{Jiang:2024viw}%
  \BibitemOpen
  \bibfield  {author} {\bibinfo {author} {\bibfnamefont {J.-Q.}\ \bibnamefont
  {Jiang}}, \bibinfo {author} {\bibfnamefont {W.}~\bibnamefont {Giar{\`e}}},
  \bibinfo {author} {\bibfnamefont {S.}~\bibnamefont {Gariazzo}}, \bibinfo
  {author} {\bibfnamefont {M.~G.}\ \bibnamefont {Dainotti}}, \bibinfo {author}
  {\bibfnamefont {E.}~\bibnamefont {Di~Valentino}}, \bibinfo {author}
  {\bibfnamefont {O.}~\bibnamefont {Mena}}, \bibinfo {author} {\bibfnamefont
  {D.}~\bibnamefont {Pedrotti}}, \bibinfo {author} {\bibfnamefont {S.~S.}\
  \bibnamefont {da~Costa}},\ and\ \bibinfo {author} {\bibfnamefont
  {S.}~\bibnamefont {Vagnozzi}},\ }\href
  {https://doi.org/10.1088/1475-7516/2025/01/153} {\bibfield  {journal}
  {\bibinfo  {journal} {JCAP}\ }\textbf {\bibinfo {volume} {01}},\ \bibinfo
  {pages} {153}},\ \Eprint {https://arxiv.org/abs/2407.18047} {arXiv:2407.18047
  [astro-ph.CO]} \BibitemShut {NoStop}%
\bibitem [{\citenamefont {Hannestad}(2005)}]{Hannestad:2005gj}%
  \BibitemOpen
  \bibfield  {author} {\bibinfo {author} {\bibfnamefont {S.}~\bibnamefont
  {Hannestad}},\ }\href {https://doi.org/10.1103/PhysRevLett.95.221301}
  {\bibfield  {journal} {\bibinfo  {journal} {Phys. Rev. Lett.}\ }\textbf
  {\bibinfo {volume} {95}},\ \bibinfo {pages} {221301} (\bibinfo {year}
  {2005})},\ \Eprint {https://arxiv.org/abs/astro-ph/0505551}
  {arXiv:astro-ph/0505551} \BibitemShut {NoStop}%
\bibitem [{\citenamefont {Lorenz}\ \emph {et~al.}(2017)\citenamefont {Lorenz},
  \citenamefont {Calabrese},\ and\ \citenamefont {Alonso}}]{Lorenz:2017fgo}%
  \BibitemOpen
  \bibfield  {author} {\bibinfo {author} {\bibfnamefont {C.~S.}\ \bibnamefont
  {Lorenz}}, \bibinfo {author} {\bibfnamefont {E.}~\bibnamefont {Calabrese}},\
  and\ \bibinfo {author} {\bibfnamefont {D.}~\bibnamefont {Alonso}},\ }\href
  {https://doi.org/10.1103/PhysRevD.96.043510} {\bibfield  {journal} {\bibinfo
  {journal} {Phys. Rev. D}\ }\textbf {\bibinfo {volume} {96}},\ \bibinfo
  {pages} {043510} (\bibinfo {year} {2017})},\ \Eprint
  {https://arxiv.org/abs/1706.00730} {arXiv:1706.00730 [astro-ph.CO]}
  \BibitemShut {NoStop}%
\bibitem [{\citenamefont {Vagnozzi}\ \emph {et~al.}(2017)\citenamefont
  {Vagnozzi}, \citenamefont {Giusarma}, \citenamefont {Mena}, \citenamefont
  {Freese}, \citenamefont {Gerbino}, \citenamefont {Ho},\ and\ \citenamefont
  {Lattanzi}}]{Vagnozzi:2017ovm}%
  \BibitemOpen
  \bibfield  {author} {\bibinfo {author} {\bibfnamefont {S.}~\bibnamefont
  {Vagnozzi}}, \bibinfo {author} {\bibfnamefont {E.}~\bibnamefont {Giusarma}},
  \bibinfo {author} {\bibfnamefont {O.}~\bibnamefont {Mena}}, \bibinfo {author}
  {\bibfnamefont {K.}~\bibnamefont {Freese}}, \bibinfo {author} {\bibfnamefont
  {M.}~\bibnamefont {Gerbino}}, \bibinfo {author} {\bibfnamefont
  {S.}~\bibnamefont {Ho}},\ and\ \bibinfo {author} {\bibfnamefont
  {M.}~\bibnamefont {Lattanzi}},\ }\href
  {https://doi.org/10.1103/PhysRevD.96.123503} {\bibfield  {journal} {\bibinfo
  {journal} {Phys. Rev. D}\ }\textbf {\bibinfo {volume} {96}},\ \bibinfo
  {pages} {123503} (\bibinfo {year} {2017})},\ \Eprint
  {https://arxiv.org/abs/1701.08172} {arXiv:1701.08172 [astro-ph.CO]}
  \BibitemShut {NoStop}%
\bibitem [{\citenamefont {Sutherland}(2018)}]{Sutherland:2018ghu}%
  \BibitemOpen
  \bibfield  {author} {\bibinfo {author} {\bibfnamefont {W.}~\bibnamefont
  {Sutherland}},\ }\href {https://doi.org/10.1093/mnras/sty687} {\bibfield
  {journal} {\bibinfo  {journal} {Mon. Not. Roy. Astron. Soc.}\ }\textbf
  {\bibinfo {volume} {477}},\ \bibinfo {pages} {1913} (\bibinfo {year}
  {2018})},\ \Eprint {https://arxiv.org/abs/1803.02298} {arXiv:1803.02298
  [astro-ph.CO]} \BibitemShut {NoStop}%
\bibitem [{\citenamefont {Sahl{\'e}n}(2019)}]{Sahlen:2018cku}%
  \BibitemOpen
  \bibfield  {author} {\bibinfo {author} {\bibfnamefont {M.}~\bibnamefont
  {Sahl{\'e}n}},\ }\href {https://doi.org/10.1103/PhysRevD.99.063525}
  {\bibfield  {journal} {\bibinfo  {journal} {Phys. Rev. D}\ }\textbf {\bibinfo
  {volume} {99}},\ \bibinfo {pages} {063525} (\bibinfo {year} {2019})},\
  \Eprint {https://arxiv.org/abs/1807.02470} {arXiv:1807.02470 [astro-ph.CO]}
  \BibitemShut {NoStop}%
\bibitem [{\citenamefont {Yang}\ \emph {et~al.}(2021)\citenamefont {Yang},
  \citenamefont {Di~Valentino}, \citenamefont {Pan},\ and\ \citenamefont
  {Mena}}]{Yang:2020ope}%
  \BibitemOpen
  \bibfield  {author} {\bibinfo {author} {\bibfnamefont {W.}~\bibnamefont
  {Yang}}, \bibinfo {author} {\bibfnamefont {E.}~\bibnamefont {Di~Valentino}},
  \bibinfo {author} {\bibfnamefont {S.}~\bibnamefont {Pan}},\ and\ \bibinfo
  {author} {\bibfnamefont {O.}~\bibnamefont {Mena}},\ }\href
  {https://doi.org/10.1016/j.dark.2020.100762} {\bibfield  {journal} {\bibinfo
  {journal} {Phys. Dark Univ.}\ }\textbf {\bibinfo {volume} {31}},\ \bibinfo
  {pages} {100762} (\bibinfo {year} {2021})},\ \Eprint
  {https://arxiv.org/abs/2007.02927} {arXiv:2007.02927 [astro-ph.CO]}
  \BibitemShut {NoStop}%
\bibitem [{\citenamefont {Zhang}(2016)}]{Zhang:2015uhk}%
  \BibitemOpen
  \bibfield  {author} {\bibinfo {author} {\bibfnamefont {X.}~\bibnamefont
  {Zhang}},\ }\href {https://doi.org/10.1103/PhysRevD.93.083011} {\bibfield
  {journal} {\bibinfo  {journal} {Phys. Rev. D}\ }\textbf {\bibinfo {volume}
  {93}},\ \bibinfo {pages} {083011} (\bibinfo {year} {2016})},\ \Eprint
  {https://arxiv.org/abs/1511.02651} {arXiv:1511.02651 [astro-ph.CO]}
  \BibitemShut {NoStop}%
\bibitem [{\citenamefont {Khalifeh}\ and\ \citenamefont
  {Jimenez}(2021)}]{Khalifeh:2021ree}%
  \BibitemOpen
  \bibfield  {author} {\bibinfo {author} {\bibfnamefont {A.~R.}\ \bibnamefont
  {Khalifeh}}\ and\ \bibinfo {author} {\bibfnamefont {R.}~\bibnamefont
  {Jimenez}},\ }\href {https://doi.org/10.1016/j.dark.2021.100897} {\bibfield
  {journal} {\bibinfo  {journal} {Phys. Dark Univ.}\ }\textbf {\bibinfo
  {volume} {34}},\ \bibinfo {pages} {100897} (\bibinfo {year} {2021})},\
  \Eprint {https://arxiv.org/abs/2105.07973} {arXiv:2105.07973 [astro-ph.CO]}
  \BibitemShut {NoStop}%
\bibitem [{\citenamefont {Nair}\ \emph {et~al.}(2025)\citenamefont {Nair},
  \citenamefont {Chakraborty}, \citenamefont {Amendola},\ and\ \citenamefont
  {Das}}]{Nair:2025uyn}%
  \BibitemOpen
  \bibfield  {author} {\bibinfo {author} {\bibfnamefont {G.~S.}\ \bibnamefont
  {Nair}}, \bibinfo {author} {\bibfnamefont {A.}~\bibnamefont {Chakraborty}},
  \bibinfo {author} {\bibfnamefont {L.}~\bibnamefont {Amendola}},\ and\
  \bibinfo {author} {\bibfnamefont {S.}~\bibnamefont {Das}},\ }\href@noop {} {\
   (\bibinfo {year} {2025})},\ \Eprint {https://arxiv.org/abs/2512.08752}
  {arXiv:2512.08752 [astro-ph.CO]} \BibitemShut {NoStop}%
\bibitem [{\citenamefont {Zhao}\ \emph {et~al.}(2007)\citenamefont {Zhao},
  \citenamefont {Xia},\ and\ \citenamefont {Zhang}}]{Zhao:2006zf}%
  \BibitemOpen
  \bibfield  {author} {\bibinfo {author} {\bibfnamefont {G.-B.}\ \bibnamefont
  {Zhao}}, \bibinfo {author} {\bibfnamefont {J.-Q.}\ \bibnamefont {Xia}},\ and\
  \bibinfo {author} {\bibfnamefont {X.-M.}\ \bibnamefont {Zhang}},\ }\href
  {https://doi.org/10.1088/1475-7516/2007/07/010} {\bibfield  {journal}
  {\bibinfo  {journal} {JCAP}\ }\textbf {\bibinfo {volume} {07}},\ \bibinfo
  {pages} {010}},\ \Eprint {https://arxiv.org/abs/astro-ph/0611227}
  {arXiv:astro-ph/0611227} \BibitemShut {NoStop}%
\bibitem [{\citenamefont {Lorenz}\ \emph {et~al.}(2021)\citenamefont {Lorenz},
  \citenamefont {Funcke}, \citenamefont {L{\"o}ffler},\ and\ \citenamefont
  {Calabrese}}]{Lorenz:2021alz}%
  \BibitemOpen
  \bibfield  {author} {\bibinfo {author} {\bibfnamefont {C.~S.}\ \bibnamefont
  {Lorenz}}, \bibinfo {author} {\bibfnamefont {L.}~\bibnamefont {Funcke}},
  \bibinfo {author} {\bibfnamefont {M.}~\bibnamefont {L{\"o}ffler}},\ and\
  \bibinfo {author} {\bibfnamefont {E.}~\bibnamefont {Calabrese}},\ }\href
  {https://doi.org/10.1103/PhysRevD.104.123518} {\bibfield  {journal} {\bibinfo
   {journal} {Phys. Rev. D}\ }\textbf {\bibinfo {volume} {104}},\ \bibinfo
  {pages} {123518} (\bibinfo {year} {2021})},\ \Eprint
  {https://arxiv.org/abs/2102.13618} {arXiv:2102.13618 [astro-ph.CO]}
  \BibitemShut {NoStop}%
\bibitem [{\citenamefont {Gariazzo}\ \emph {et~al.}(2015)\citenamefont
  {Gariazzo}, \citenamefont {Giunti},\ and\ \citenamefont
  {Laveder}}]{Gariazzo:2014dla}%
  \BibitemOpen
  \bibfield  {author} {\bibinfo {author} {\bibfnamefont {S.}~\bibnamefont
  {Gariazzo}}, \bibinfo {author} {\bibfnamefont {C.}~\bibnamefont {Giunti}},\
  and\ \bibinfo {author} {\bibfnamefont {M.}~\bibnamefont {Laveder}},\ }\href
  {https://doi.org/10.1088/1475-7516/2015/04/023} {\bibfield  {journal}
  {\bibinfo  {journal} {JCAP}\ }\textbf {\bibinfo {volume} {04}},\ \bibinfo
  {pages} {023}},\ \Eprint {https://arxiv.org/abs/1412.7405} {arXiv:1412.7405
  [astro-ph.CO]} \BibitemShut {NoStop}%
\bibitem [{\citenamefont {Esteban}\ \emph {et~al.}(2024)\citenamefont
  {Esteban}, \citenamefont {Gonzalez-Garcia}, \citenamefont {Maltoni},
  \citenamefont {Martinez-Soler}, \citenamefont {Pinheiro},\ and\ \citenamefont
  {Schwetz}}]{Esteban:2024eli}%
  \BibitemOpen
  \bibfield  {author} {\bibinfo {author} {\bibfnamefont {I.}~\bibnamefont
  {Esteban}}, \bibinfo {author} {\bibfnamefont {M.~C.}\ \bibnamefont
  {Gonzalez-Garcia}}, \bibinfo {author} {\bibfnamefont {M.}~\bibnamefont
  {Maltoni}}, \bibinfo {author} {\bibfnamefont {I.}~\bibnamefont
  {Martinez-Soler}}, \bibinfo {author} {\bibfnamefont {J.~P.}\ \bibnamefont
  {Pinheiro}},\ and\ \bibinfo {author} {\bibfnamefont {T.}~\bibnamefont
  {Schwetz}},\ }\href {https://doi.org/10.1007/JHEP12(2024)216} {\bibfield
  {journal} {\bibinfo  {journal} {JHEP}\ }\textbf {\bibinfo {volume} {12}},\
  \bibinfo {pages} {216}},\ \Eprint {https://arxiv.org/abs/2410.05380}
  {arXiv:2410.05380 [hep-ph]} \BibitemShut {NoStop}%
\bibitem [{\citenamefont {de~Salas}\ \emph {et~al.}(2021)\citenamefont
  {de~Salas}, \citenamefont {Forero}, \citenamefont {Gariazzo}, \citenamefont
  {Mart{\'\i}nez-Mirav{\'e}}, \citenamefont {Mena}, \citenamefont {Ternes},
  \citenamefont {T{\'o}rtola},\ and\ \citenamefont {Valle}}]{deSalas:2020pgw}%
  \BibitemOpen
  \bibfield  {author} {\bibinfo {author} {\bibfnamefont {P.~F.}\ \bibnamefont
  {de~Salas}}, \bibinfo {author} {\bibfnamefont {D.~V.}\ \bibnamefont
  {Forero}}, \bibinfo {author} {\bibfnamefont {S.}~\bibnamefont {Gariazzo}},
  \bibinfo {author} {\bibfnamefont {P.}~\bibnamefont
  {Mart{\'\i}nez-Mirav{\'e}}}, \bibinfo {author} {\bibfnamefont
  {O.}~\bibnamefont {Mena}}, \bibinfo {author} {\bibfnamefont {C.~A.}\
  \bibnamefont {Ternes}}, \bibinfo {author} {\bibfnamefont {M.}~\bibnamefont
  {T{\'o}rtola}},\ and\ \bibinfo {author} {\bibfnamefont {J.~W.~F.}\
  \bibnamefont {Valle}},\ }\href {https://doi.org/10.1007/JHEP02(2021)071}
  {\bibfield  {journal} {\bibinfo  {journal} {JHEP}\ }\textbf {\bibinfo
  {volume} {02}},\ \bibinfo {pages} {071}},\ \Eprint
  {https://arxiv.org/abs/2006.11237} {arXiv:2006.11237 [hep-ph]} \BibitemShut
  {NoStop}%
\bibitem [{\citenamefont {Capozzi}\ \emph {et~al.}(2021)\citenamefont
  {Capozzi}, \citenamefont {Di~Valentino}, \citenamefont {Lisi}, \citenamefont
  {Marrone}, \citenamefont {Melchiorri},\ and\ \citenamefont
  {Palazzo}}]{Capozzi:2021fjo}%
  \BibitemOpen
  \bibfield  {author} {\bibinfo {author} {\bibfnamefont {F.}~\bibnamefont
  {Capozzi}}, \bibinfo {author} {\bibfnamefont {E.}~\bibnamefont
  {Di~Valentino}}, \bibinfo {author} {\bibfnamefont {E.}~\bibnamefont {Lisi}},
  \bibinfo {author} {\bibfnamefont {A.}~\bibnamefont {Marrone}}, \bibinfo
  {author} {\bibfnamefont {A.}~\bibnamefont {Melchiorri}},\ and\ \bibinfo
  {author} {\bibfnamefont {A.}~\bibnamefont {Palazzo}},\ }\href
  {https://doi.org/10.1103/PhysRevD.104.083031} {\bibfield  {journal} {\bibinfo
   {journal} {Phys. Rev. D}\ }\textbf {\bibinfo {volume} {104}},\ \bibinfo
  {pages} {083031} (\bibinfo {year} {2021})},\ \Eprint
  {https://arxiv.org/abs/2107.00532} {arXiv:2107.00532 [hep-ph]} \BibitemShut
  {NoStop}%
\bibitem [{\citenamefont {Aker}\ \emph {et~al.}(2025)\citenamefont {Aker} \emph
  {et~al.}}]{KATRIN:2024cdt}%
  \BibitemOpen
  \bibfield  {author} {\bibinfo {author} {\bibfnamefont {M.}~\bibnamefont
  {Aker}} \emph {et~al.} (\bibinfo {collaboration} {KATRIN}),\ }\href
  {https://doi.org/10.1126/science.adq9592} {\bibfield  {journal} {\bibinfo
  {journal} {Science}\ }\textbf {\bibinfo {volume} {388}},\ \bibinfo {pages}
  {adq9592} (\bibinfo {year} {2025})},\ \Eprint
  {https://arxiv.org/abs/2406.13516} {arXiv:2406.13516 [nucl-ex]} \BibitemShut
  {NoStop}%
\bibitem [{\citenamefont {Dvali}\ and\ \citenamefont
  {Funcke}(2016)}]{Dvali:2016uhn}%
  \BibitemOpen
  \bibfield  {author} {\bibinfo {author} {\bibfnamefont {G.}~\bibnamefont
  {Dvali}}\ and\ \bibinfo {author} {\bibfnamefont {L.}~\bibnamefont {Funcke}},\
  }\href {https://doi.org/10.1103/PhysRevD.93.113002} {\bibfield  {journal}
  {\bibinfo  {journal} {Phys. Rev. D}\ }\textbf {\bibinfo {volume} {93}},\
  \bibinfo {pages} {113002} (\bibinfo {year} {2016})},\ \Eprint
  {https://arxiv.org/abs/1602.03191} {arXiv:1602.03191 [hep-ph]} \BibitemShut
  {NoStop}%
\bibitem [{\citenamefont {Lorenz}\ \emph {et~al.}(2019)\citenamefont {Lorenz},
  \citenamefont {Funcke}, \citenamefont {Calabrese},\ and\ \citenamefont
  {Hannestad}}]{Lorenz:2018fzb}%
  \BibitemOpen
  \bibfield  {author} {\bibinfo {author} {\bibfnamefont {C.~S.}\ \bibnamefont
  {Lorenz}}, \bibinfo {author} {\bibfnamefont {L.}~\bibnamefont {Funcke}},
  \bibinfo {author} {\bibfnamefont {E.}~\bibnamefont {Calabrese}},\ and\
  \bibinfo {author} {\bibfnamefont {S.}~\bibnamefont {Hannestad}},\ }\href
  {https://doi.org/10.1103/PhysRevD.99.023501} {\bibfield  {journal} {\bibinfo
  {journal} {Phys. Rev. D}\ }\textbf {\bibinfo {volume} {99}},\ \bibinfo
  {pages} {023501} (\bibinfo {year} {2019})},\ \Eprint
  {https://arxiv.org/abs/1811.01991} {arXiv:1811.01991 [astro-ph.CO]}
  \BibitemShut {NoStop}%
\bibitem [{\citenamefont {Dvali}\ \emph {et~al.}(2023)\citenamefont {Dvali},
  \citenamefont {Funcke},\ and\ \citenamefont {Vachaspati}}]{Dvali:2021uvk}%
  \BibitemOpen
  \bibfield  {author} {\bibinfo {author} {\bibfnamefont {G.}~\bibnamefont
  {Dvali}}, \bibinfo {author} {\bibfnamefont {L.}~\bibnamefont {Funcke}},\ and\
  \bibinfo {author} {\bibfnamefont {T.}~\bibnamefont {Vachaspati}},\ }\href
  {https://doi.org/10.1103/PhysRevLett.130.091601} {\bibfield  {journal}
  {\bibinfo  {journal} {Phys. Rev. Lett.}\ }\textbf {\bibinfo {volume} {130}},\
  \bibinfo {pages} {091601} (\bibinfo {year} {2023})},\ \Eprint
  {https://arxiv.org/abs/2112.02107} {arXiv:2112.02107 [hep-ph]} \BibitemShut
  {NoStop}%
\bibitem [{\citenamefont {Franca}\ \emph {et~al.}(2009)\citenamefont {Franca},
  \citenamefont {Lattanzi}, \citenamefont {Lesgourgues},\ and\ \citenamefont
  {Pastor}}]{Franca:2009xp}%
  \BibitemOpen
  \bibfield  {author} {\bibinfo {author} {\bibfnamefont {U.}~\bibnamefont
  {Franca}}, \bibinfo {author} {\bibfnamefont {M.}~\bibnamefont {Lattanzi}},
  \bibinfo {author} {\bibfnamefont {J.}~\bibnamefont {Lesgourgues}},\ and\
  \bibinfo {author} {\bibfnamefont {S.}~\bibnamefont {Pastor}},\ }\href
  {https://doi.org/10.1103/PhysRevD.80.083506} {\bibfield  {journal} {\bibinfo
  {journal} {Phys. Rev. D}\ }\textbf {\bibinfo {volume} {80}},\ \bibinfo
  {pages} {083506} (\bibinfo {year} {2009})},\ \Eprint
  {https://arxiv.org/abs/0908.0534} {arXiv:0908.0534 [astro-ph.CO]}
  \BibitemShut {NoStop}%
\bibitem [{\citenamefont {Huang}\ \emph {et~al.}(2022)\citenamefont {Huang},
  \citenamefont {Lindner}, \citenamefont {Mart{\'\i}nez-Mirav{\'e}},\ and\
  \citenamefont {Sen}}]{Huang:2022wmz}%
  \BibitemOpen
  \bibfield  {author} {\bibinfo {author} {\bibfnamefont {G.-y.}\ \bibnamefont
  {Huang}}, \bibinfo {author} {\bibfnamefont {M.}~\bibnamefont {Lindner}},
  \bibinfo {author} {\bibfnamefont {P.}~\bibnamefont
  {Mart{\'\i}nez-Mirav{\'e}}},\ and\ \bibinfo {author} {\bibfnamefont
  {M.}~\bibnamefont {Sen}},\ }\href
  {https://doi.org/10.1103/PhysRevD.106.033004} {\bibfield  {journal} {\bibinfo
   {journal} {Phys. Rev. D}\ }\textbf {\bibinfo {volume} {106}},\ \bibinfo
  {pages} {033004} (\bibinfo {year} {2022})},\ \Eprint
  {https://arxiv.org/abs/2205.08431} {arXiv:2205.08431 [hep-ph]} \BibitemShut
  {NoStop}%
\bibitem [{\citenamefont {Bert{\'o}lez-Mart{\'\i}nez}\ \emph
  {et~al.}(2025{\natexlab{b}})\citenamefont {Bert{\'o}lez-Mart{\'\i}nez},
  \citenamefont {L{\'o}pez-Sarri{\'o}n},\ and\ \citenamefont
  {Salvado}}]{Bertolez-Martinez:2025pgm}%
  \BibitemOpen
  \bibfield  {author} {\bibinfo {author} {\bibfnamefont {T.}~\bibnamefont
  {Bert{\'o}lez-Mart{\'\i}nez}}, \bibinfo {author} {\bibfnamefont
  {J.}~\bibnamefont {L{\'o}pez-Sarri{\'o}n}},\ and\ \bibinfo {author}
  {\bibfnamefont {J.}~\bibnamefont {Salvado}},\ }\href@noop {} {\  (\bibinfo
  {year} {2025}{\natexlab{b}})},\ \Eprint {https://arxiv.org/abs/2509.22867}
  {arXiv:2509.22867 [hep-ph]} \BibitemShut {NoStop}%
\bibitem [{\citenamefont {Hung}(2000)}]{Hung:2000yg}%
  \BibitemOpen
  \bibfield  {author} {\bibinfo {author} {\bibfnamefont {P.~Q.}\ \bibnamefont
  {Hung}},\ }\href@noop {} {\  (\bibinfo {year} {2000})},\ \Eprint
  {https://arxiv.org/abs/hep-ph/0010126} {arXiv:hep-ph/0010126} \BibitemShut
  {NoStop}%
\bibitem [{\citenamefont {Gu}\ \emph {et~al.}(2003)\citenamefont {Gu},
  \citenamefont {Wang},\ and\ \citenamefont {Zhang}}]{Gu:2003er}%
  \BibitemOpen
  \bibfield  {author} {\bibinfo {author} {\bibfnamefont {P.}~\bibnamefont
  {Gu}}, \bibinfo {author} {\bibfnamefont {X.}~\bibnamefont {Wang}},\ and\
  \bibinfo {author} {\bibfnamefont {X.}~\bibnamefont {Zhang}},\ }\href
  {https://doi.org/10.1103/PhysRevD.68.087301} {\bibfield  {journal} {\bibinfo
  {journal} {Phys. Rev. D}\ }\textbf {\bibinfo {volume} {68}},\ \bibinfo
  {pages} {087301} (\bibinfo {year} {2003})},\ \Eprint
  {https://arxiv.org/abs/hep-ph/0307148} {arXiv:hep-ph/0307148} \BibitemShut
  {NoStop}%
\bibitem [{\citenamefont {Fardon}\ \emph {et~al.}(2004)\citenamefont {Fardon},
  \citenamefont {Nelson},\ and\ \citenamefont {Weiner}}]{Fardon:2003eh}%
  \BibitemOpen
  \bibfield  {author} {\bibinfo {author} {\bibfnamefont {R.}~\bibnamefont
  {Fardon}}, \bibinfo {author} {\bibfnamefont {A.~E.}\ \bibnamefont {Nelson}},\
  and\ \bibinfo {author} {\bibfnamefont {N.}~\bibnamefont {Weiner}},\ }\href
  {https://doi.org/10.1088/1475-7516/2004/10/005} {\bibfield  {journal}
  {\bibinfo  {journal} {JCAP}\ }\textbf {\bibinfo {volume} {10}},\ \bibinfo
  {pages} {005}},\ \Eprint {https://arxiv.org/abs/astro-ph/0309800}
  {arXiv:astro-ph/0309800} \BibitemShut {NoStop}%
\bibitem [{\citenamefont {Brookfield}\ \emph
  {et~al.}(2006{\natexlab{b}})\citenamefont {Brookfield}, \citenamefont {van~de
  Bruck}, \citenamefont {Mota},\ and\ \citenamefont
  {Tocchini-Valentini}}]{Brookfield:2005bz}%
  \BibitemOpen
  \bibfield  {author} {\bibinfo {author} {\bibfnamefont {A.~W.}\ \bibnamefont
  {Brookfield}}, \bibinfo {author} {\bibfnamefont {C.}~\bibnamefont {van~de
  Bruck}}, \bibinfo {author} {\bibfnamefont {D.~F.}\ \bibnamefont {Mota}},\
  and\ \bibinfo {author} {\bibfnamefont {D.}~\bibnamefont
  {Tocchini-Valentini}},\ }\href {https://doi.org/10.1103/PhysRevD.73.083515}
  {\bibfield  {journal} {\bibinfo  {journal} {Phys. Rev. D}\ }\textbf {\bibinfo
  {volume} {73}},\ \bibinfo {pages} {083515} (\bibinfo {year}
  {2006}{\natexlab{b}})},\ \bibinfo {note} {[Erratum: Phys.Rev.D 76, 049901
  (2007)]},\ \Eprint {https://arxiv.org/abs/astro-ph/0512367}
  {arXiv:astro-ph/0512367} \BibitemShut {NoStop}%
\bibitem [{\citenamefont {Ghedini}\ \emph {et~al.}(2026)\citenamefont
  {Ghedini}, \citenamefont {Hajjar},\ and\ \citenamefont {Mena}}]{Ghedini2026}%
  \BibitemOpen
  \bibfield  {author} {\bibinfo {author} {\bibfnamefont {P.}~\bibnamefont
  {Ghedini}}, \bibinfo {author} {\bibfnamefont {R.}~\bibnamefont {Hajjar}},\
  and\ \bibinfo {author} {\bibfnamefont {O.}~\bibnamefont {Mena}},\ }\href@noop
  {} {\  (\bibinfo {year} {2026})},\ \bibinfo {note} {to appear}\BibitemShut
  {NoStop}%
\bibitem [{\citenamefont {Lesgourgues}(2011)}]{Lesgourgues:2011re}%
  \BibitemOpen
  \bibfield  {author} {\bibinfo {author} {\bibfnamefont {J.}~\bibnamefont
  {Lesgourgues}},\ }\href@noop {} {\  (\bibinfo {year} {2011})},\ \Eprint
  {https://arxiv.org/abs/1104.2932} {arXiv:1104.2932 [astro-ph.IM]}
  \BibitemShut {NoStop}%
\bibitem [{\citenamefont {Blas}\ \emph {et~al.}(2011)\citenamefont {Blas},
  \citenamefont {Lesgourgues},\ and\ \citenamefont {Tram}}]{Blas:2011rf}%
  \BibitemOpen
  \bibfield  {author} {\bibinfo {author} {\bibfnamefont {D.}~\bibnamefont
  {Blas}}, \bibinfo {author} {\bibfnamefont {J.}~\bibnamefont {Lesgourgues}},\
  and\ \bibinfo {author} {\bibfnamefont {T.}~\bibnamefont {Tram}},\ }\href
  {https://doi.org/10.1088/1475-7516/2011/07/034} {\bibfield  {journal}
  {\bibinfo  {journal} {JCAP}\ }\textbf {\bibinfo {volume} {07}},\ \bibinfo
  {pages} {034}},\ \Eprint {https://arxiv.org/abs/1104.2933} {arXiv:1104.2933
  [astro-ph.CO]} \BibitemShut {NoStop}%
\bibitem [{\citenamefont {Torrado}\ and\ \citenamefont
  {Lewis}(2021)}]{Torrado:2020dgo}%
  \BibitemOpen
  \bibfield  {author} {\bibinfo {author} {\bibfnamefont {J.}~\bibnamefont
  {Torrado}}\ and\ \bibinfo {author} {\bibfnamefont {A.}~\bibnamefont
  {Lewis}},\ }\href {https://doi.org/10.1088/1475-7516/2021/05/057} {\bibfield
  {journal} {\bibinfo  {journal} {JCAP}\ }\textbf {\bibinfo {volume} {05}},\
  \bibinfo {pages} {057}},\ \Eprint {https://arxiv.org/abs/2005.05290}
  {arXiv:2005.05290 [astro-ph.IM]} \BibitemShut {NoStop}%
\bibitem [{\citenamefont {{Torrado}}\ and\ \citenamefont
  {{Lewis}}(2019)}]{2019ascl.soft10019T}%
  \BibitemOpen
  \bibfield  {author} {\bibinfo {author} {\bibfnamefont {J.}~\bibnamefont
  {{Torrado}}}\ and\ \bibinfo {author} {\bibfnamefont {A.}~\bibnamefont
  {{Lewis}}},\ }\href@noop {} {\bibinfo {title} {{Cobaya: Bayesian analysis in
  cosmology}}},\ \bibinfo {howpublished} {Astrophysics Source Code Library,
  record ascl:1910.019} (\bibinfo {year} {2019})\BibitemShut {NoStop}%
\bibitem [{\citenamefont {Gelman}\ and\ \citenamefont
  {Rubin}(1992)}]{Gelman:1992zz}%
  \BibitemOpen
  \bibfield  {author} {\bibinfo {author} {\bibfnamefont {A.}~\bibnamefont
  {Gelman}}\ and\ \bibinfo {author} {\bibfnamefont {D.~B.}\ \bibnamefont
  {Rubin}},\ }\href {https://doi.org/10.1214/ss/1177011136} {\bibfield
  {journal} {\bibinfo  {journal} {Statist. Sci.}\ }\textbf {\bibinfo {volume}
  {7}},\ \bibinfo {pages} {457} (\bibinfo {year} {1992})}\BibitemShut {NoStop}%
\bibitem [{\citenamefont {Lewis}(2025)}]{Lewis:2019xzd}%
  \BibitemOpen
  \bibfield  {author} {\bibinfo {author} {\bibfnamefont {A.}~\bibnamefont
  {Lewis}},\ }\href {https://doi.org/10.1088/1475-7516/2025/08/025} {\bibfield
  {journal} {\bibinfo  {journal} {JCAP}\ }\textbf {\bibinfo {volume} {08}},\
  \bibinfo {pages} {025}},\ \Eprint {https://arxiv.org/abs/1910.13970}
  {arXiv:1910.13970 [astro-ph.IM]} \BibitemShut {NoStop}%
\bibitem [{\citenamefont {Aghanim}\ \emph
  {et~al.}(2020{\natexlab{a}})\citenamefont {Aghanim} \emph
  {et~al.}}]{Planck:2018nkj}%
  \BibitemOpen
  \bibfield  {author} {\bibinfo {author} {\bibfnamefont {N.}~\bibnamefont
  {Aghanim}} \emph {et~al.} (\bibinfo {collaboration} {Planck}),\ }\href
  {https://doi.org/10.1051/0004-6361/201833880} {\bibfield  {journal} {\bibinfo
   {journal} {Astron. Astrophys.}\ }\textbf {\bibinfo {volume} {641}},\
  \bibinfo {pages} {A1} (\bibinfo {year} {2020}{\natexlab{a}})},\ \Eprint
  {https://arxiv.org/abs/1807.06205} {arXiv:1807.06205 [astro-ph.CO]}
  \BibitemShut {NoStop}%
\bibitem [{\citenamefont {Aghanim}\ \emph
  {et~al.}(2020{\natexlab{b}})\citenamefont {Aghanim} \emph
  {et~al.}}]{Planck:2018vyg}%
  \BibitemOpen
  \bibfield  {author} {\bibinfo {author} {\bibfnamefont {N.}~\bibnamefont
  {Aghanim}} \emph {et~al.} (\bibinfo {collaboration} {Planck}),\ }\href
  {https://doi.org/10.1051/0004-6361/201833910} {\bibfield  {journal} {\bibinfo
   {journal} {Astron. Astrophys.}\ }\textbf {\bibinfo {volume} {641}},\
  \bibinfo {pages} {A6} (\bibinfo {year} {2020}{\natexlab{b}})},\ \bibinfo
  {note} {[Erratum: Astron.Astrophys. 652, C4 (2021)]},\ \Eprint
  {https://arxiv.org/abs/1807.06209} {arXiv:1807.06209 [astro-ph.CO]}
  \BibitemShut {NoStop}%
\bibitem [{\citenamefont {Aghanim}\ \emph
  {et~al.}(2020{\natexlab{c}})\citenamefont {Aghanim} \emph
  {et~al.}}]{Planck:2019nip}%
  \BibitemOpen
  \bibfield  {author} {\bibinfo {author} {\bibfnamefont {N.}~\bibnamefont
  {Aghanim}} \emph {et~al.} (\bibinfo {collaboration} {Planck}),\ }\href
  {https://doi.org/10.1051/0004-6361/201936386} {\bibfield  {journal} {\bibinfo
   {journal} {Astron. Astrophys.}\ }\textbf {\bibinfo {volume} {641}},\
  \bibinfo {pages} {A5} (\bibinfo {year} {2020}{\natexlab{c}})},\ \Eprint
  {https://arxiv.org/abs/1907.12875} {arXiv:1907.12875 [astro-ph.CO]}
  \BibitemShut {NoStop}%
\bibitem [{\citenamefont {Aghanim}\ \emph
  {et~al.}(2020{\natexlab{d}})\citenamefont {Aghanim} \emph
  {et~al.}}]{Planck:2018lbu}%
  \BibitemOpen
  \bibfield  {author} {\bibinfo {author} {\bibfnamefont {N.}~\bibnamefont
  {Aghanim}} \emph {et~al.} (\bibinfo {collaboration} {Planck}),\ }\href
  {https://doi.org/10.1051/0004-6361/201833886} {\bibfield  {journal} {\bibinfo
   {journal} {Astron. Astrophys.}\ }\textbf {\bibinfo {volume} {641}},\
  \bibinfo {pages} {A8} (\bibinfo {year} {2020}{\natexlab{d}})},\ \Eprint
  {https://arxiv.org/abs/1807.06210} {arXiv:1807.06210 [astro-ph.CO]}
  \BibitemShut {NoStop}%
\bibitem [{\citenamefont {Louis}\ \emph {et~al.}(2025)\citenamefont {Louis}
  \emph {et~al.}}]{ACT:2025fju}%
  \BibitemOpen
  \bibfield  {author} {\bibinfo {author} {\bibfnamefont {T.}~\bibnamefont
  {Louis}} \emph {et~al.} (\bibinfo {collaboration} {ACT}),\ }\href@noop {} {\
  (\bibinfo {year} {2025})},\ \Eprint {https://arxiv.org/abs/2503.14452}
  {arXiv:2503.14452 [astro-ph.CO]} \BibitemShut {NoStop}%
\bibitem [{\citenamefont {Qu}\ \emph {et~al.}(2024)\citenamefont {Qu} \emph
  {et~al.}}]{ACT:2023dou}%
  \BibitemOpen
  \bibfield  {author} {\bibinfo {author} {\bibfnamefont {F.~J.}\ \bibnamefont
  {Qu}} \emph {et~al.} (\bibinfo {collaboration} {ACT}),\ }\href
  {https://doi.org/10.3847/1538-4357/acfe06} {\bibfield  {journal} {\bibinfo
  {journal} {Astrophys. J.}\ }\textbf {\bibinfo {volume} {962}},\ \bibinfo
  {pages} {112} (\bibinfo {year} {2024})},\ \Eprint
  {https://arxiv.org/abs/2304.05202} {arXiv:2304.05202 [astro-ph.CO]}
  \BibitemShut {NoStop}%
\bibitem [{\citenamefont {Brout}\ \emph {et~al.}(2022)\citenamefont {Brout}
  \emph {et~al.}}]{Brout:2022vxf}%
  \BibitemOpen
  \bibfield  {author} {\bibinfo {author} {\bibfnamefont {D.}~\bibnamefont
  {Brout}} \emph {et~al.},\ }\href {https://doi.org/10.3847/1538-4357/ac8e04}
  {\bibfield  {journal} {\bibinfo  {journal} {Astrophys. J.}\ }\textbf
  {\bibinfo {volume} {938}},\ \bibinfo {pages} {110} (\bibinfo {year}
  {2022})},\ \Eprint {https://arxiv.org/abs/2202.04077} {arXiv:2202.04077
  [astro-ph.CO]} \BibitemShut {NoStop}%
\bibitem [{\citenamefont {Silva}\ and\ \citenamefont
  {Nunes}(2025)}]{Silva:2025twg}%
  \BibitemOpen
  \bibfield  {author} {\bibinfo {author} {\bibfnamefont {E.}~\bibnamefont
  {Silva}}\ and\ \bibinfo {author} {\bibfnamefont {R.~C.}\ \bibnamefont
  {Nunes}},\ }\href {https://doi.org/10.1088/1475-7516/2025/11/078} {\bibfield
  {journal} {\bibinfo  {journal} {JCAP}\ }\textbf {\bibinfo {volume} {11}},\
  \bibinfo {pages} {078}},\ \Eprint {https://arxiv.org/abs/2507.13989}
  {arXiv:2507.13989 [astro-ph.CO]} \BibitemShut {NoStop}%
\bibitem [{\citenamefont {Li}\ \emph {et~al.}(2025{\natexlab{b}})\citenamefont
  {Li}, \citenamefont {Du}, \citenamefont {Zhou}, \citenamefont {Li},
  \citenamefont {Zhang},\ and\ \citenamefont {Zhang}}]{Li:2025vuh}%
  \BibitemOpen
  \bibfield  {author} {\bibinfo {author} {\bibfnamefont {T.-N.}\ \bibnamefont
  {Li}}, \bibinfo {author} {\bibfnamefont {G.-H.}\ \bibnamefont {Du}}, \bibinfo
  {author} {\bibfnamefont {S.-H.}\ \bibnamefont {Zhou}}, \bibinfo {author}
  {\bibfnamefont {Y.-H.}\ \bibnamefont {Li}}, \bibinfo {author} {\bibfnamefont
  {J.-F.}\ \bibnamefont {Zhang}},\ and\ \bibinfo {author} {\bibfnamefont
  {X.}~\bibnamefont {Zhang}},\ }\href@noop {} {\  (\bibinfo {year}
  {2025}{\natexlab{b}})},\ \Eprint {https://arxiv.org/abs/2511.22512}
  {arXiv:2511.22512 [astro-ph.CO]} \BibitemShut {NoStop}%
\bibitem [{\citenamefont {Capozziello}\ \emph {et~al.}(2025)\citenamefont
  {Capozziello}, \citenamefont {Chaudhary}, \citenamefont {Harko},\ and\
  \citenamefont {Mustafa}}]{Capozziello:2025qmh}%
  \BibitemOpen
  \bibfield  {author} {\bibinfo {author} {\bibfnamefont {S.}~\bibnamefont
  {Capozziello}}, \bibinfo {author} {\bibfnamefont {H.}~\bibnamefont
  {Chaudhary}}, \bibinfo {author} {\bibfnamefont {T.}~\bibnamefont {Harko}},\
  and\ \bibinfo {author} {\bibfnamefont {G.}~\bibnamefont {Mustafa}},\
  }\href@noop {} {\  (\bibinfo {year} {2025})},\ \Eprint
  {https://arxiv.org/abs/2512.10585} {arXiv:2512.10585 [astro-ph.CO]}
  \BibitemShut {NoStop}%
\bibitem [{\citenamefont {Vagnozzi}\ \emph {et~al.}(2018)\citenamefont
  {Vagnozzi}, \citenamefont {Dhawan}, \citenamefont {Gerbino}, \citenamefont
  {Freese}, \citenamefont {Goobar},\ and\ \citenamefont
  {Mena}}]{Vagnozzi:2018jhn}%
  \BibitemOpen
  \bibfield  {author} {\bibinfo {author} {\bibfnamefont {S.}~\bibnamefont
  {Vagnozzi}}, \bibinfo {author} {\bibfnamefont {S.}~\bibnamefont {Dhawan}},
  \bibinfo {author} {\bibfnamefont {M.}~\bibnamefont {Gerbino}}, \bibinfo
  {author} {\bibfnamefont {K.}~\bibnamefont {Freese}}, \bibinfo {author}
  {\bibfnamefont {A.}~\bibnamefont {Goobar}},\ and\ \bibinfo {author}
  {\bibfnamefont {O.}~\bibnamefont {Mena}},\ }\href
  {https://doi.org/10.1103/PhysRevD.98.083501} {\bibfield  {journal} {\bibinfo
  {journal} {Phys. Rev. D}\ }\textbf {\bibinfo {volume} {98}},\ \bibinfo
  {pages} {083501} (\bibinfo {year} {2018})},\ \Eprint
  {https://arxiv.org/abs/1801.08553} {arXiv:1801.08553 [astro-ph.CO]}
  \BibitemShut {NoStop}%
\bibitem [{\citenamefont {Sharma}\ \emph {et~al.}(2022)\citenamefont {Sharma},
  \citenamefont {Pandey},\ and\ \citenamefont {Das}}]{Sharma:2022ifr}%
  \BibitemOpen
  \bibfield  {author} {\bibinfo {author} {\bibfnamefont {R.~K.}\ \bibnamefont
  {Sharma}}, \bibinfo {author} {\bibfnamefont {K.~L.}\ \bibnamefont {Pandey}},\
  and\ \bibinfo {author} {\bibfnamefont {S.}~\bibnamefont {Das}},\ }\href
  {https://doi.org/10.3847/1538-4357/ac7a33} {\bibfield  {journal} {\bibinfo
  {journal} {Astrophys. J.}\ }\textbf {\bibinfo {volume} {934}},\ \bibinfo
  {pages} {113} (\bibinfo {year} {2022})},\ \Eprint
  {https://arxiv.org/abs/2202.01749} {arXiv:2202.01749 [astro-ph.CO]}
  \BibitemShut {NoStop}%
\bibitem [{\citenamefont {Du}\ \emph {et~al.}(2025)\citenamefont {Du},
  \citenamefont {Wu}, \citenamefont {Li},\ and\ \citenamefont
  {Zhang}}]{Du:2024pai}%
  \BibitemOpen
  \bibfield  {author} {\bibinfo {author} {\bibfnamefont {G.-H.}\ \bibnamefont
  {Du}}, \bibinfo {author} {\bibfnamefont {P.-J.}\ \bibnamefont {Wu}}, \bibinfo
  {author} {\bibfnamefont {T.-N.}\ \bibnamefont {Li}},\ and\ \bibinfo {author}
  {\bibfnamefont {X.}~\bibnamefont {Zhang}},\ }\href
  {https://doi.org/10.1140/epjc/s10052-025-14094-0} {\bibfield  {journal}
  {\bibinfo  {journal} {Eur. Phys. J. C}\ }\textbf {\bibinfo {volume} {85}},\
  \bibinfo {pages} {392} (\bibinfo {year} {2025})},\ \Eprint
  {https://arxiv.org/abs/2407.15640} {arXiv:2407.15640 [astro-ph.CO]}
  \BibitemShut {NoStop}%
\bibitem [{\citenamefont {Roy~Choudhury}\ and\ \citenamefont
  {Naskar}(2019)}]{RoyChoudhury:2018vnm}%
  \BibitemOpen
  \bibfield  {author} {\bibinfo {author} {\bibfnamefont {S.}~\bibnamefont
  {Roy~Choudhury}}\ and\ \bibinfo {author} {\bibfnamefont {A.}~\bibnamefont
  {Naskar}},\ }\href {https://doi.org/10.1140/epjc/s10052-019-6762-z}
  {\bibfield  {journal} {\bibinfo  {journal} {Eur. Phys. J. C}\ }\textbf
  {\bibinfo {volume} {79}},\ \bibinfo {pages} {262} (\bibinfo {year} {2019})},\
  \Eprint {https://arxiv.org/abs/1807.02860} {arXiv:1807.02860 [astro-ph.CO]}
  \BibitemShut {NoStop}%
\bibitem [{\citenamefont {Roy~Choudhury}\ and\ \citenamefont
  {Hannestad}(2020)}]{RoyChoudhury:2019hls}%
  \BibitemOpen
  \bibfield  {author} {\bibinfo {author} {\bibfnamefont {S.}~\bibnamefont
  {Roy~Choudhury}}\ and\ \bibinfo {author} {\bibfnamefont {S.}~\bibnamefont
  {Hannestad}},\ }\href {https://doi.org/10.1088/1475-7516/2020/07/037}
  {\bibfield  {journal} {\bibinfo  {journal} {JCAP}\ }\textbf {\bibinfo
  {volume} {07}},\ \bibinfo {pages} {037}},\ \Eprint
  {https://arxiv.org/abs/1907.12598} {arXiv:1907.12598 [astro-ph.CO]}
  \BibitemShut {NoStop}%
\end{thebibliography}%

\clearpage
\onecolumngrid
\appendix
\renewcommand\thefigure{\thesection\arabic{figure}}

\setcounter{figure}{0}

\section{PCHIP formalism}
\label{sec:AppA_pchipformalism}

In this Appendix, we introduce the PCHIP functional form and show why it is considered a method that achieves an effective data-driven functional result. 

Given a set of $N$ nodes $x_j$ with corresponding function values $f(x_j)=y_j$ ($j=1, ...,N$), a piecewise cubic interpolation is performed with $N-1$ cubic polynomials between the nodes, requiring the determination of $4(N-1)$ coefficients. Following the methodology of~\cite{Gariazzo:2014dla}, we write the cubic interpolating polynomial between the nodes $x_j$ and $x_{j+1}$ in the Hermite form 

\begin{equation}
    f(x;y_1,...,y_N) = \frac{(h_j+2t)(h_j-t)^2}{h_j^3}y_j+\frac{(3h_j-2t)t^2}{h_j^3}y_{j+1}+\frac{t(h_j-t)^2}{h_j^2}d_j+\frac{t^2(h_j-t)}{h_j^2}d_{j+1} ~,
\end{equation}

where $t = x - x_j$, $h_j = x_{j+1} - x_j$ and $d_{j}$ and $d_{j+1}$ are the derivatives in the nodes $j$ and $j+1$ respectively, chosen in order to preserve the local monotonicity of the interpolated points. The $d_j$ factor depends on the slopes between the nodes $x_j$ and $x_{j+1}$.

The slopes at each $x_j$ are chosen in such a way that the interpolant preserves the shape of the data and respects monotonicity, achieved by imposing that the first derivative vanishes if there is a change of monotonicity in the function between two nodes.

\begin{figure}[h]
\centering
\includegraphics[width=0.5\linewidth]{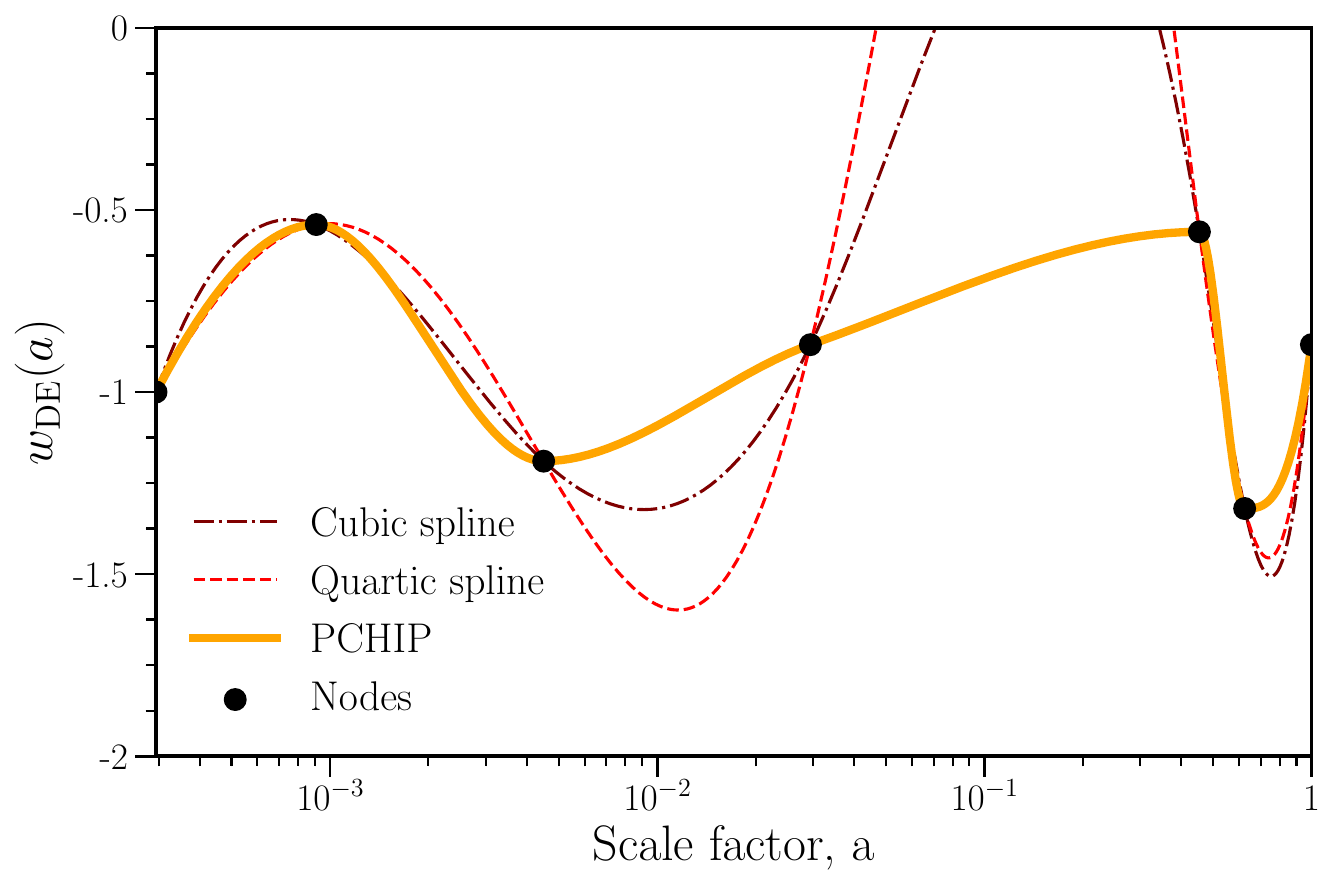}
\caption{Plot comparing different interpolating techniques to highlight the efficiency of the PCHIP method in removing oscillations. In particular, if we would consider one of the other methods considered here, we could get also values of $w_\mathrm{DE}>0$, for some combinations of values of the nodes, due to the high oscillations introduced by those interpolating methods.}
\label{fig:pchip_comparison}
\end{figure}

\Cref{fig:pchip_comparison} shows the functional form of different interpolation methods for given function values at some nodes. The PCHIP interpolator does not introduce spurious oscillatory behavior.

\newpage

\section{Neutrinos and tensions}
\label{sec:nutensions}

In this Appendix we present some additional plots in the $H_0-\sigma_8$ plane, for the two cases in which we vary $\sum m_\nu$ and we reconstruct $\sum m_\nu$, highlighting the correlations of the two parameters with $\sum m_\nu$.

\begin{figure}[h]
\centering
\includegraphics[width=0.7\linewidth]{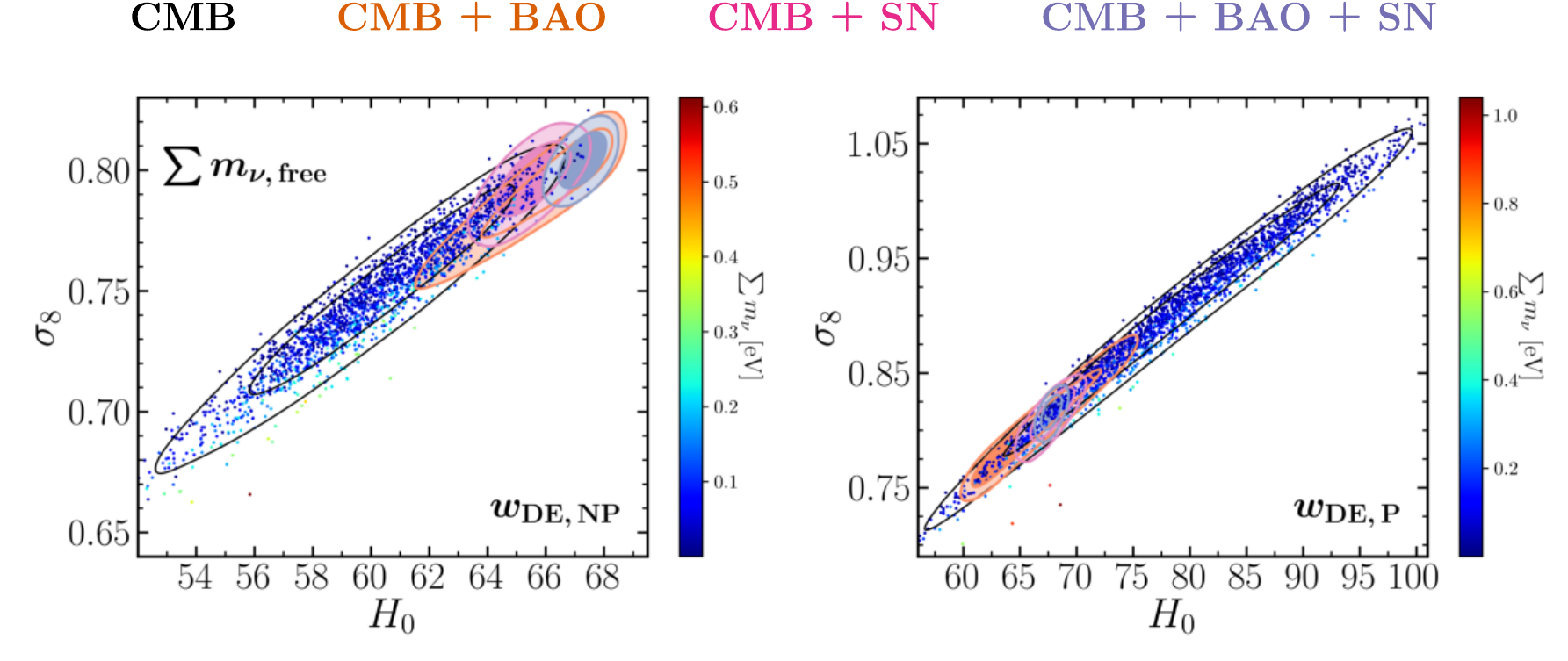}
\caption{Contour plot showing the 68 and 95 \% C.L. in the $H_0$ and $\sigma_8$ plane, for the case in which $\sum m_\nu$ is a free parameter in the MCMC. We highlight the bounds on $\sum m_\nu$ only for the analysis with the CMB measurements.}
\label{fig:H0sigma8nu2}
\end{figure}

\begin{figure}[h]
\centering
\includegraphics[width=\linewidth]{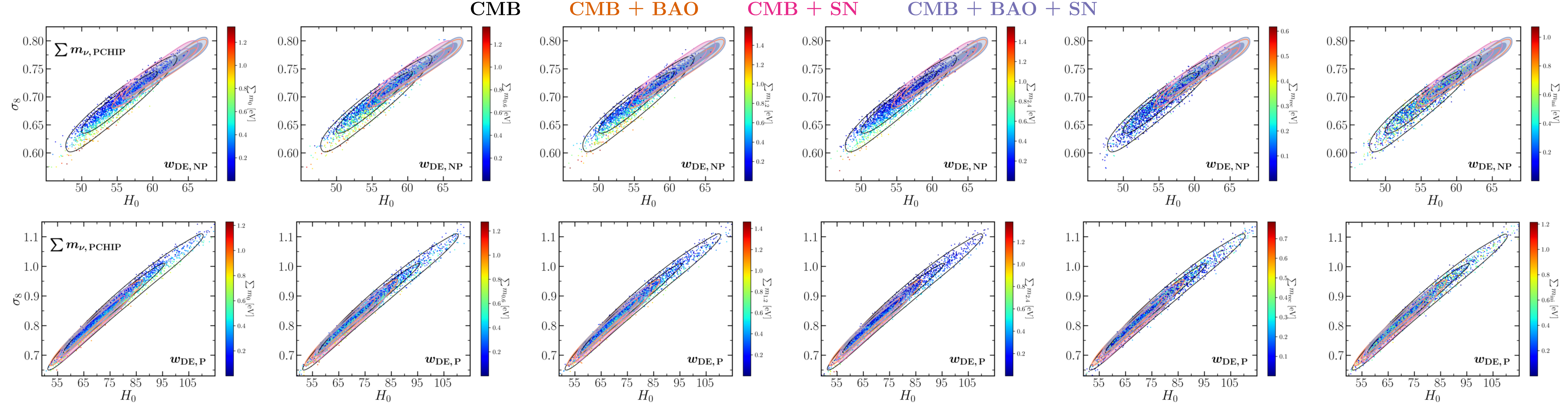}
\caption{Contour plot showing the 68 and 95 \% C.L. in the $H_0$ and $\sigma_8$ plane, for the case in which $\sum m_\nu$ is reconstructed via the PCHIP method. We highlight the bounds on each $\sum m_\nu$ node only for the analysis with the CMB measurements.}
\label{fig:H0sigma8nu3}
\end{figure}

\Cref{fig:H0sigma8nu2} and \Cref{fig:H0sigma8nu3} show the values of the $\sum m_\nu$ parameter or node for the case in which we only take into account CMB data.
In both scenarios, we see that, for a given value of $H_0$, $\sum m_\nu$ increases as $\sigma_8$ decreases. This reflects the well-known suppression of the small-scale structure caused by massive neutrinos.

It is still true that when we allow for the phantom case the bounds get relaxed, even if this is not true for all nodes in the reconstructed $\sum m_\nu$ case (see Fig. \ref{fig:H0sigma8nu3}). This could be related to the fact that in general, as $H_0$ increases, we loose the possibility to have higher masses for neutrinos because of the positive correlation between $H_0$ and $\sigma_8$ and the anti-correlation between $\sigma_8$ and $\sum m_\nu$.

\newpage

\section{Full chains}
\label{sec:FullResult}

In this Appendix we present the triangular plots showing the 68 \% and 95 \% C.L. of the parameters used in the reconstruction of $w_\mathrm{DE}$ and on the neutrino parameters: depending on the case considered, we show $\sum m_\nu$ when we leave it free or $\sum m_{\nu, i}$ when we reconstruct it.

Additionally, we present the tables with the 95 \% C.L. bounds on the same parameters, plus $H_0$, $\sigma_8$ and $\Omega_\mathrm{m}$.

\begin{figure}[b]
\centering
\includegraphics[width=\linewidth]{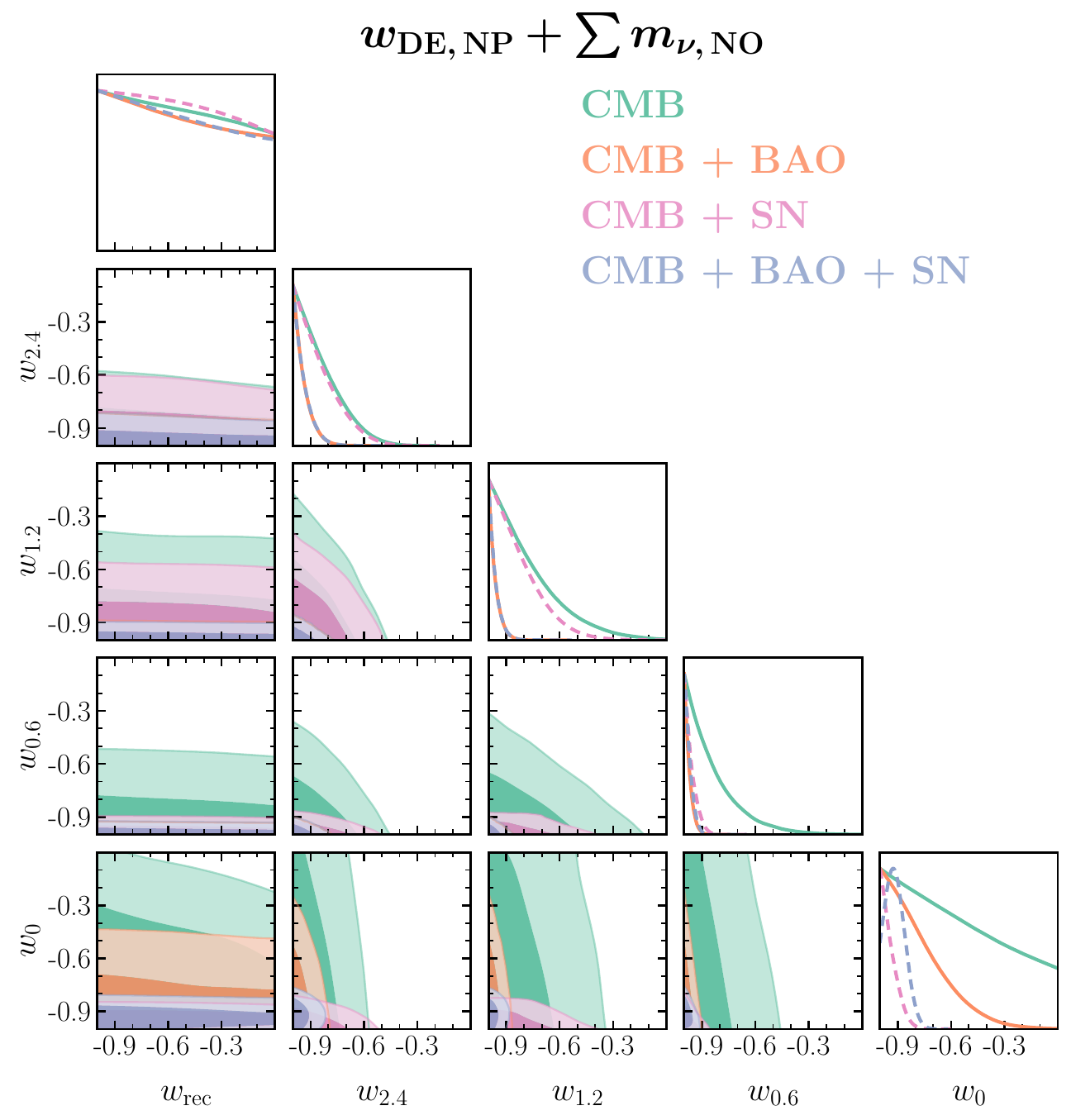}
\caption{Triangular plots for the nodes used in the reconstruction of $w_\mathrm{DE}$ in the case of non-phantom dark energy and $\sum m_\nu$ fixed.}
\label{fig:triangularcase1NP}
\end{figure}

\begin{figure}[h!]
\centering
\includegraphics[width=\linewidth]{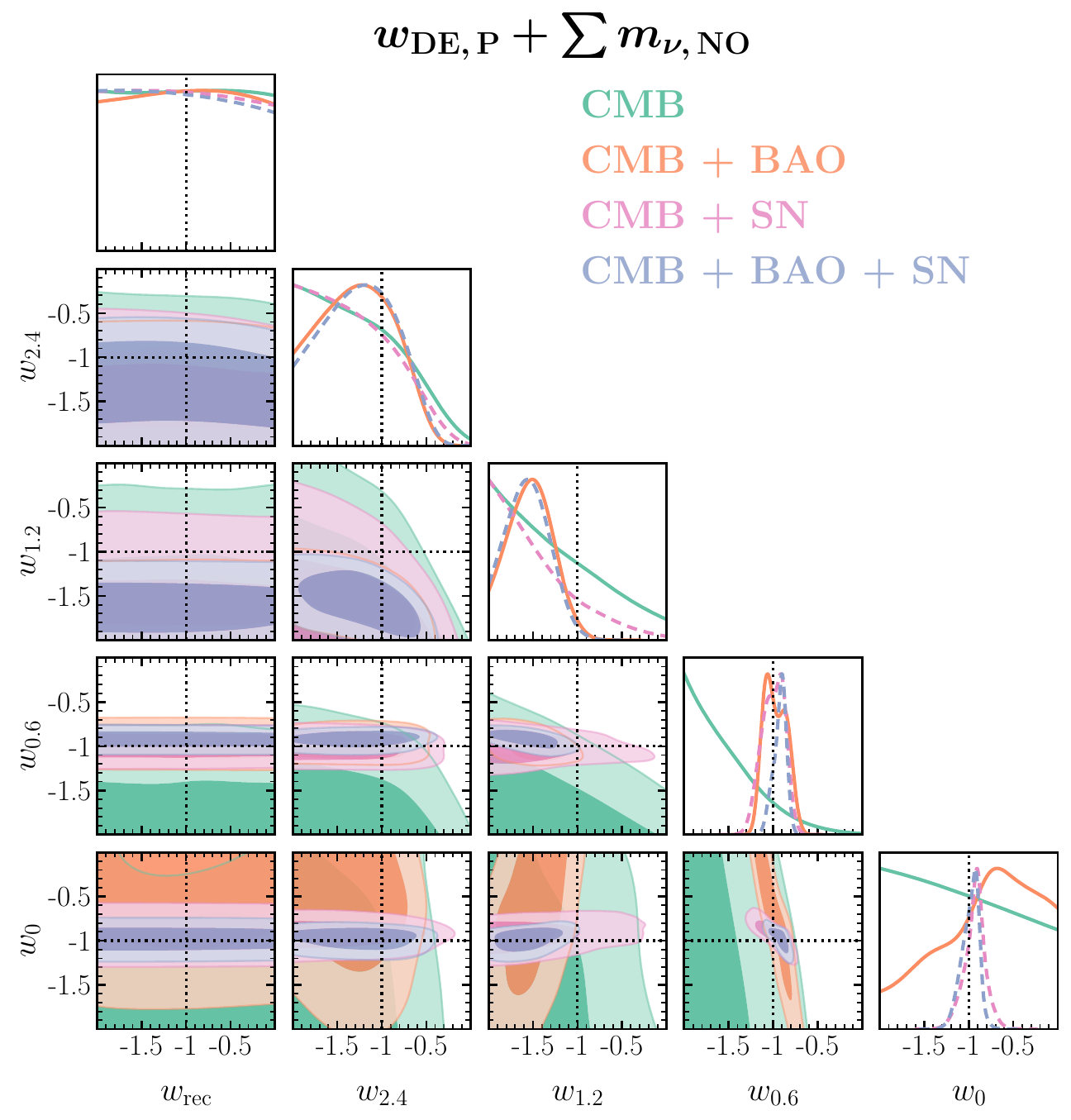}
\caption{Triangular plots for the nodes used in the reconstruction of $w_\mathrm{DE}$ in the case of phantom dark energy and $\sum m_\nu$ fixed.}
\label{fig:triangularcase1P}
\end{figure}

\begin{figure}[h!]
\centering
\includegraphics[width=\linewidth]{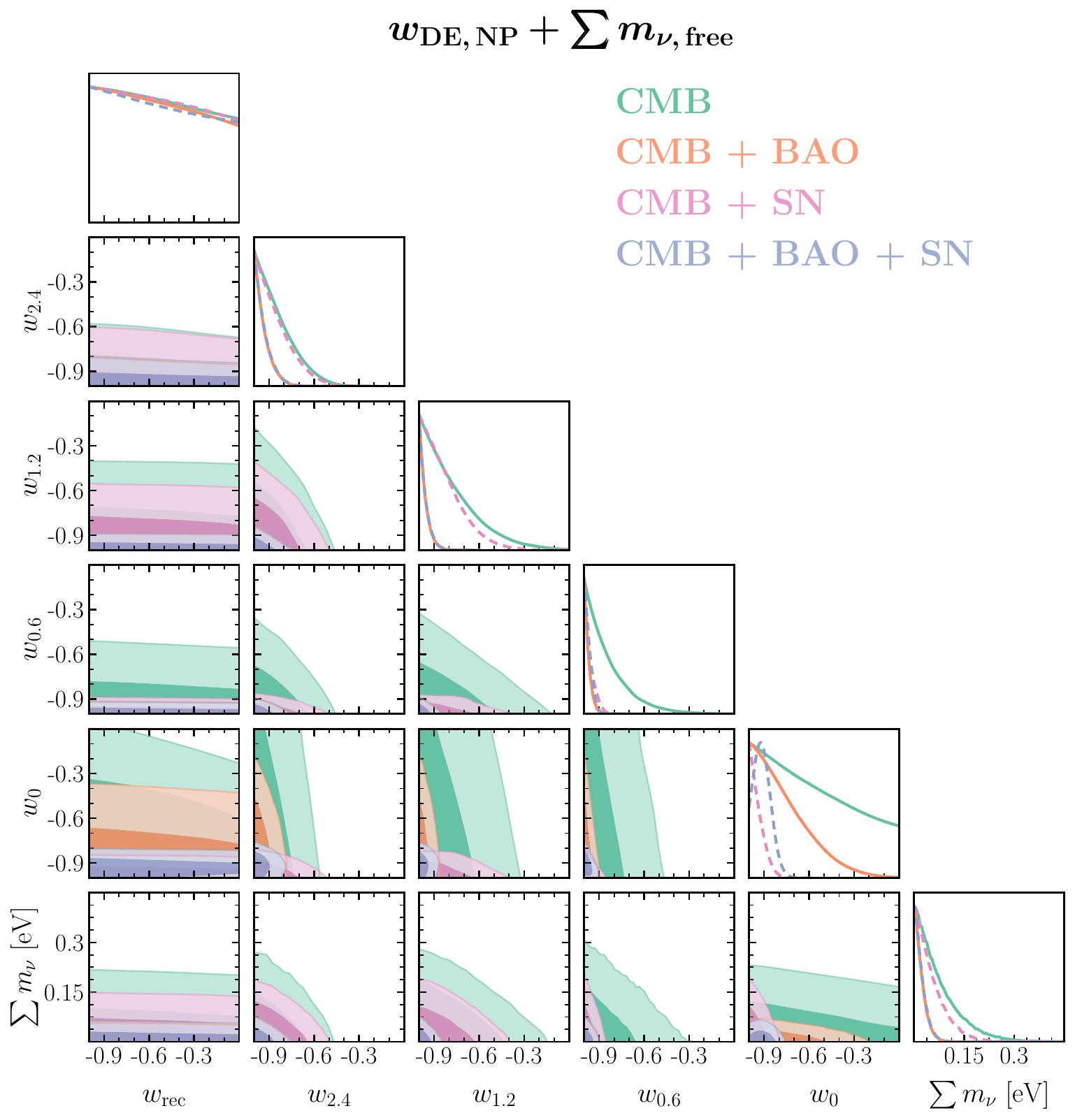}
\caption{Triangular plots for the nodes used in the reconstruction of $w_\mathrm{DE}$ in the case of non-phantom dark energy, plus the bounds on $\sum m_\nu$.}
\label{fig:triangularcase2NP}
\end{figure}

\begin{figure}[h!]
\centering
\includegraphics[width=\linewidth]{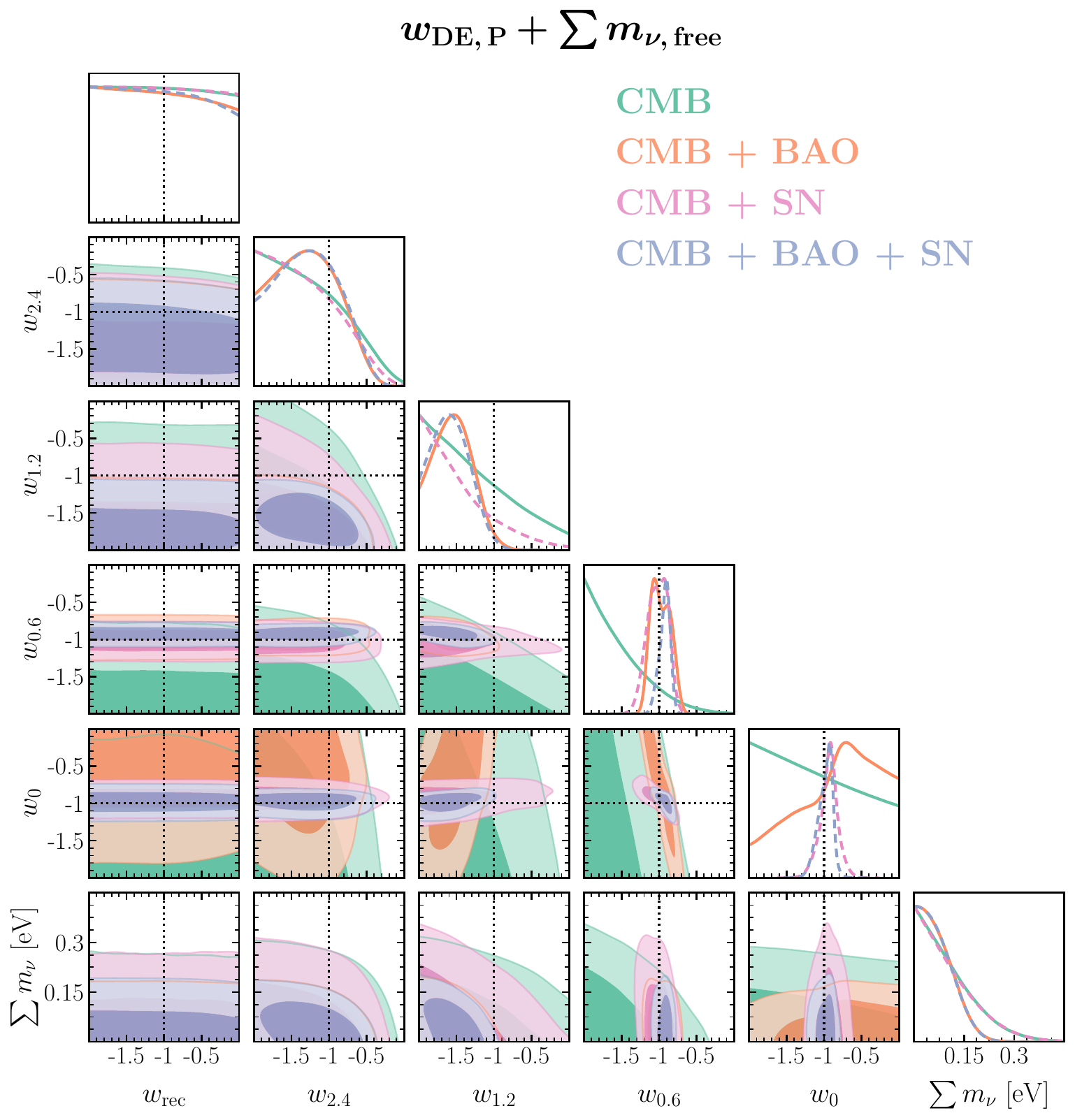}
\caption{Triangular plots for the nodes used in the reconstruction of $w_\mathrm{DE}$ in the case of phantom dark energy, plus the bounds on $\sum m_\nu$.}
\label{fig:triangularcase2P}
\end{figure}

\begin{figure}[h!]
\centering
\includegraphics[width=\linewidth]{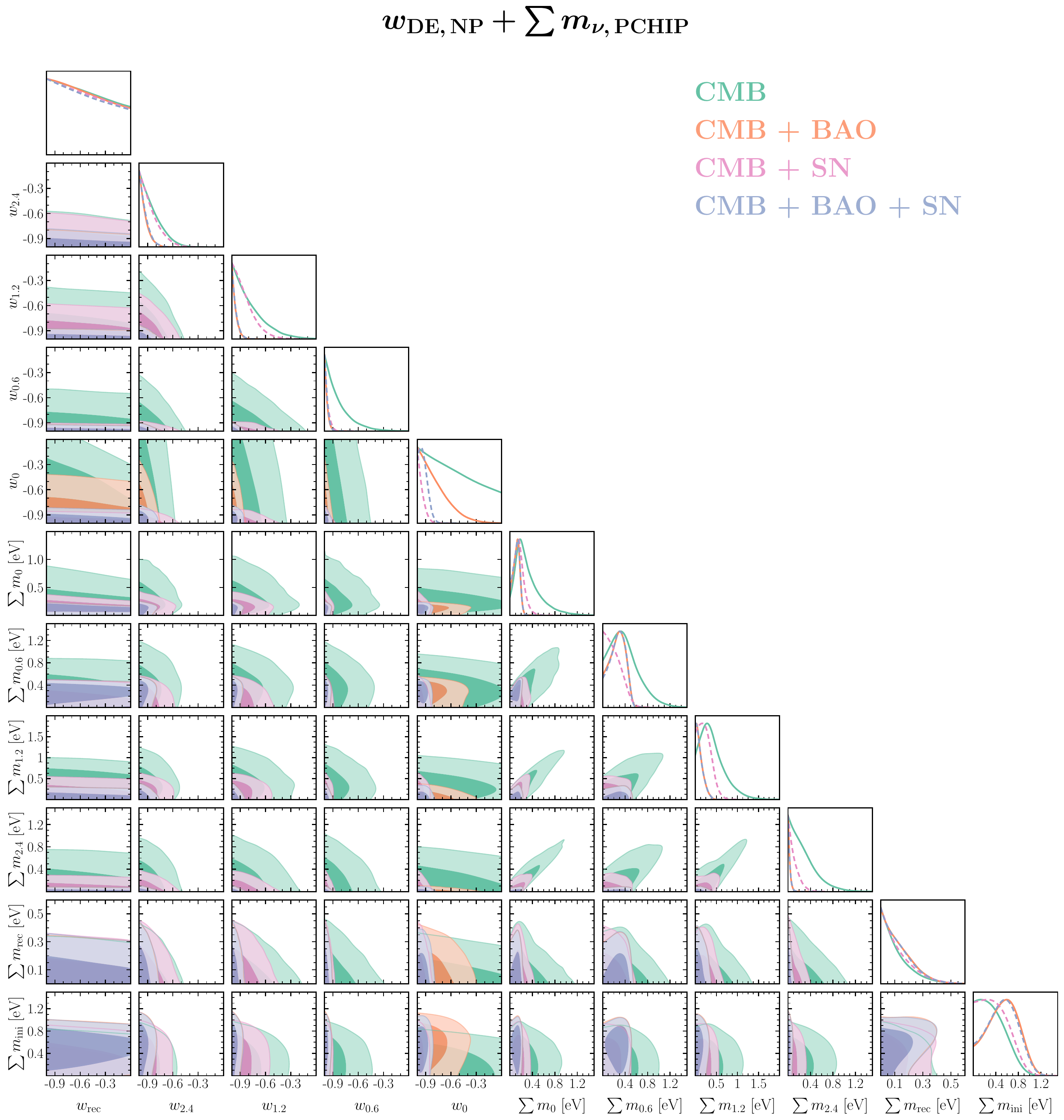}
\caption{Triangular plots for the nodes used in the reconstruction of both $w_\mathrm{DE}$ and $\sum m_\nu$, in the case of non-phantom dark energy.}
\label{fig:triangularcase3NP}
\end{figure}

\begin{figure}[h!]
\centering
\includegraphics[width=\linewidth]{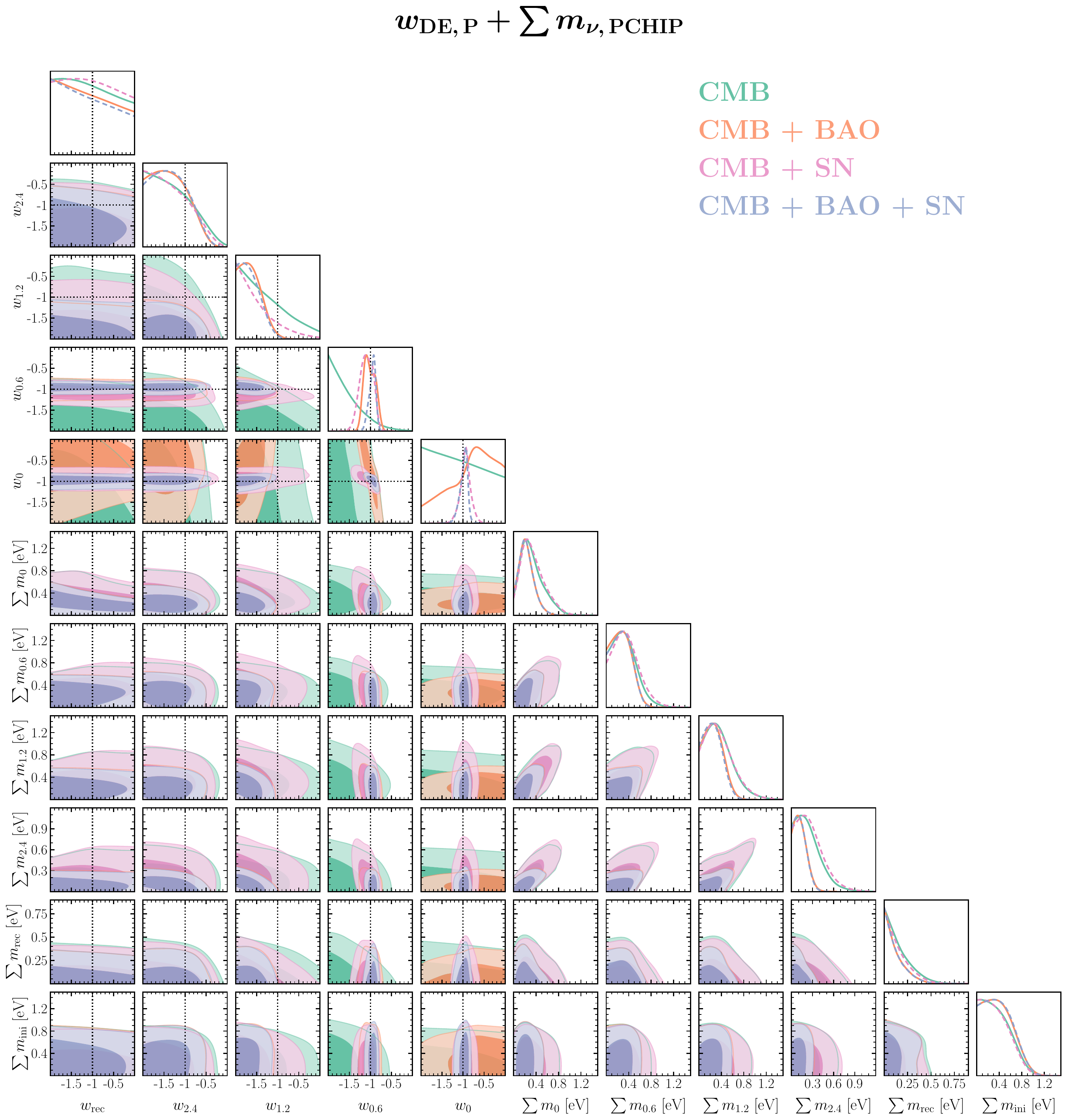}
\caption{Triangular plots for the nodes used in the reconstruction of both $w_\mathrm{DE}$ and $\sum m_\nu$, in the case of phantom dark energy.}
\label{fig:triangularcase3P}
\end{figure}

\begin{table}[t]
\centering
\small\addtolength{\tabcolsep}{+3pt}
\def\arraystretch{1.25}
\begin{tabular}{ccccc}
\multicolumn{5}{c}{$\boldsymbol{w_\mathrm{DE,\,NP} + \sum m_{\nu,\,\mathrm{NO}}}$} \\ \hline
 & \textbf{CMB} & \textbf{CMB+BAO} & \textbf{CMB+SN} & \textbf{CMB+BAO+SN} \\
\hline
{\boldmath$w_{rec}$} & --- & --- & --- & --- \\
{\boldmath$w_{2.4}$} & $< -0.618$ & $< -0.847$ & $< -0.637$ & $< -0.846$ \\
{\boldmath$w_{1.2}$} & $< -0.412$ & $< -0.902$ & $< -0.576$ & $< -0.901$ \\
{\boldmath$w_{0.6}$} & $< -0.538$ & $< -0.932$ & $< -0.900$ & $< -0.929$ \\
{\boldmath$w_{0}$} & --- & $< -0.458$ & $< -0.858$ & $< -0.817$ \\
{\boldmath$H_0$} & $60.3^{+5.4}_{-5.5}$ & $66.1^{+2.3}_{-3.0}$ & $65.5^{+1.5}_{-1.7}$ & $67.16^{+0.99}_{-1.1}$ \\
{\boldmath$\sigma_8$} &  $0.750^{+0.050}_{-0.052}$ & $0.789^{+0.025}_{-0.029}$ & $0.795^{+0.016}_{-0.017}$ & $0.798\pm0.014$ \\
{\boldmath$\Omega_m$} & $0.398^{+0.080}_{-0.071}$ & $0.322^{+0.030}_{-0.023}$ & $0.334^{+0.020}_{-0.018}$ & $0.311^{+0.011}_{-0.010}$ \\
\hline
\multicolumn{5}{c}{$\boldsymbol{w_\mathrm{DE,\,P} + \sum m_{\nu,\,\mathrm{NO}}}$} \\ \hline
{\boldmath$w_{rec}$} & --- & --- & --- & --- \\
{\boldmath$w_{2.4}$} & $< -0.417$ & $< -0.619$ & $< -0.513$ & $< -0.598$ \\
{\boldmath$w_{1.2}$} & $< -0.295$ & $< -1.11$ & $< -0.572$ & $< -1.15$ \\
{\boldmath$w_{0.6}$} & $< -0.765$ & $-0.97^{+0.23}_{-0.22}$ & $-1.00^{+0.22}_{-0.25}$ & $-0.92^{+0.13}_{-0.16}$ \\
{\boldmath$w_{0}$} & --- & --- & $-0.93^{+0.24}_{-0.27}$ & $-0.98^{+0.17}_{-0.22}$ \\
{\boldmath$H_0$} & $79\pm20$ & $66.3^{+7.1}_{-5.7}$ & $68.0^{+2.2}_{-2.6}$ & $67.8\pm1.2$ \\
{\boldmath$\sigma_8$} & $0.90\pm0.14$ & $0.804^{+0.062}_{-0.052}$ & $0.819^{+0.022}_{-0.025}$ & $0.817\pm0.017$ \\
{\boldmath$\Omega_m$} & $0.240^{+0.12}_{-0.098}$ & $0.328^{+0.058}_{-0.066}$ & $0.309^{+0.025}_{-0.023}$ & $0.311\pm0.011$ \\
\hline
\end{tabular}
\caption{95 \% C.L. bounds on the nodes used for the reconstruction of $w_\mathrm{DE}$ using the PCHIP method, in the case in which $\sum m_\nu = 0.06$ eV. We also show the bounds on the parameters that exhibit tensions in the base $\Lambda$CDM model. We fix in all cases $w_\mathrm{ini} = w_\mathrm{mre}=-1$.}
\label{tab:case1}
\end{table}

\begin{table}[t]
\centering
\small\addtolength{\tabcolsep}{+3pt}
\def\arraystretch{1.25}
\begin{tabular}{ccccc}
\multicolumn{5}{c}{$\boldsymbol{w_\mathrm{DE,\,NP} + \sum m_{\nu,\,\mathrm{free}}}$} \\ \hline
 & \textbf{CMB} & \textbf{CMB+BAO} & \textbf{CMB+SN} & \textbf{CMB+BAO+SN} \\
\hline
{\boldmath$w_{rec}$} & --- & --- & --- & --- \\
{\boldmath$w_{2.4}$} & $< -0.615$ & $< -0.831$ & $< -0.640$ & $< -0.827$ \\
{\boldmath$w_{1.2}$} & $< -0.414$ & $< -0.894$ & $< -0.571$ & $< -0.890$ \\
{\boldmath$w_{0.6}$} & $< -0.545$ & $< -0.925$ & $< -0.897$ & $< -0.923$ \\
{\boldmath$w_{0}$} & --- & $< -0.413$ & $< -0.855$ & $< -0.814$ \\
{\boldmath$\Sigma m_\nu$} \textbf{[eV]} & $< 0.202$ & $< 0.0532$ & $< 0.138$ & $< 0.0545$ \\
{\boldmath$H_0$} & $60.2^{+5.5}_{-5.6}$ & $66.0^{+2.5}_{-3.2}$ & $65.5^{+1.6}_{-1.8}$ & $67.2^{+1.0}_{-1.1}$ \\
{\boldmath$\sigma_8$} & $0.748^{+0.053}_{-0.055}$ & $0.793^{+0.027}_{-0.031}$ & $0.796^{+0.020}_{-0.021}$ & $0.804^{+0.015}_{-0.016}$ \\
{\boldmath$\Omega_m$} & $0.400^{+0.082}_{-0.073}$ & $0.322^{+0.033}_{-0.025}$ & $0.333^{+0.021}_{-0.020}$ & $0.310^{+0.011}_{-0.010}$ \\
\hline
\multicolumn{5}{c}{$\boldsymbol{w_\mathrm{DE,\,P} + \sum m_{\nu,\,\mathrm{free}}}$} \\ \hline
{\boldmath$w_{rec}$} & --- & --- & --- & --- \\
{\boldmath$w_{2.4}$} & $< -0.441$ & $< -0.623$ & $< -0.531$ & $< -0.609$ \\
{\boldmath$w_{1.2}$} & $< -0.310$ & $< -1.11$ & $< -0.586$ & $< -1.15$ \\
{\boldmath$w_{0.6}$} & $< -0.791$ & $-0.98^{+0.24}_{-0.22}$ & $-1.02^{+0.24}_{-0.26}$ & $-0.93^{+0.13}_{-0.15}$ \\
{\boldmath$w_{0}$} & --- & --- & $-0.94^{+0.24}_{-0.27}$ & $-0.98^{+0.16}_{-0.21}$ \\
{\boldmath$\Sigma m_\nu$} \textbf{[eV]} & $< 0.264$ & $< 0.163$ & $< 0.267$ & $< 0.163$ \\
{\boldmath$H_0$} & $78\pm20$ & $66.3^{+7.2}_{-5.8}$ & $67.8^{+2.4}_{-2.7}$ & $67.8\pm1.2$ \\
{\boldmath$\sigma_8$} & $0.90\pm0.14$ & $0.803^{+0.064}_{-0.054}$ & $0.813^{+0.029}_{-0.032}$ & $0.815^{+0.019}_{-0.021}$ \\
{\boldmath$\Omega_m$} & $0.243^{+0.12}_{-0.099}$ & $0.327^{+0.059}_{-0.067}$ & $0.313^{+0.029}_{-0.027}$ & $0.311^{+0.012}_{-0.011}$ \\
\hline
\end{tabular}
\caption{95 \% C.L. bounds on the nodes used for the reconstruction of $w_\mathrm{DE}$ using the PCHIP method, in the case in which $\sum m_\nu$ is free in the MCMC. We also show the bounds on the parameters that exhibit tensions in the base $\Lambda$CDM model. We fix in all cases $w_\mathrm{ini} = w_\mathrm{mre}=-1$.}
\label{tab:case2}
\end{table}

\begin{table}[t]
\centering
\small\addtolength{\tabcolsep}{+3pt}
\def\arraystretch{1.25}
\begin{tabular}{ccccc}
\multicolumn{5}{c}{$\boldsymbol{w_\mathrm{DE,\,NP} + \sum m_{\nu,\,\mathrm{PCHIP}}}$} \\ \hline
 & \textbf{CMB} & \textbf{CMB+BAO} & \textbf{CMB+SN} & \textbf{CMB+BAO+SN} \\
\hline
{\boldmath$w_{rec}$} & --- & --- & --- & --- \\
{\boldmath$w_{2.4}$} & $< -0.619$ & $< -0.823$ & $< -0.651$ & $< -0.819$ \\
{\boldmath$w_{1.2}$} & $< -0.421$ & $< -0.887$ & $< -0.605$ & $< -0.882$ \\
{\boldmath$w_{0.6}$} & $< -0.524$ & $< -0.928$ & $< -0.909$ & $< -0.931$ \\
{\boldmath$w_{0}$} & --- & $< -0.462$ & $< -0.870$ & $< -0.835$ \\
{\boldmath$\Sigma m_\mathrm{ini}$} \textbf{[eV]} & $< 0.763$ & $< 0.958$ & $< 0.812$ & $< 0.946$ \\
{\boldmath$\Sigma m_\mathrm{rec}$} \textbf{[eV]} & $< 0.323$ & $< 0.326$ & $< 0.338$ & $< 0.327$ \\
{\boldmath$\Sigma m_{2.4}$} \textbf{[eV]} & $< 0.732$ & $< 0.0764$ & $< 0.281$ & $< 0.0768$ \\
{\boldmath$\Sigma m_{1.2}$} \textbf{[eV]} & $< 0.972$ & $< 0.285$ & $< 0.493$ & $< 0.287$ \\
{\boldmath$\Sigma m_{0.6}$} \textbf{[eV]} & $< 0.863$ & $<0.469$ & $< 0.454$ & $0.27^{+0.20}_{-0.26}$ \\
{\boldmath$\Sigma m_{0}$} \textbf{[eV]} & $< 0.778$ & $< 0.203$ & $< 0.323$ & $< 0.204$ \\
{\boldmath$H_0$} & $55^{+7}_{-6}$ & $61.0\pm5.6$ & $60.4\pm5.0$ & $62.0^{+5.0}_{-5.1}$ \\
{\boldmath$\sigma_8$} & $0.687\pm0.069$ & $0.743^{+0.050}_{-0.051}$ & $0.743^{+0.048}_{-0.047}$ & $0.752^{+0.045}_{-0.047}$ \\
{\boldmath$\Omega_m$} & $0.430^{+0.10}_{-0.090}$ & $0.325^{+0.031}_{-0.024}$ & $0.339^{+0.024}_{-0.023}$ & $0.313^{+0.011}_{-0.010}$ \\
\hline
\multicolumn{5}{c}{$\boldsymbol{w_\mathrm{DE,\,P} + \sum m_{\nu,\,\mathrm{PCHIP}}}$} \\ \hline
{\boldmath$w_{rec}$} & --- & --- & --- & --- \\
{\boldmath$w_{2.4}$} & $< -0.466$ & $< -0.656$ & $< -0.534$ & $< -0.640$ \\
{\boldmath$w_{1.2}$} & $< -0.337$ & $< -1.19$ & $< -0.614$ & $< -1.22$ \\
{\boldmath$w_{0.6}$} & $< -0.823$ & $-1.01^{+0.25}_{-0.23}$ & $-1.10^{+0.28}_{-0.29}$ & $-0.95^{+0.13}_{-0.17}$ \\
{\boldmath$w_{0}$} & --- & --- & $-0.94^{+0.26}_{-0.27}$ & $-0.99^{+0.15}_{-0.21}$ \\
{\boldmath$\Sigma m_\mathrm{ini}$} \textbf{[eV]} & $< 0.792$ & $< 0.823$ & $< 0.773$ & $< 0.818$ \\
{\boldmath$\Sigma m_\mathrm{rec}$} \textbf{[eV]} & $< 0.414$ & $< 0.338$ & $< 0.388$ & $< 0.338$ \\
{\boldmath$\Sigma m_{2.4}$} \textbf{[eV]} & $< 0.577$ & $< 0.289$ & $< 0.620$ & $< 0.281$ \\
{\boldmath$\Sigma m_{1.2}$} \textbf{[eV]} & $< 0.819$ & $< 0.547$ & $< 0.835$ & $< 0.527$ \\
{\boldmath$\Sigma m_{0.6}$} \textbf{[eV]} & $< 0.704$ & $< 0.565$ & $< 0.769$ & $< 0.562$ \\
{\boldmath$\Sigma m_{0}$} \textbf{[eV]} & $< 0.719$ & $< 0.479$ & $< 0.737$ & $< 0.477$ \\
{\boldmath$H_0$} & $78^{+30}_{-20}$ & $68^{+20}_{-10}$ & $68\pm10$ & $70\pm10$ \\
{\boldmath$\sigma_8$} & $0.88^{+0.20}_{-0.18}$ & $0.80^{+0.12}_{-0.11}$ & $0.80^{+0.11}_{-0.10}$ & $0.817^{+0.097}_{-0.10}$ \\
{\boldmath$\Omega_m$} & $0.26^{+0.13}_{-0.11}$ & $0.333^{+0.059}_{-0.069}$ & $0.327^{+0.038}_{-0.034}$ & $0.315\pm0.013$ \\
\hline
\end{tabular}
\caption{95 \% C.L. bounds on the nodes used for the reconstruction of both $w_\mathrm{DE}$ and $\sum m_\nu$ using the PCHIP method. We also show the bounds on the parameters that exhibit tensions in the base $\Lambda$CDM model. We fix in all cases $w_\mathrm{ini} = w_\mathrm{mre}=-1$ and $\sum m_\mathrm{\nu,\,ini} = \sum m_\mathrm{\nu,\,mre}$.}
\label{tab:case3}
\end{table}

\end{document}